%% file: manuscript.tex
\documentclass[11pt,technote,draftclsnofoot,onecolumn,oneside]{IEEEtran}
\usepackage{amsmath}
\usepackage{graphicx}
\usepackage{subfigure}
\usepackage{amsfonts}
\usepackage{amssymb}
\usepackage{url}
\usepackage{color}
\usepackage{cite}

\def\b#1{\mathchoice{\mbox{\boldmath$\displaystyle#1$}}
                    {\mbox{\boldmath$\textstyle#1$}}
                    {\mbox{\boldmath$\scriptstyle#1$}}
                    {\mbox{\boldmath$\scriptscriptstyle#1$}}}

\newcommand{\bm}[1]{\mbox{\boldmath $#1$}}
\newcommand{\bo}[1]{{\bf #1}}

\title{Hyperspectral Unmixing Overview: \\Geometrical, Statistical, and Sparse Regression-Based Approaches}

\author{Jos\'e M. Bioucas-Dias,  Antonio Plaza, Nicolas Dobigeon, \\Mario Parente,  Qian Du, Paul Gader and Jocelyn Chanussot\\
\thanks{J. M. Bioucas-Dias is with Instituto de Telecomunica\c{c}\~oes, Instituto Superior T\'ecnico, Technical University of Lisbon, 1049-1 Lisbon, Portugal (e-mail: bioucas@lx.it.pt).}
\thanks{A. Plaza is with Hyperspectral Computing Laboratory, Department of Technology of Computers and Communications, University of Extremadura, 10003 Caceres, Spain (email: aplaza@unex.es).}
\thanks{N. Dobigeon is with University of Toulouse, IRIT/INP-ENSEEIHT/TeSA, 31071 Toulouse Cedex 7, France (email: Nicolas.Dobigeon@enseeiht.fr).}
\thanks{M. Parente is with Department of  Electrical \& Computer Engineering, University of Massachusetts, Amherst, MA 01030 USA (email: mparente@ecs.umass.edu).}
\thanks{Q. Du is with Department of  Electrical \& Computer Engineering, Mississippi State University, Mississippi State, MS 39762 USA (email:du@ece.msstate.edu).}
\thanks{P. Gader is with Department of Computer \& Information Science \& Engineering, University of Florida, Gainesville, FL 32611 USA and GIPSA-Lab, Grenoble Institute of Technology, 38402 Grenoble, France
(email: pgader@cise.ufl.edu).}
\thanks{J. Chanussot is with GIPSA-Lab, Grenoble Institute of Technology, 38402 Grenoble, France (email:jocelyn.chanussot@gipsa-lab.grenoble-inp.fr).}
 }

\begin{document}
 \maketitle

\begin{abstract}

Imaging spectrometers measure  electromagnetic energy scattered  in  their instantaneous field view in hundreds or thousands of spectral channels with higher spectral resolution than \textit{multispectral} cameras. Imaging spectrometers are therefore often referred to as \textit{hyperspectral cameras} (HSCs).  Higher spectral resolution enables material identification via spectroscopic analysis, which facilitates countless applications that require identifying materials in scenarios unsuitable for classical spectroscopic analysis.  Due to low spatial resolution of HSCs, microscopic material mixing, and multiple scattering, spectra measured by HSCs are mixtures of spectra of materials in a scene.  Thus, accurate estimation requires unmixing.  Pixels are assumed to be mixtures of a few materials, called \textit{endmembers}.  Unmixing involves estimating all or some of:  the number of endmembers, their spectral signatures, and their \textit{abundances} at each pixel.  Unmixing is a challenging, ill-posed inverse problem because of model inaccuracies, observation noise, environmental conditions, endmember variability, and data set size.  Researchers have devised and investigated many models searching for robust, stable, tractable, and accurate unmixing algorithms.   This paper presents an overview of unmixing methods from the time of Keshava and Mustard's unmixing tutorial \cite{bi:Keshava_02}  to the present. Mixing models are first discussed. Signal-subspace, geometrical, statistical,  sparsity-based, and spatial-contextual unmixing algorithms are described.  Mathematical problems and potential solutions are described.  Algorithm characteristics are illustrated experimentally.
\end{abstract}


\input{introduction.tex}
\input{lmm.tex}

\input{subspaceIdentification}
\input{geo_unmixing}
\input{stat_unmixing_ND}
\input{sparse_unmixing}
\input{spatial_ND_v2}

\section{Summary}
More than one decade after Keshava and Mustard's tutorial paper on spectral unmixing published in the IEEE Signal Processing Magazine \cite{bi:Keshava_02}, effective spectral unmixing still remains an elusive exploitation goal and a very active research topic in the remote sensing community. Regardless of the available spatial resolution of remotely sensed data sets, the spectral signals collected in natural environments are invariably a mixture of the signatures of the various materials found within the spatial extent of the ground instantaneous field view of the remote sensing imaging instrument. The availability of hyperspectral imaging instruments with increasing spectral resolution (exceeding the number of spectral mixture components) has fostered many developments in recent years. In order to present the state-of-the-art and the most recent developments in this area, this paper provides an overview of recent developments in hyperspectral unmixing. Several main aspects are covered, including mixing models (linear versus nonlinear), signal subspace identification, geometrical-based spectral unmixing, statistical-based spectral unmixing, sparse regression-based unmixing and the integration of spatial and spectral information for unmixing purposes. In each topic, we describe the physical or mathematical problems involved and many widely used algorithms to address these problems. Because of the high level of activity and limited space, there are many methods that have not been addressed directly in this manuscript.  However, combined, the topics mentioned here provide a snapshot of the state-of-the-art in the area of spectral unmixing, offering a perspective on the potential and emerging challenges in this strategy for hyperspectral data interpretation. The compendium of techniques presented in this work reflects the increasing sophistication of a field that is rapidly maturing at the intersection of many different disciplines, including signal and image processing, physical modeling, linear algebra and computing developments.

In this regard, a recent trend in hyperspectral imaging in general (and spectral unmixing in particular) has been the computationally efficient implementation of techniques using high performance computing (HPC) architectures \cite{plazaspm2011}. This is particularly important to address applications of spectral unmixing with high societal impact such as, monitoring of natural disasters (e.g., earthquakes and floods) or tracking of man-induced hazards (e.g., oil spills and other types of chemical contamination). Many of these applications require timely responses for swift decisions which depend upon (near) real-time performance of algorithm analysis \cite{plaza2007hpc}. Although the role of different types of HPC architectures depends heavily on the considered application, cluster-based parallel computing has been used for efficient information extraction from very large data archives using spectral unmixing technniques \cite{plazacbir2011}, while on-board and real-time hardware architectures such as field programmable gate arrays (FPGAs) \cite{plazafpga2012} and graphics processing units (GPUs) \cite{plazagpu2011} have also been used for efficient implementation and exploitation of spectral unmixing techniques.  The HPC techniques, together with the recent discovery of theoretically correct methods for parallel Gibbs samplers and further coupled with the potential of the fully stochastic models represents an opportunity for huge advances in multi-modal unmixing.  That is, these developments offer the possibility that complex hyperspectral images that contain that can be piecewise linear and nonlinear mixtures of endmembers that are represented by distributions and for which the number of endmembers in each piece varies, may be accurately processed in a practical time.

There is a great deal of work yet to be done; the list of ideas could be several pages long!  A few directions are mentioned here.  Proper representations of endmember distributions need to be identified.  Researchers have considered some distributions but not all.  Furthermore, it may become necessary to include distributions or tree structured representations into sparse processing with libraries.  As images cover larger and larger areas, piecewise processing will become more important since such images will cover several different types of areas.  Furthermore, in many of these cases, linear and nonlinear mixing will both occur.  Random fields that combine spatial and spectral information, manifold approximations by mixtures of low rank Gaussians, and model clustering are all methods that can be investigated for this purpose.   Finally, software tools and measurements for large scale quantitative analysis are needed to perform meaningful statistical analyses of algorithm performance.

\section{Acknowledgements}

The authors acknowledge Robert O. Green and the AVIRIS team for making the
Rcuprite hyperspectral data set available to the community, and the United
States Geological Survey (USGS) for their publicly available library of
mineral signatures. The authors also acknowledge the Army Geospatial Center,
US Army Corps of Engineers, for making the HYDICE Rterrain data set
available to the community.

%
%
%
%

\vfill \vfill \vfill

\bibliographystyle{IEEEtran}
\bibliography{shortcuts,references}

\end{document}

%% file: introduction.tex

\section{Introduction}
Hyperspectral cameras \cite{bi:Smith_85, bi:Adams_86, bi:Gillespie_90, bi:Vane_93, art:clark:93, art:Green:98, bi:Shaw_02, bi:Landgrebe_02,bi:Keshava_02, bi:Manolakis_02, bi:Stein_02} contribute significantly to earth observation and remote sensing\cite{bi:Plazarse_09, bi:Schaepmanrse_09}. Their potential motivates the development of small, commercial,  high spatial and spectral resolution instruments. They have also been used in food safety \cite{art:Berman:02, art:Gowen07, art:Mahest:08, larsen2009kernel},  pharmaceutical process monitoring and quality control \cite{art:Kim01, art:Rodionova05, art:Gendrin08, art:Juan:09, art:Lopes:10}, and biomedical, industrial, and biometric, and forensic applications \cite{bi:Begelman_09, bi:Akbari_10, bi:Picon_09, bi:Chang_10, bi:Brewer_08}.

\begin{figure*}
\centering
\includegraphics[width=10cm]{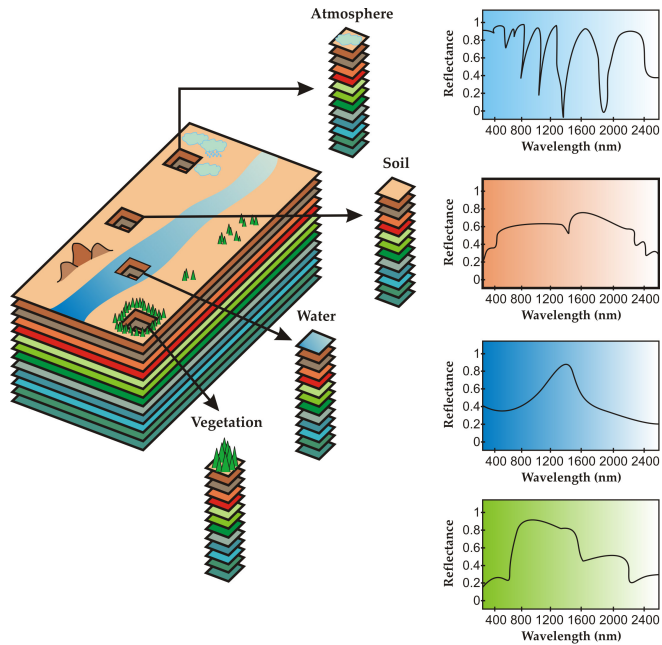}
\caption{Hyperspectral imaging concept.}
\label{fig:hy_imgaing_concept}
\end{figure*}

HSCs can be built to function in many regions of the electro-magnetic spectrum.  The focus here is on those covering the visible, near-infrared, and shortwave infrared spectral bands (in the range $0.3{\mu}m$ to $2.5{\mu}m$ \cite{bi:Vane_93}). Disregarding atmospheric effects, the signal recorded by an HSC at a pixel is a mixture of light scattered by substances located in the field of view \cite{bi:Adams_86}. Fig. \ref{fig:hy_imgaing_concept} illustrates the measured data. They are organized into planes forming a data cube. Each plane corresponds to radiance acquired over a spectral band for  all pixels. Each spectral vector corresponds to the radiance acquired at a given location for all spectral bands.

\begin{figure*}
\centering
\includegraphics[width=12cm]{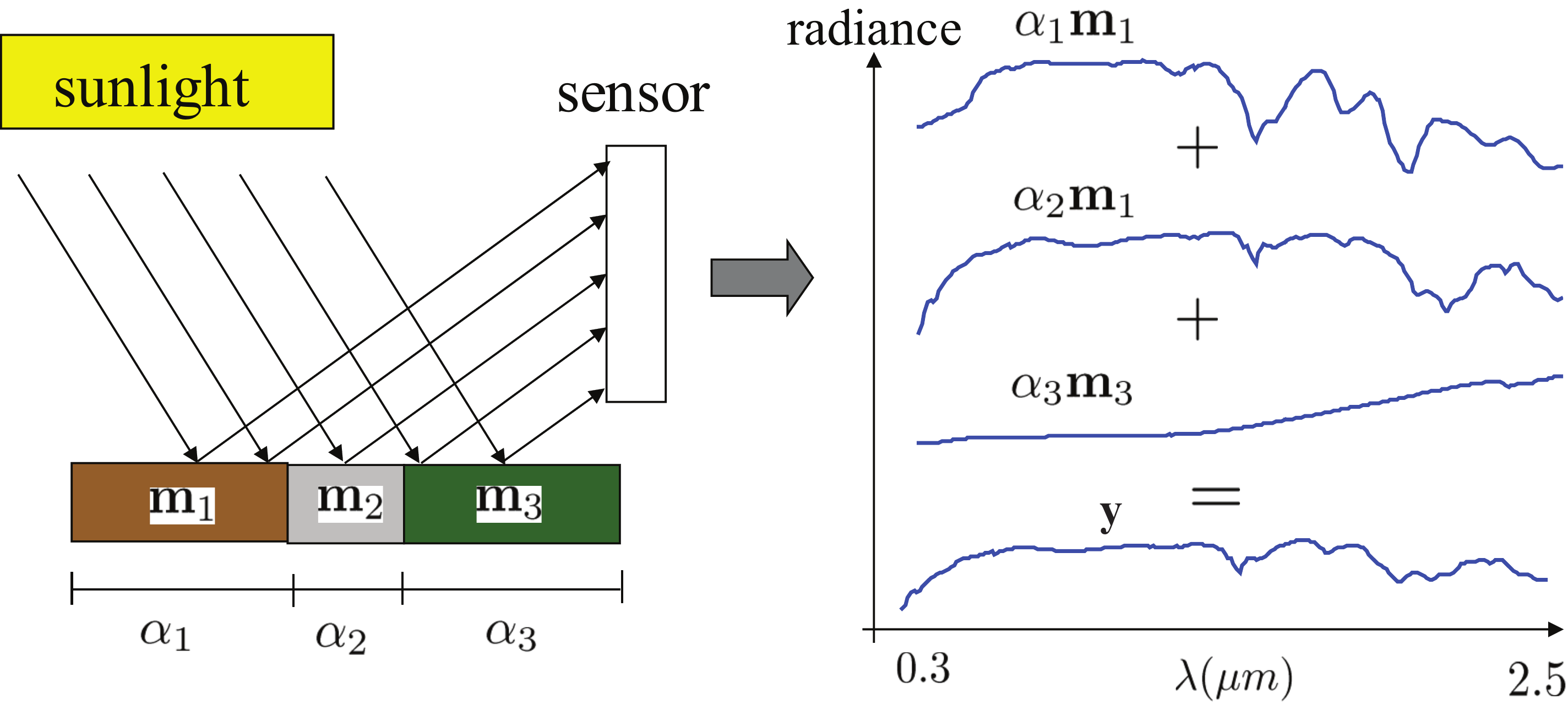}
\caption{Linear mixing . The measured radiance at a pixel
is a weighted average of the radiances of the materials present at the pixel.}
\label{fig:squematic_mix}
\end{figure*}

\subsection{Linear and nonlinear mixing models}

\begin{figure*}
\centering
\includegraphics[width=12cm]{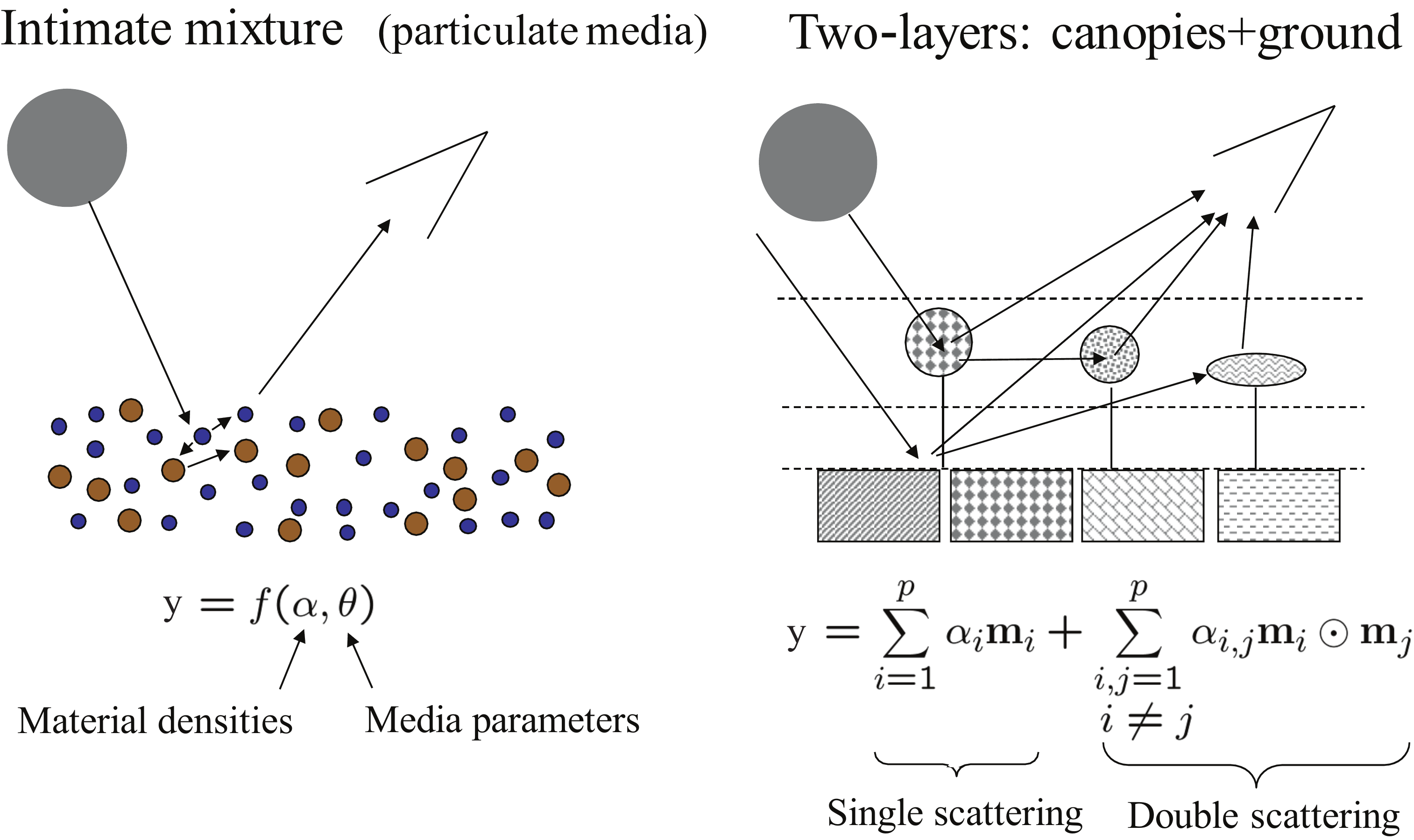}
\caption{Two nonlinear mixing scenarios. Left hand: intimate mixture; Right
hand:  multilayered scene.}
\label{fig:non_lin_mix}
\end{figure*}

Hyperspectral unmixing (HU) refers to any process that separates the pixel spectra from a hyperspectral image into
a collection of constituent spectra, or spectral signatures, called \textit{endmembers} and
a set of fractional \textit{abundances}, one set per pixel.  The endmembers are generally assumed to represent the \textit{pure} materials present in the image and the set of abundances, or simply abundances, at each pixel to represent the percentage of each endmember that is present in the pixel.

There are a number of subtleties in this definition.  First, the
notion of a pure material can be subjective and problem dependent.
For example, suppose a hyperspectral image contains spectra measured
from bricks laid on the ground, the mortar between the bricks, and
two types of plants that are growing through cracks in the brick.
One may suppose then that there are four endmembers.  However, if
the percentage of area that is covered by the mortar is very small
then we may not want to have an endmember for mortar.  We may just
want an endmember for ``brick''.  It depends on if we have a need to
directly measure the proportion of mortar present.  If we have need to measure the mortar, then we
may not care to distinguish between the plants since they may have
similar signatures.  On the other hand, suppose that one type of
plant is desirable and the other is an invasive plant that needs to
be removed.  Then we may want two plant endmembers.  Furthermore,
one may only be interested in the chlorophyll present in the entire
scene.  Obviously, this discussion can be continued \textit{ad
nauseum} but it is clear that the definition of the endmembers can depend upon the application.

The second subtlety is with the proportions.  Most researchers
assume that a proportion represents the percentage of material
associated with an endmember present in the part of the scene imaged
by a particular pixel.  Indeed, Hapke \cite{bi:Hapke} states that
the abundances in a linear mixture represent the relative area of
the corresponding endmember in an imaged region.  Lab experiments
conducted by some of the authors have confirmed this in a laboratory
setting.  However, in the nonlinear case, the situation is not as
straightforward.   For example, calibration objects can sometimes be
used to map hyperspectral measurements to reflectance, or at least
to relative reflectance.  Therefore, the coordinates of the
endmembers are approximations to the reflectance of the material,
which we may assume for the sake of argument to be accurate.   The
reflectance is usually not a linear function of the mass of the
material nor is it a linear function of the cross-sectional area of
the material.    A highly reflective, yet small object may
dominate a much larger but dark object at a pixel, which may lead to
inaccurate estimates of the amount of material present in the region
imaged by a pixel, but accurate estimates of  the contribution of
each material to the reflectivity measured at the pixel.  Regardless
of these subtleties, the large number of applications of
hyperspectral research in the past ten years indicates that current
models have value.

Unmixing algorithms currently rely on the expected type of mixing.  Mixing models can be  characterized as either linear or nonlinear
\cite{bi:Liangrocapart_98,bi:Keshava_02}. Linear mixing  holds when
the mixing scale is macroscopic \cite{bi:Singer_79} and the incident
light interacts with just one material, as is the case in
checkerboard type scenes \cite{bi:Hapke_81,bi:Clark_84}. In this
case, the mixing occurs within the instrument itself.  It is due to
the fact that the resolution of the instrument is not fine enough.
The light from the materials, although almost completely separated,
is mixed within the measuring instrument.  Fig. \ref{fig:squematic_mix} depicts linear mixing: Light scattered by three materials in a scene is incident on a detector that measures radiance in $B$ bands. The measured spectrum ${\bf y}\in \mathbb{R}^B$ is a weighted average of the material spectra. The relative amount of each material is represented by the associated weight.

Conversely, nonlinear mixing is usually due to physical interactions
between the light scattered by multiple materials in the scene.
These interactions can be at a \textit{classical}, or
\textit{multilayered}, level or at a \textit{microscopic}, or
\textit{intimate}, level.  Mixing at the classical level occurs when
light is scattered from one or more objects, is reflected off
additional objects, and eventually is measured by hyperspectral
imager.  A nice illustrative
derivation of a multilayer model is given by Borel and Gerstl
\cite{bi:Borel_94} who show that the model results in an infinite
sequence of powers of products of reflectances.  Generally, however,
the first order terms are sufficient and this leads to the bilinear
model.  Microscopic mixing occurs when two materials are
homogeneously mixed \cite{bi:Hapke}.  In this case, the interactions
consist of photons emitted from molecules of one material are
absorbed by molecules of another material, which may in turn emit
more photons.    The mixing is modeled by Hapke as occurring at the
\textit{albedo} level and not at the reflectance level.  The
apparent albedo of the mixture is a linear average of the albedos of
the individual substances but the reflectance is a nonlinear
function of albedo, thus leading to a different type of nonlinear
model.

Fig. \ref{fig:non_lin_mix} illustrates two
non-linear mixing scenarios: the left-hand panel represents an
intimate mixture, meaning that the materials are in close proximity;
the right-hand panel illustrates a multilayered scene, where there
are multiple interactions among the scatterers at the different
layers.

Most of this overview is devoted to the linear mixing model. The
reason is that, despite its simplicity, it  is an acceptable
approximation of the light scattering mechanisms in many real
scenarios. Furthermore, in contrast to nonlinear mixing, the linear mixing model is the basis of a plethora of unmixing models and algorithms spanning back at least 25 years.  A sampling can be found in \cite{bioucasplaza, art:PlazaRecentDevelop10,conf:Parente:Wispers:10,art:plaza:MPP:TGRS:04,
bi:Shaw_03,bi:Keshava_02,bi:Keshava_00,bi:Hu_99,bi:Petrou_99, bi:Settle_96,bi:Mazer_88,bi:Yuhas_92,bi:Chang_94,bi:Chang_98,bi:Heinz_99}).  Others will be discussed throughout the rest of this paper.

\subsection{Brief overview of nonlinear approaches}

Radiative transfer theory (RTT)  \cite{book:Chandrasekhar} is a well established mathematical model for the transfer of energy
as photons interacts with the materials in the scene. A complete physics based approach to nonlinear unmixing would require inferring the spectral
signatures and  material densities based on the RTT.  Unfortunately, this is an extremely complex
ill-posed problem, relying on scene parameters very hard or impossible to
obtain. The Hapke \cite{bi:Hapke_81}, Kulbelka-Munk
\cite{art:kulberka31} and Shkuratov \cite{art:shkuratov99}
scattering  formulations  are three approximations  for the
analytical solution to the  RTT.  The former has been widely used to
study diffuse reflection spectra in chemistry \cite{art:myrick11}
whereas the  later two have been used, for example, in mineral
unmixing applications \cite{bi:Keshava_02, art:poulet10}.

One wide class of strategies is aimed at avoiding the complex physical models using simpler but physics inspired models, such kernel methods. In
\cite{Broadwater2007igarss} and following works
\cite{art:broadwater09,Broadwater2009whispers,Broadwater2010whispers,Broadwater2011whispers},
Broadwater \emph{et al.} have proposed several kernel-based unmixing
algorithms to specifically account for intimate mixtures. Some of
these kernels are designed to be sufficiently flexible to allow
several nonlinearity degrees (using, e.g., radial basis functions or
polynomials expansions) while others are physics-inspired kernels
\cite{Broadwater2009whispers}.

Conversely, bilinear models have been successively proposed in
\cite{art:du08,Somers2009,art:fan09,art:nascimentoSPIE09,Halimi2011}
to handle scattering effects, e.g., occurring in the multilayered
scene. These models generalize the standard linear model by
introducing additional interaction terms. They mainly differ from
each other by the additivity constraints imposed on the mixing
coefficients \cite{Altmann2011whispers}.

However, limitations inherent to the unmixing algorithms that
explicitly rely on both models are twofold. Firstly, they are not
multipurpose in the sense that those developed to process intimate
mixtures are inefficient in the multiple interaction scenario (and
vice versa). Secondly, they generally require the prior knowledge of
the endmember signatures. If such information is not available,
these signatures have to be estimated from the data by using an
endmember extraction algorithm.

To achieve flexibility, some have resorted to machine learning
strategies such as neural networks
\cite{Guilfoyle2001,art:liu05,art:plaza09nl,art:plaza10nl,Altmann2011igarssRBF,
Licciardi_TGRS11, Licciardi_GRSL11}, to nonlinearly reduce dimensionality or learn model parameters
in a supervised fashion from a collection of examples
(see \cite{art:PlazaRecentDevelop10} and references therein). The
polynomial post nonlinear mixing model introduced in
\cite{Altmann2012ip} seems also to be sufficiently versatile to
cover a wide class of nonlinearities. However, again, these
algorithms assumes the prior knowledge or extraction of the
endmembers.

Mainly due to the difficulty of the issue, very few attempts have
been conducted to address the problem of fully unsupervised
nonlinear unmixing. One must still concede that a significant
contribution has been carried by Heylen \emph{et al} in
\cite{Heylen2011jstsp} where a strategy is introduced to extract
endmembers that have been nonlinearly mixed. The algorithmic scheme
is similar in many respects to the well-known N-FINDR algorithm \cite{bi:Winter_99}.
The key idea is to maximize the simplex volume computed with geodesic
measures on the data manifold. In this work, exact geodesic
distances are approximated by shortest-path distances in a
nearest-neighbor graph. Even more recently, same authors have shown
in \cite{Heylen2012grsl} that exact geodesic distances can be
derived on any data manifold induced by a nonlinear mixing model,
such as the generalized bilinear model introduced in
\cite{Halimi2011}.

Quite recently, Close and Gader have devised two methods for fully
unsupervised nonlinear unmixing in the case of intimate mixtures
\cite{RCloseTGRS11,RClose11}  based on Hapke's average albedo model
cited above.  One method assumes that each pixel is either linearly
or nonlinearly mixed.  The other assumes that there can be both
nonlinear and linear mixing present in a single pixel.  The methods
were shown to more accurately estimate physical mixing parameters
using measurements made by Mustard et al. \cite{  MustardPieters,
Guilfoyle2001, Broadwater2011whispers, Broadwater2010whispers} than
existing techniques.  There is still a great deal of work to be
done, including evaluating the usefulness of combining bilinear
models with average albedo models.

In summary, although researchers are beginning to expand more
aggressively into nonlinear mixing, the research is  immature
compared with linear mixing.  There has been a tremendous effort in
the past decade to solve linear unmixing problems and that is what
will be discussed in the rest of this paper.

\subsection{Hyperspectral unmixing processing chain}

\begin{figure}
\centering
\includegraphics[width=12cm]{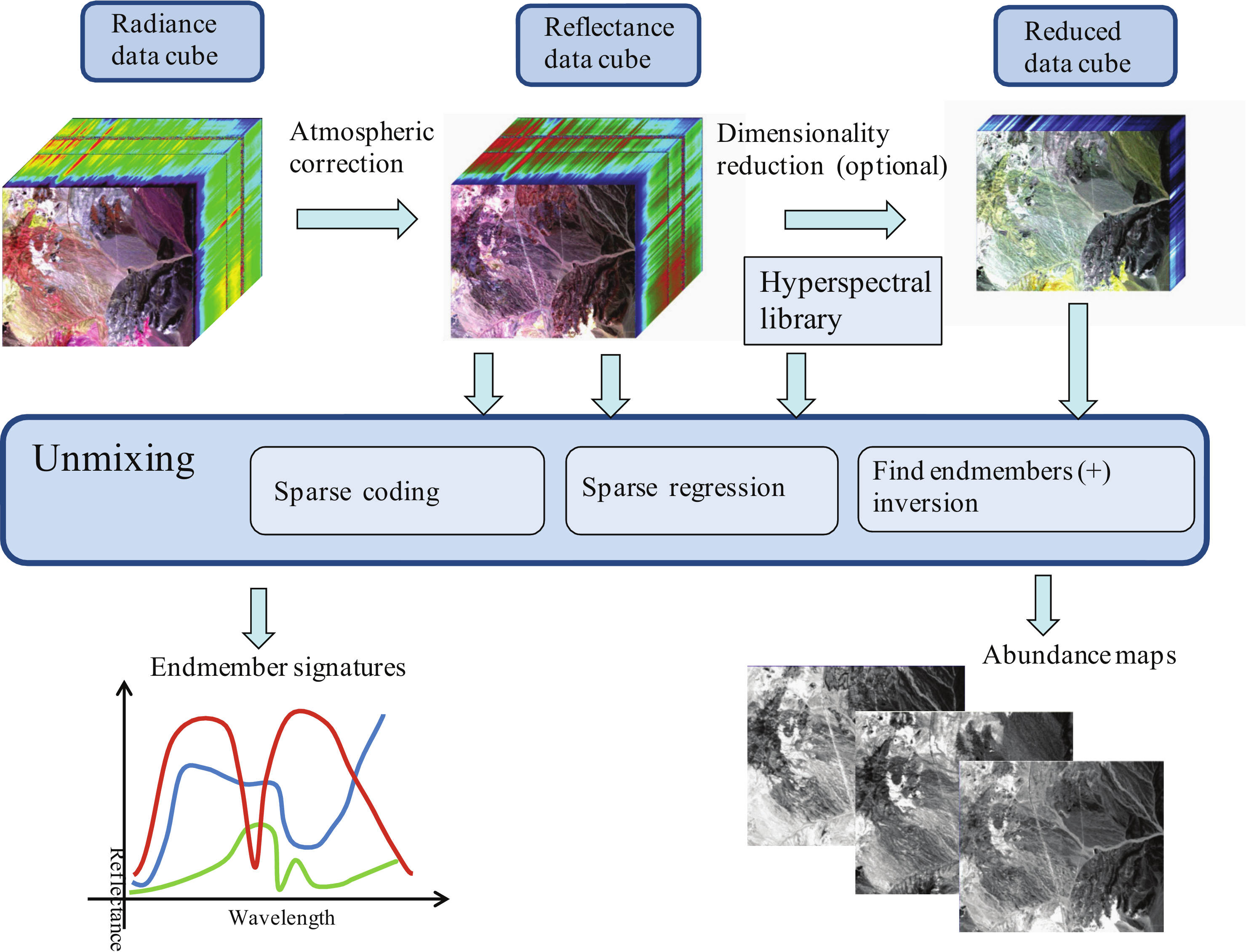}
\caption{Schematic diagram of the hyperspectral unmixing process.}
\label{fig:squemtic_view_lin_unmix}
\end{figure}

Fig. \ref{fig:squemtic_view_lin_unmix} shows the processing steps usually involved in the hyperspectral unmixing chain: atmospheric correction, dimensionality reduction, and unmixing, which may be tackled via the classical  endmember determination plus inversion,
or via sparse regression or sparse coding approaches. Often, endmember determination and inversion are implemented simultaneously. Below, we provide a brief characterization of  each of these steps:

\begin{enumerate}

\item \textbf{Atmospheric correction}. The atmosphere attenuates and scatterers the light and therefore affects the  radiance at the sensor. The atmospheric correction compensates for {\color{black} these} effects by converting radiance into reflectance,  which is an intrinsic property of the materials. We stress, however, that linear unmixing can
be carried out directly on radiance data.

\item \textbf{Data reduction}. The dimensionality of the space spanned by spectra from an image is generally much lower than available number of bands. Identifying appropriate subspaces facilitates dimensionality reduction, improving algorithm performance and complexity and data storage. Furthermore, if the linear mixture model is accurate, the signal subspace dimension is one less than equal to the number of endmembers, a crucial figure in hyperspectral unmixing.

\item \textbf{Unmixing}. The unmixing step consists of identifying the endmembers in the scene and the fractional abundances at each pixel.  Three general approaches will be discussed here.  \textit{Geometrical} approaches exploit the fact that linearly mixed vectors are in a simplex set or in a positive cone.  \textit{Statistical} approaches focus on using parameter estimation techniques to determine endmember and abundance parameters.  \textit{Sparse regression} approaches, which formulates unmixing as a linear sparse regression problem, in a fashion similar to that of \emph{compressive sensing} \cite{art:candes06, art:donoho06}.  This framework relies on the existence of spectral libraries usually acquired in laboratory. A step forward, termed \emph{sparse coding} \cite{art:olshausen96},  consists of learning the dictionary from the data and, thus, avoiding not only the need of libraries but also calibration issues related to different conditions under which the libraries and the data were acquired.

\item \textbf{Inversion}. Given the observed spectral vectors and the identified endmembers, the inversion step consists of solving a constrained optimization problem which minimizes the residual between the observed spectral vectors and {\color{black} the} linear space spanned by the inferred spectral signatures; the implicit fractional abundances are, very often, constrained to be nonnegative and to sum to one ({\em i.e.}, they belong to the probability simplex).  There are, however, many  hyperspectral unmixing approaches in which the  endmember determination and inversion steps are implemented simultaneously.

\end{enumerate}

The remainder of the paper is organized as follows. Section 2 describes the linear spectral mixture model adopted as the baseline model in this contribution. Section 3 describes techniques for subspace identification. Sections 4, 5, 6, 7 describe  four classes of  techniques for endmember and fractional abundances estimation under the linear spectral unmixing. Sections 4 and 5 are devoted to
the longstanding geometrical and statistical  based  approaches, respectively.  Sections 6 and 7 are devoted to
the recently introduced  sparse regression based unmixing and  to the exploitation of the spatial contextual information, respectively. Each of these sections introduce the underlying  mathematical problem and summarizes state-of-the-art algorithms to address such problem.

Experimental results obtained from simulated and real data sets illustrating the potential and limitations of each class of algorithms are described. The experiments do not constitute an exhaustive comparison.  Both code and data for all the experiments described here are available at \url{http://www.lx.it.pt/~bioucas/code/unmixing_overview.zip}.
The paper concludes with a summary and discussion of plausible future developments in the area of spectral unmixing.

%% file: lmm.tex

\section{Linear mixture model}

\label{sec:lmm}

If the multiple scattering among distinct endmembers is negligible and
the surface is partitioned according to the fractional abundances, as illustrated in
Fig. \ref{fig:squematic_mix}, then the spectrum of each pixel is well approximated by a linear mixture of endmember
spectra weighted by the corresponding fractional abundances
\cite{bi:Adams_86,bi:Liangrocapart_98, bi:Keshava_00,bi:Keshava_02}.  In this case, the spectral measurement\footnote{Although the type of spectral quantity (radiance, reflectance, etc.) is important when processing data, specification is not necessary to derive the mathematical approaches.} at channel $i\in\{1,\dots,B \}$ ($B$ is the total number of channels) from a given pixel, denoted by $y_i$, is given by the \emph{linear mixing model} (LMM)
\begin{equation}
  \label{eq:channel_i_out}
  y_i=\sum_{j=1}^p\rho_{ij}\alpha_j+w_i,
\end{equation}
where $\rho_{ij}\geq 0$ denotes the spectral measurement of endmember  $j\in\{1,\dots,p\}$ at
$i^{th}$ the spectral band, $\alpha_j\geq 0$ denotes the fractional
abundance of endmember $j$, $w_i$ denotes an additive perturbation
({\em e.g.}, noise and modeling errors),  and $p$ denotes the number of endmembers.  At a given pixel, the fractional abundance $\alpha_j$, as the name suggests, represents the fractional area occupied by the $j$th endmember.  Therefore, the fractional abundances are subject to the following constraints:
\begin{equation}
    \label{eq:constraint}
    \begin{array}{ll}
      \text{Nonnegativity} & \alpha_j\geq 0, \: j=1,\, \dots, p\\
      \text{Sum-to-one} & \sum_{j=1}^p\alpha_j=1;
    \end{array}
\end{equation}
{\em i.e.,} the fractional abundance vector $\b{\alpha} \equiv [\alpha_1,
\alpha_2, \ldots ,\alpha_p]^T$ (the notation $(\cdot)^T$ indicates vector transposed)  is in
the standard  $(p-1)$-simplex (or unit $(p-1)$-simplex). In HU jargon, the
nonnegativity and the sum-to-one constraints are termed {\em abundance nonnegativity constraint} (ANC)
and {\em abundance sum constraint} (ASC), respectively.  Researchers may sometimes expect that the abundance fractions sum to less than one since an algorithm may not be able to account for every material in a pixel; it is not clear whether it is better to relax the constraint or to simply consider that part of the modeling error.


Let ${\bf y}\in\mathbb{R}^B$ denote a  $B$-dimensional column vector, and ${\bf m}_j \equiv [\rho_{1j}, \rho_{2j}, \ldots ,\rho_{Bj}]^T$ denote the spectral signature of the $j$th endmember. Expression (\ref{eq:channel_i_out}) can then be written as
\begin{equation}
      {\bf y}={\bf M} {\b\alpha} + {\bf w},
      \label{eq:sensor_M}
\end{equation}
where ${\bf M} \equiv [{\bf m}_1, {\bf m}_2, \ldots ,{\bf m}_p]$ is the
mixing  matrix containing the signatures of the endmembers present
in the covered area, and ${\bf w}\equiv [w_1,\ldots,w_B]^T$. Assuming that the
columns of $\bo{M}$ are affinely independent, {\em i.e.}, ${\bf m}_2-{\bf
m}_1,{\bf m}_3-{\bf m}_1, \dots,{\bf m}_p-{\bf m}_1$ are linearly
independent, then  the set
$$
   C\equiv\{{\bf y}={\bf M} {\b\alpha}\,:\,\sum_{j=1}^p\alpha_j=1, \:\:\alpha_j\geq 0, \:\: j=1,\, \ldots,
  ,\,p\}
$$
{\em i.e.}, the convex hull of the columns of $\bf M$, is a $(p-1)$-simplex in $\mathbb{R}^B$.
\begin{figure}[h]
    \centering
  \includegraphics[width=5cm,angle=0]{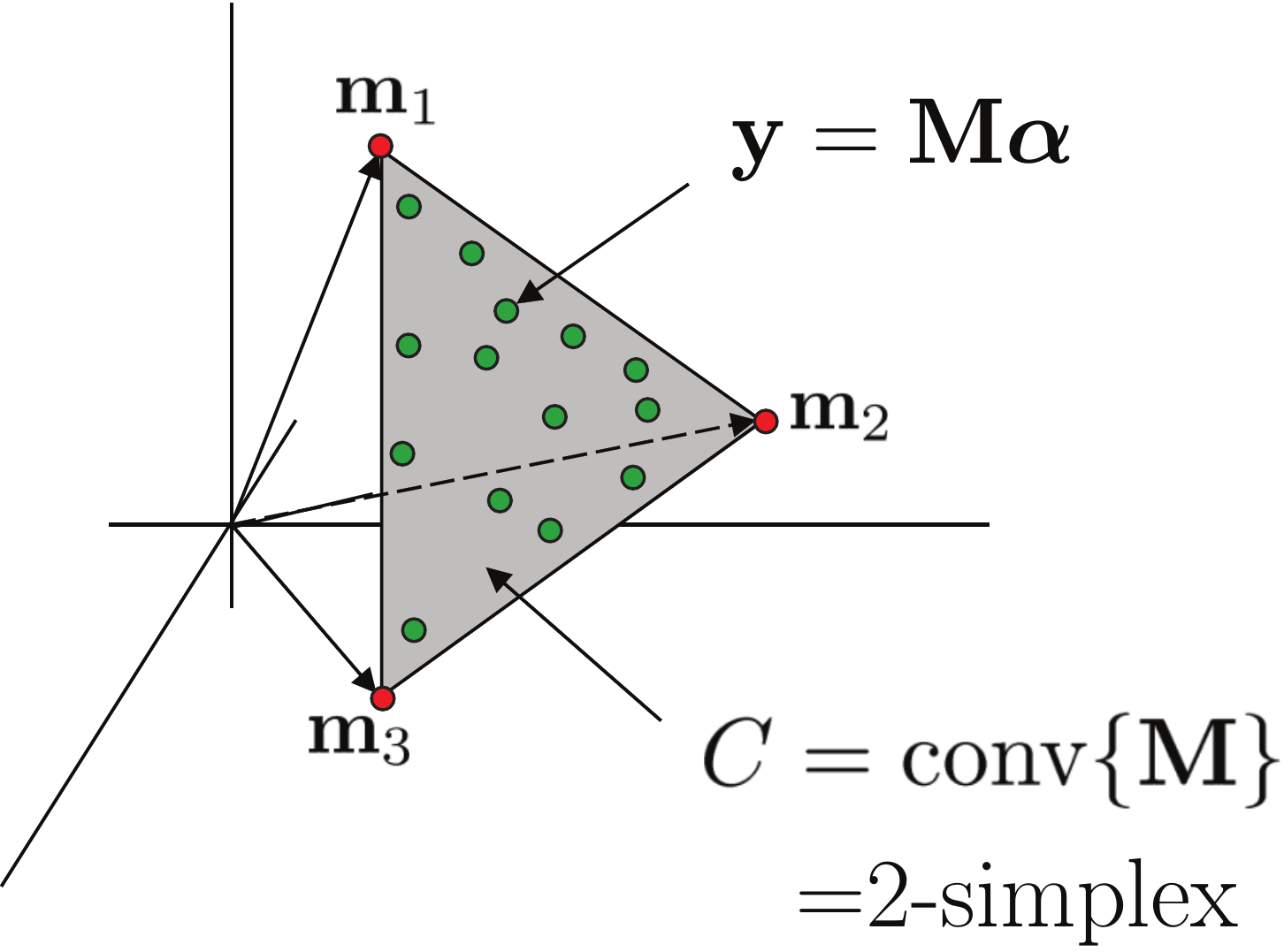}
  \caption{Illustration of the simplex set $C$ for p=3 ($C$ is the convex hull of the columns
  of $\bf M$, $C=\text{conv}\{{\bf M}\}$). Green circles  represent spectral vectors.  Red circles represent vertices  of the simplex and correspond to the endmembers.}
  \label{fig:simplex}
\end{figure}
Fig. \ref{fig:simplex} illustrates a 2-simplex  $C$ for a  hypothetical mixing
matrix $\bo{M}$ containing three endmembers. The points in green denote spectral vectors, whereas the points in red are vertices of the simplex and correspond to the endmembers.  Note that the inference of the mixing matrix $\bo{M}$ is equivalent to identifying the vertices of the simplex $C$. This geometrical  point of view, exploited by many unmixing algorithms, will be further {\color{black} developed} in Sections \ref{sec:geo_unmixing}.

\begin{figure*}[h]
\centering
\includegraphics[width=12cm]{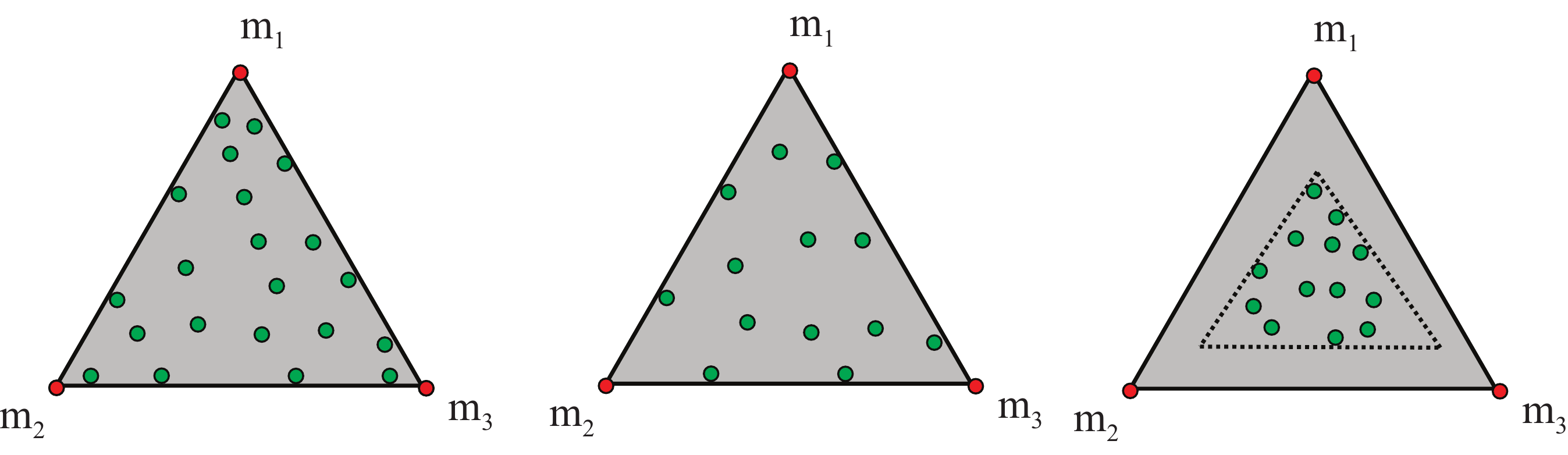}
\caption{Illustration of the concept of simplex of minimum volume
         containing the data for three  data sets. The endmembers in the left
          hand side and in the middle are identifiable by fitting a simplex of minimum volume to the data, whereas this is not applicable to the right hand side data set. The former data set correspond to a highly mixed scenario.}
\label{fig:simplices}
\end{figure*}

Since many algorithms  adopt
either a geometrical or a statistical framework \cite{bioucasplaza,conf:Parente:Wispers:10}, they are a focus of this paper.
To motivate these two directions, let us consider  the three data sets shown in
Fig. \ref{fig:simplices} generated under the linear model given in Eq. \ref{eq:sensor_M} where the noise is assumed to be negligible. The
spectral vectors generated according to Eq. (\ref{eq:sensor_M}) are in a simplex whose vertices
correspond to the endmembers.  The left hand side data set  contains pure pixels, {\em
i.e},  for any of the $p$ endmembers there is at least one pixel containing only the
correspondent material; the data set in the middle does not contain  pure pixels but
contains at least $p-1$ spectral vectors on each facet.  In both data sets (left and
middle), the endmembers may by inferred by fitting a minimum volume (MV)  simplex  to the
data; this  rather simple and yet powerful idea, introduced by Craig in his seminal work
\cite{bi:Craig_94}, underlies several geometrical based unmixing  algorithms. A similar
idea was  introduced in  1989 by Perczel in the area of Chemometrics
{\em et al.}\cite{Perczel89dichroism}.

The MV simplex shown in the right hand side example of Fig. \ref{fig:simplices} is
smaller than the true one. This situation corresponds to a highly mixed  data set where there are
no  spectral vectors near the facets. For these classes of problems, we usually resort to the statistical framework in which the estimation of the mixing matrix and of the fractional abundances are formulated as a statistical
 inference problem by adopting suitable probability models for the variables and
parameters involved, namely for the fractional abundances and for the mixing matrix.

\subsection{Characterization of the  Spectral Unmixing Inverse Problem}

Given the data set ${\bf Y}\equiv [{\bf y}_1,\dots,{\bf y}_n]\in \mathbb{R}^{B\times n}$ containing $n$ $B$-dimensional spectral vectors, the linear HU problem is, with reference to the linear model (\ref{eq:sensor_M}),  the estimation of the mixing matrix $\bf M$ and of the fractional abundances vectors $\bm{\alpha}_i$ corresponding to pixels $i=1,\dots,n$. This is often a difficult inverse problem, because the spectral signatures tend to be strongly correlated,  yielding badly-conditioned mixing matrices and, thus, HU estimates can be
highly sensitive to noise.
\begin{figure}[h]
    \centering
  \includegraphics[width=6cm,angle=0]{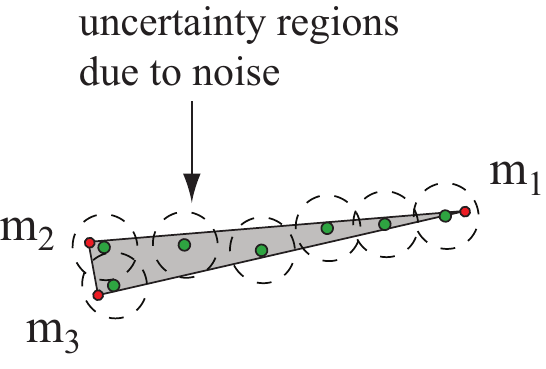}
  \caption{Illustration of a badly-conditioned mixing matrices and noise (represented by uncertainty  regions) centered on clean spectral vectors represented by green circles.}
  \label{fig:bad_cond_simplex}
\end{figure}
This scenario is illustrated in Fig. \ref{fig:bad_cond_simplex}, where endmembers ${\bf m}_2$ and ${\bf m}_3$ are very close, thus yielding a badly-conditioned matrix $\bf M$, and the effect of noise is represented by uncertainty  regions.

To characterize the linear HU inverse problem, we use the  \emph{signal-to-noise-ratio} (SNR)
\begin{equation*}
     \text{SNR} \equiv \frac{\mathbb{E}[\|{\bf x}\|^2]}{\mathbb{E}[\|{\bf w}\|^2]}
               =\frac{\text{trace}({\bf R}_x)}{\text{trace}({\bf R}_w)},
\end{equation*}
where  ${\bf R}_x$ and ${\bf R}_w$ are, respectively, the  signal  ({\em i.e.,} ${\bf x\equiv M}\bm{\alpha}$) and noise correlation matrices and $\mathbb{E}$ denotes expected value.  Besides SNR,  we  introduce  the
\emph{signal-to-noise-ratio spectral distribution} (SNR-SD) defined as
\begin{equation}
     \label{eq:SNR_SD}
     \text{SNR-SD}(i) = \frac{\lambda_{i,x}}{{\bf e}_{i,x}^T{\bf R}_w{\bf e}_{i,x}},\; i=1,\dots,p,
\end{equation}
where $(\lambda_{i,x}, {\bf e}_{i,x})$ is the $i^{th}$ eigenvalue-eigenvector couple of ${\bf R}_x$ ordered by  decreasing value of $\lambda_{i,x}$. The ratio $ \text{SNR-SD}(i)$ yields the signal-to-noise ratio (SNR) along the signal direction ${\bf e}_{i,x}$. Therefore, we  must have  $\text{SNR-SD}(i) \gg 1$ for $i=1,\dots,p$, in order to obtain acceptable unmixing results. Otherwise, there are directions in the signal subspace significantly corrupted by noise.


\begin{figure}[h]
    \centering
  \includegraphics[width=8cm,angle=0]{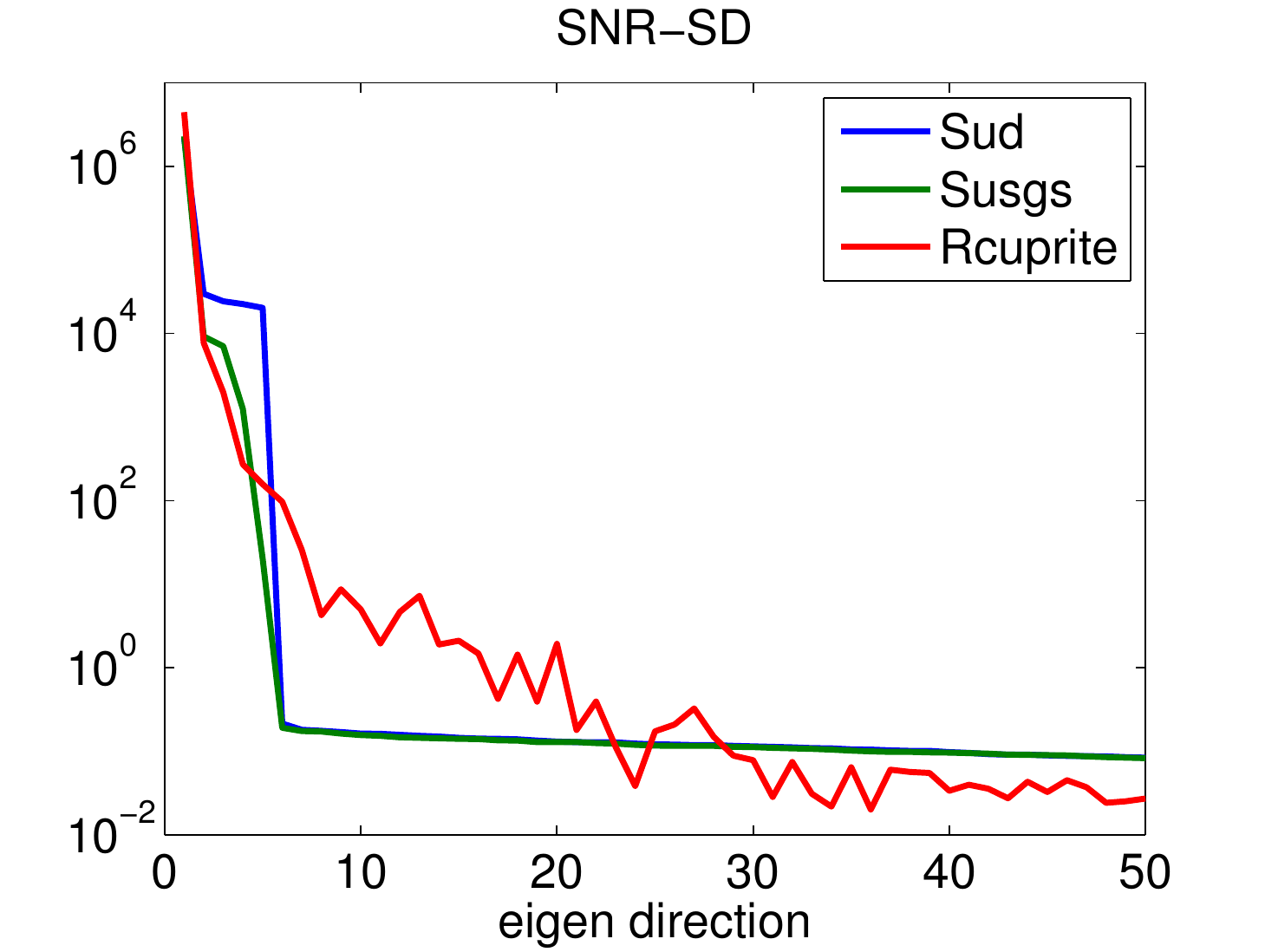}
  \caption{Signal-to-noise-ratio spectral distribution (SNR-SD)
   for the data sets SudP5SNR40, SusgsP5SNR40, and Rcuprite. The first two  are simulated and contain $p=5$ endmembers and the third  is a subset of the AVIRIS Cuprite data set.}
  \label{fig:snr_sd}
\end{figure}

Fig. \ref{fig:snr_sd}  plots  $\text{SNR-SD}(i)$, in the interval $i=1,\dots,50$, for the following data sets:
\begin{itemize}
  \item  {\bf SudP5SNR40}: simulated; mixing matrix $\bf M$ sampled from a uniformly distributed random variable in the interval $[0,1]$;  $p=5$; $n=5000$; fractional abundances distributed uniformly on the 4-unit simplex; SNR$\,=\,$40dB.
  \item {\bf SusgsP5SNR40}: simulated; mixing matrix $\bf M$ sampled from the United States Geological Survey (USGS) spectral library\footnote{Available online from: http://speclab.cr.usgs.gov/spectral-lib.html}; $p=5$; $n=5000$;  fractional abundances distributed uniformly on the 4-unit simplex;  SNR$\,=\,$40dB.
  \item {\bf Rcuprite}: real; subset of the well-known AVIRIS cuprite data cube\footnote{Available online from: http://aviris.jpl.nasa.gov/data/free\_data.html} with size 250 lines by 191 columns by 188 bands (noisy bands were removed).
  \end{itemize}
The signal and noise correlation matrices were obtained with the algorithms and code distributed with  HySime \cite{hysime}.   From those plots, we read  that, for  SudP5SNR40 data set, $\text{SNR-SD}(i) \gg 1$ for $i \leq 5$ and
$\text{SNR-SD}(i) \ll 1$ for $i > 5$, indicating that the SNR
is high  in the signal subspace.  For SusgsP5SNR40, the   singular values of the mixing matrix decay faster due to the  high correlation of the USGS spectral signatures. Nevertheless
the  ``big picture'' is similar to that of SudP5SNR40 data set.  The Rcuprite data set yields the more difficult inverse problem because  $\text{SNR-SD}(i)$  has ``close to convex shape'' slowly approaching  the value 1. This is a clear indication of a badly-conditioned inverse problem
\cite{book:berteroBoccacci}.\\

%% file: subspaceIdentification.tex
\section{Signal subspace identification}

%

The number of endmembers present in a given scene is, very often, much smaller than the
number of bands $B$. Therefore, assuming that the linear model is a good approximation,
spectral vectors lie in or very close to a low-dimensional linear
subspace. The identification of this subspace enables low-dimensional yet accurate representation of spectral
vectors, thus yielding gains in computational time and
complexity,  data storage, and  SNR.  It is usually advantageous and sometimes necessary to operate on data represented in the signal subspace.  Therefore, a signal subspace identification algorithm is required as a first processing step.


Unsupervised subspace identification has been approached in many ways. Band selection or
band extraction, as the name suggests, exploits the high correlation existing between
adjacent bands to select a few spectral components among those with higher  SNR
\cite{bi:chang_wang_06,bi:Shen_02}.  Projection techniques seek for the best subspaces to
represent data by optimizing objective functions. For example, \emph{principal component
analysis} (PCA) \cite{bi:Ian_86} minimizes sums of squares; \emph{singular value
decomposition} (SVD) \cite{bi:Scharf} maximizes power; projections on the first $p$ eigenvectors of the empirical correlation matrix maximize likelihood, if the noise is additive and white and the subspace dimension
is known to be $p$ \cite{bi:Scharf}; \emph{maximum noise fraction} (MNF)\cite{bi:Green_88} and
\emph{noise adjusted principal components} (NAPC)\cite{bi:Lee_90} minimize the ratio of noise power to signal power. NAPC is mathematically
equivalent to MNF \cite{bi:Lee_90} and can be interpreted as a sequence of two principal
component transforms: the first applies to the noise and the second applies to the
transformed data set.   MNF  is related to SNR-SD introduced in  (\ref{eq:SNR_SD}). In fact,
both metrics are equivalent in the case of white noise, {\em i.e}, ${\bf R}_w = \sigma^2{\bf I}$, where
$\bf I$ denotes the identity matrix. However, they differ when ${\bf R}_w \neq \sigma^2{\bf I}$.


The optical real-time adaptive spectral identification system (ORASIS) \cite{bi:Bowles_97}  framework, developed by U. S. Naval Research Laboratory aiming at real-time data processing,
has been used both for dimensionality reduction and endmember
extraction.  This framework consists of several modules, where   the dimension reduction is achieved by identifying a subset of exemplar pixels that convey the variability in a scene.
Each new pixel collected from the scene is compared to each exemplar pixel by
using an angle metric. The new pixel is added to the exemplar set if it is
sufficiently different from each of the existing exemplars. An orthogonal basis
is periodically created from the current set of exemplars using a modified
Gram-Schmidt procedure \cite{bi:Keshava_LLJ_00}.

The identification of the signal subspace is a model order inference problem
to which information theoretic criteria like the {\em minimum description
length} (MDL) \cite{bi:Schwarz_78, bi:Rissanen_78} or the {\em Akaike
information criterion} (AIC) \cite{bi:Akaike_74} comes to mind. These criteria
have in fact been used in  hyperspectral applications \cite{bi:Chang_04}
adopting the approach introduced by Wax and Kailath in
\cite{bi:Wax_Kailath_85}. {\color{black} In turn, } Harsanyi, Farrand, and Chang \cite{bi:Harsanyi_93} developed a Neyman-Pearson detection theory-based thresholding method (HFC) to determine the number of spectral endmembers in hyperspectral data, referred to in \cite{bi:Chang_04} as \emph{virtual dimensionality} (VD). The HFC method is based on a detector built on  the eigenvalues of the sample correlation and covariance matrices. A
modified version, termed noise-whitened HFC (NWHFC),  includes a
noise-whitening step \cite{bi:Chang_04}.
HySime (\emph{hyperspectral signal identification by minimum error}) \cite{hysime} adopts a minimum mean squared error based approach to infer the signal subspace. The method is eigendecomposition
based, unsupervised, and fully-automatic ({\em i.e.}, it does not depend on any
tuning parameters). It first estimates the signal and noise correlation
matrices and then selects the subset of eigenvalues that best represents the
signal subspace in the least square error sense.

When the spectral mixing is nonlinear,  the low dimensional subspace of the linear
case is often replaced with a low dimensional manifold, a concept defined in the mathematical subject of topology
\cite{bi:Bruske_98}. A variety of local methods exist for estimating manifolds.  For example, curvilinear component analysis \cite{bi:Demartines_97}, curvilinear distance
analysis \cite{bi:Lennon_cca_01}, manifold learning
\cite{conf:kim2003hyperspectral, bi:Bachmann_etal_06, bi:Bachmann_etal_05, bi:Yangchi_etal_05,
bi:Gillis_05, art:Mohan07} are non-linear projections based on the preservation of the local
topology.  Independent component analysis \cite{bi:Wang_Chang_06,
bi:Lennon_ICA_01}, projection pursuit \cite{bi:Chang_00,
bi:Bachmann_Donato_00}, and wavelet decomposition
\cite{bi:Othman_06,bi:Kaewpijit_03} have also been considered.

\subsection{Projection on the signal subspace}
\label{sect:background}

Assume that  the signal subspace, denoted by $\cal S$, has been identified by using one of the above referred to methods and let the columns of  ${\bf E}\equiv [{\bf e}_1,\dots,{\bf e}_p]$  be an orthonormal basis for $\cal S$, where ${\bf e}_i\in\mathbb{R}^B$, for $i=1,\dots,p$. The coordinates of the orthogonal  projection of a spectral vector ${\bf y}\in\mathbb{R}^B$ onto $\cal S$, with respect to the basis  ${\bf E}$, are given by ${\bf
y}_{\cal S}={\bf E}^T{\bf y}\in\mathbb{R}^p$. Replacing $\bf y$ by the observation
model (\ref{eq:sensor_M}), we have
\[
  {\bf y}_{\cal S} = {\bf E}^T{\bf M}\bm{\alpha} + {\bf E}^T{\bf w}.
\]
As referred to before, projecting onto a
signal subspace can yield  large computational, storage, and SNR gains. The first
two are a direct consequence of the fact that $p\ll B$ in most applications; to briefly explain the
latter, let us assume that the noise $\bf w$ is zero-mean and has covariance $\sigma^2\bf
I$. The mean power of the projected noise term
${\bf E}^T{\bf w}$ is then $\mathbb{E}\|{\bf E}^T{\bf w}\|^2=\sigma^2 p$
($\mathbb{E}(\cdot)$ denotes mean value). The relative attenuation of the noise power
implied by the projection is then $p/B$.

\begin{figure}[h]
    \centering
  \includegraphics[width=7cm,angle=0]{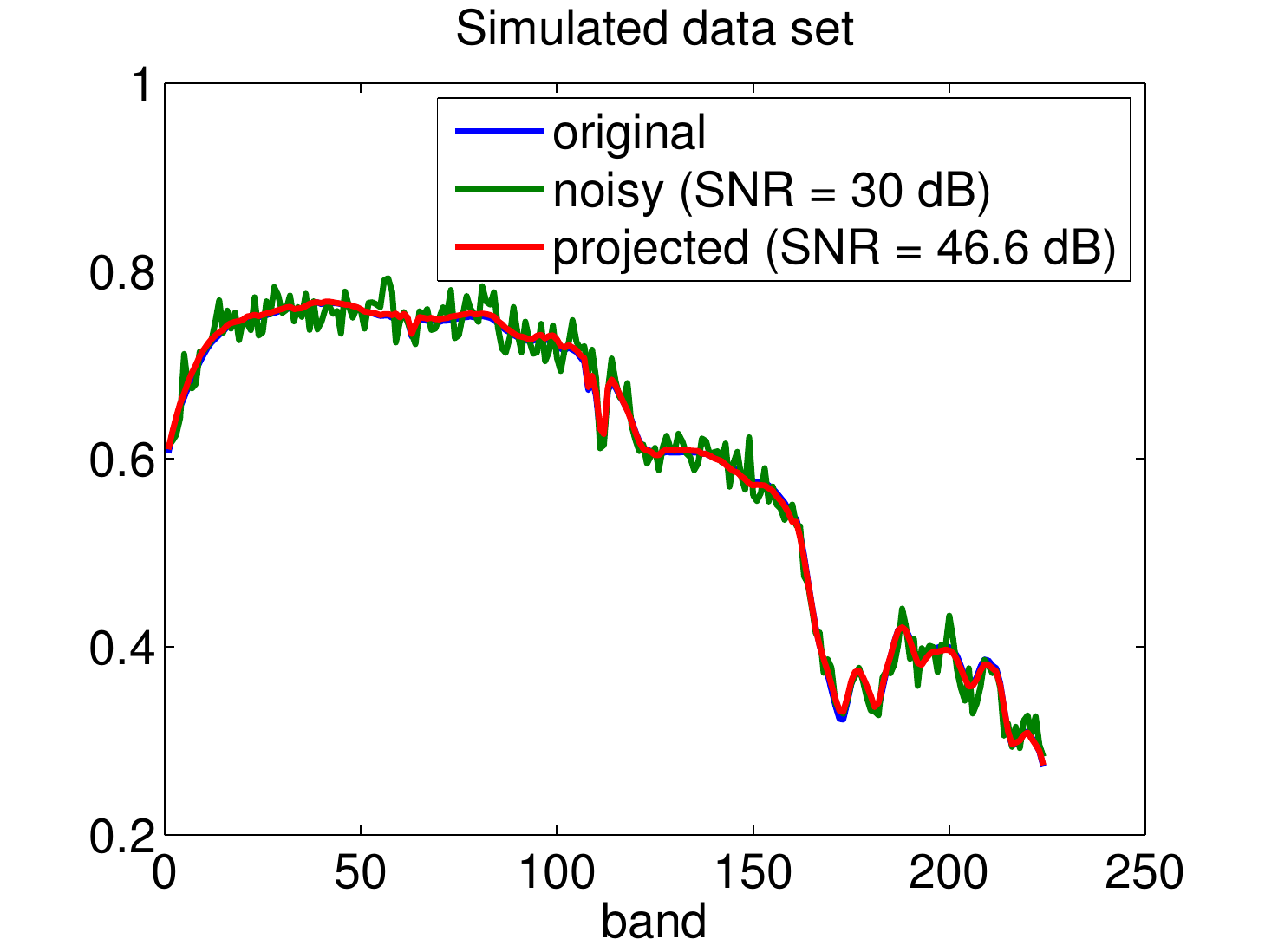}
  \includegraphics[width=7cm,angle=0]{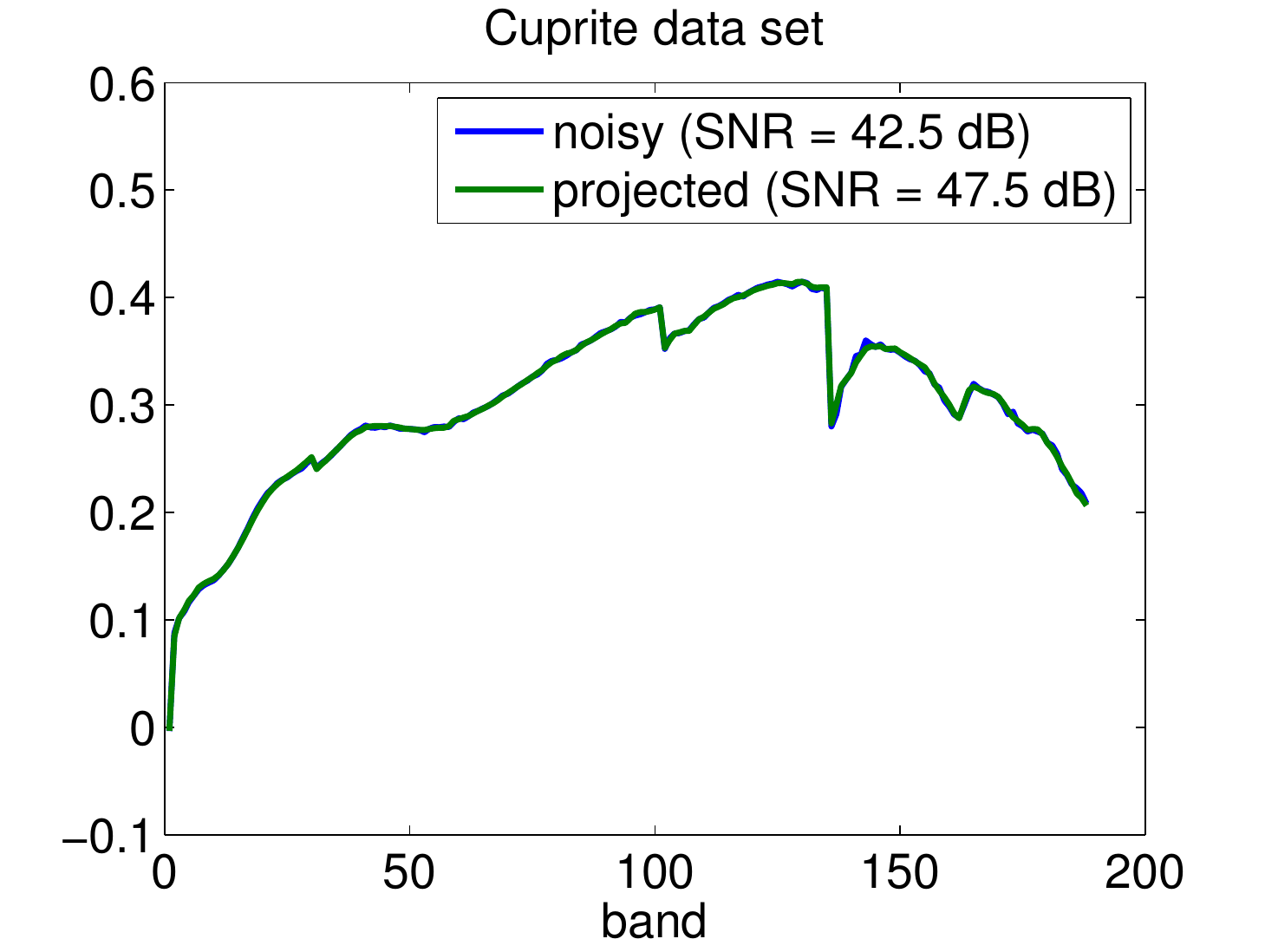}
  \caption{Left: Noisy and projected spectra from the simulates data set SusgsP5SNR30.
           Right:  Noisy and projected spectra from the real data set  Rcuprite}
  \label{fig:projection_advantages}
\end{figure}
Fig. (\ref{fig:projection_advantages}) illustrates the advantages of projecting  the data sets on
the signal subspace.  The  noise and the signal subspace were estimated  with HySime \cite{hysime}. The  plot on the  left hand side shows a  noisy and  the corresponding projected spectra taken from the simulated data set $\textbf{SusgsP5SNR30}$\footnote{Parameters of the simulated data set {\bf SusgsP5SNR30}: mixing matrix $\bf M$ sampled from a uniformly distributed random variable in the interval $[0,1]$;  $p=5$; $n=5000$; fractional abundances distributed uniformly on the 4-unit simplex; SNR$\,=\,$40dB.}. The  subspace dimension was  correctly identified. The SNR of the projected data set is $46.6\,$dB, which  is $16.6\, \text{dB} \simeq (B/p)\,\text{dB}$  above to that of the noisy  data set. The  plot on the  right hand side shows a  noisy and the corresponding projected spectra from the \textbf{Rcuprite} data set. The identified subspace dimension has   dimension 18. The SNR of the projected data set is $47.5\,$dB, which  is $5\,$dB
above to that of the noisy  data set.  The colored nature of the additive noise explains the   difference $(B/p)\,\text{dB} - 5\,\text{dB} \simeq 5\, $dB.

\begin{figure}[h]
    \centering
  \includegraphics[width=8cm,angle=0]{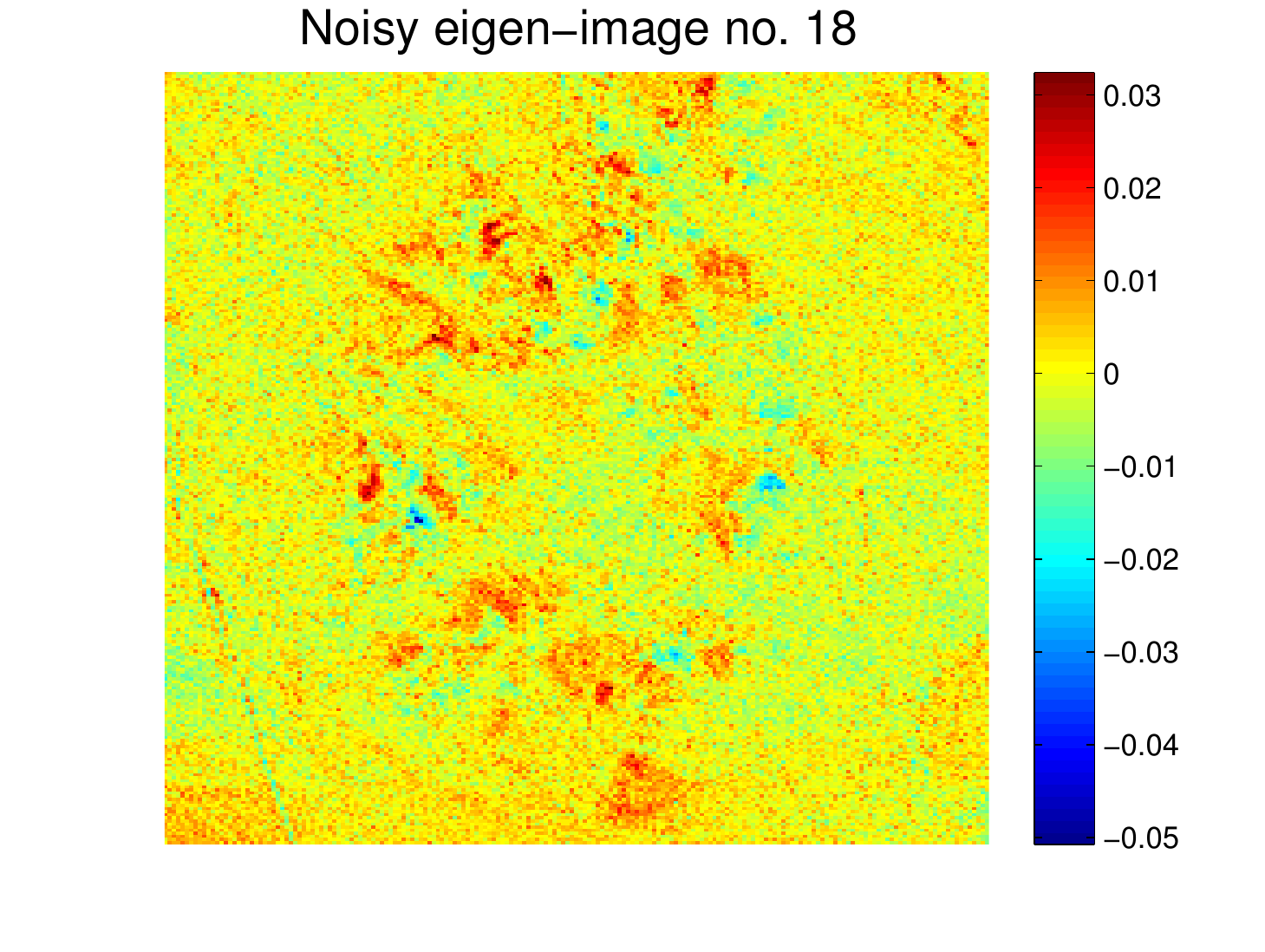}
  \includegraphics[width=8cm,angle=0]{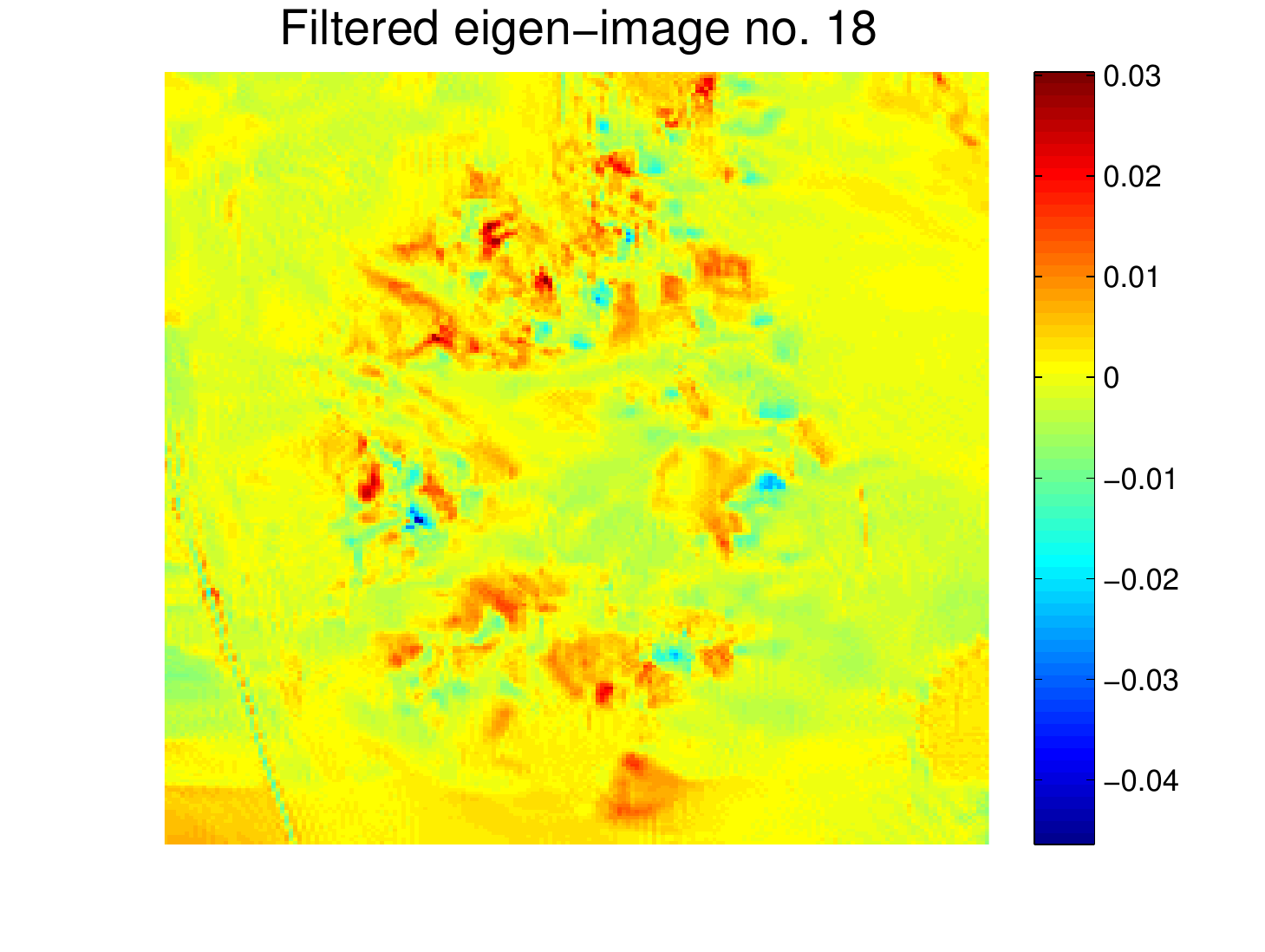}
  \includegraphics[width=8cm,angle=0]{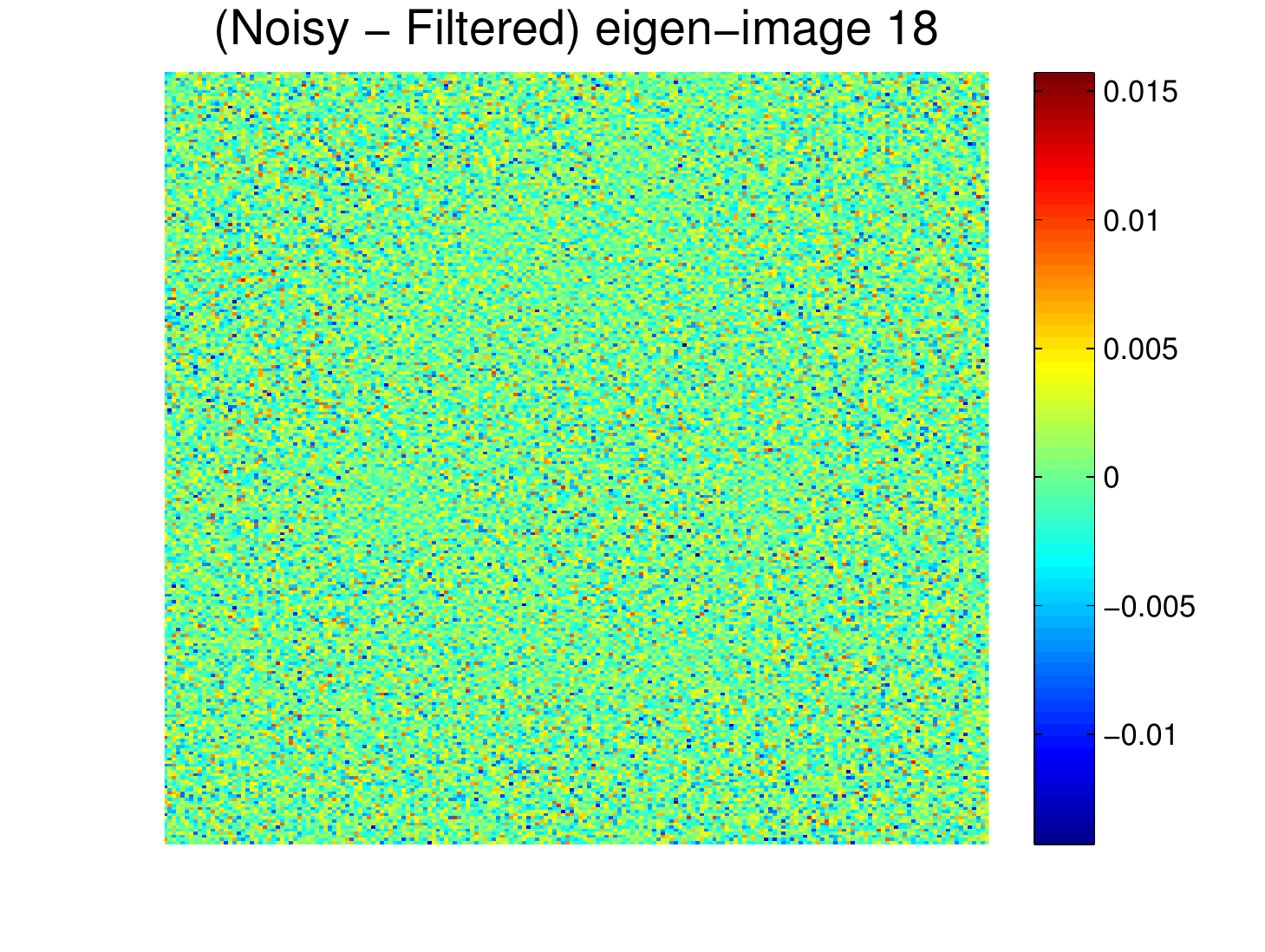}
  \includegraphics[width=8cm,angle=0]{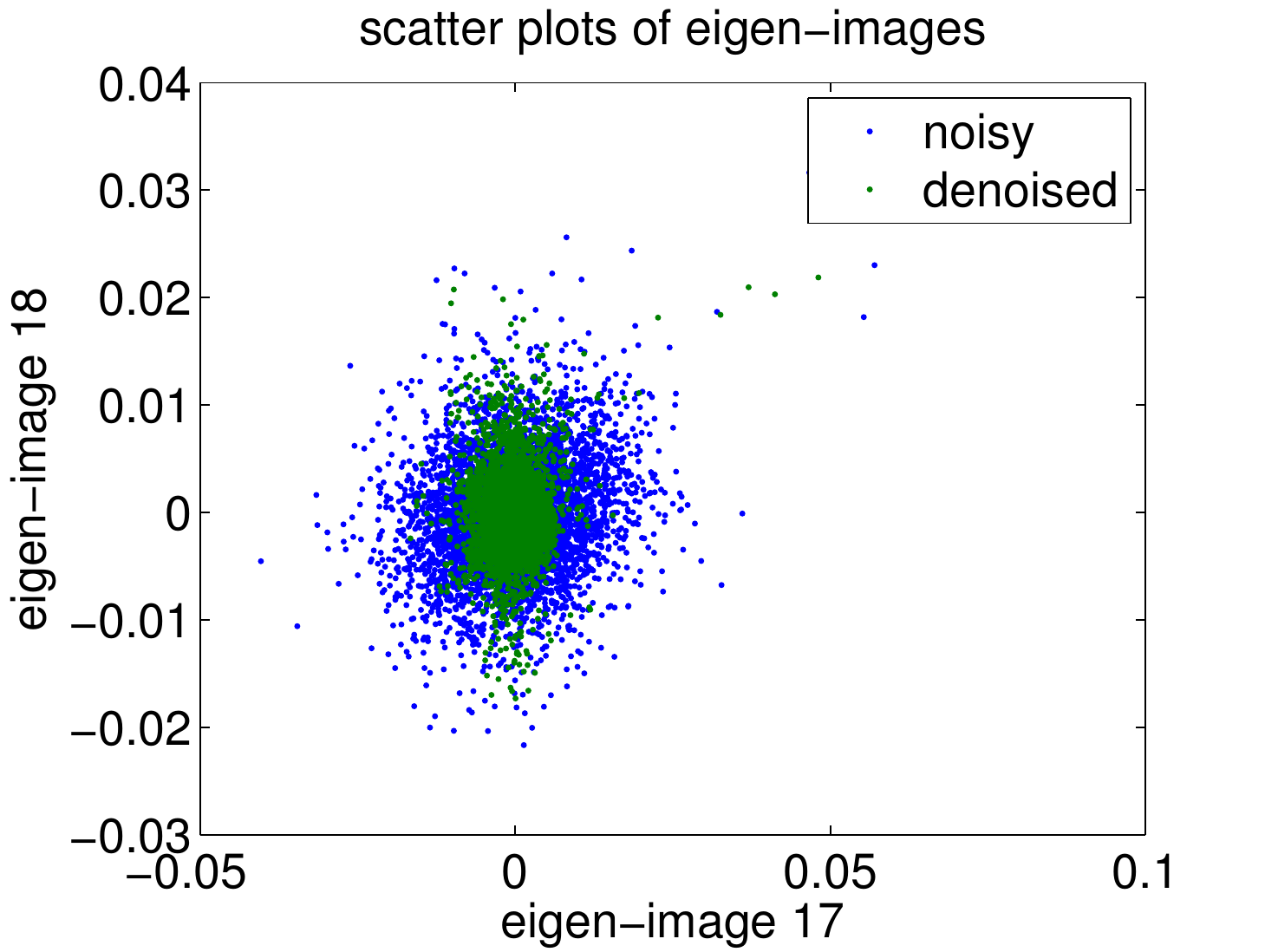}
  \caption{Top left: noisy eigen-image no. 18 of the \textbf{Rcuprite} data set. Top right:   denoised  no. 18; Bottom left: difference between  noisy  and denoised  images. Botton right:
  scatter plots of the Eigen-image no. 17 and no. 18  of the \textbf{Rcuprite} data set          (blue dots: noisy data; Green dots: denoised  data).}
  \label{fig:eigen_images}
\end{figure}

A final word of warning: although the projection of the data set onto the signal subspace often removes a large percentage of the noise,  it does not  improve the conditioning of the HU inverse problem, as this projection does not change the values of SNR-SD($i$) for the signal subspace eigen-components.

A possible line of attack to further reduce the noise in the signal subspace is  to exploit  spectral and spatial  contextual information. We give a brief illustration in the spatial domain. Fig. \ref{fig:eigen_images},   on the  the top left hand side,  shows the eigen-image no. 18, {\em i.e.}, the image obtained  from ${\bf e}_i^T {\bf y}_{\cal S}$ for $i=18$, of the \textbf{Rcuprite} data set.  The  basis of the signal subspace were obtained with the HySime algorithm.  A filtered version using then BM3D \cite{art:dabov:KE:07}
is shown on  the top right hand side. The denoising algorithm is quite effective in this example,
as confirmed by the absence of structure in the noise estimate (the difference between the noisy and the denoised images) shown in the bottom left hand side image.
This effectiveness can also be perceived  from the  scatter plots  of the noisy (blue dots) and denoised (green dots) eigen-images 17 and 18 shown in the bottom right hand side figure.
The scatter plot corresponding to the denoised image is much more dense, reflecting the lower variance.

\subsection{Affine set projection}

\label{sec:aff_proj}

From now on,  we assume that  the  observed data set has been projected onto the signal
subspace and, for simplicity of notation, we still represent the projected vectors as in
(\ref{eq:sensor_M}), that is
\begin{equation}
  {\bf y} = {\bf M}\bm{\alpha} + {\bf w},
  \label{eq:sub_proj}
\end{equation}
where ${\bf y,w\in\mathbb{R}}^p$ and  ${\bf M}\in \mathbb{R}^{p\times p}$.  Since the
columns of $\bf M$ belong to the signal subspace, the original mixing matrix is simply
given  by the matrix product ${\bf E}{\bf M}$.

Model (\ref{eq:sub_proj}) is a simplification of reality, as
it does not model pixel-to-pixel signature variability. Signature variability has been
studied and accounted for in a few unmixing algorithms (see, e.g.,
\cite{bi:Bateson_00,bi:Kruse_98,bi:Boardman_94,SongEndMemVar}), including all statistical algorithms that treat endmembers as distributions. Some of this variability is amplitude-based and therefore primarily characterized by spectral shape invariance \cite{bi:Shaw_03}; {\em i.e.}, while the spectral shapes of the endmembers are fairly consistent, their amplitudes are variable. This implies that the endmember signatures are
affected by a positive scale factor that varies from pixel to pixel.  Hence, instead of one matrix of endmember spectra for the entire scene, there is a matrix of endmember spectra for each pixel $[s(i, 1)\mathbf{m}_1, \dots,  s(i, p)\mathbf{m}_p]$ = $\mathbf{M}\mathbf{s}_i$ for $i=1,\dots,n$.  In this case, and in the absence of noise,  the observed spectral vectors
are no longer in a simplex defined by a fixed set of endmembers but rather in the set

\begin{equation}
{\{\mathbf{y}_i | \mathbf{y}_i = \sum_{j=1}^p  {\alpha_js(i,j)\mathbf{m}_j}\}},
\end{equation}
as illustrated in  Fig.  \ref{fig:perspective_proj}.
Therefore,  the coefficients of the endmember spectra $\bm{s}_i^t\bm{\alpha}$  need not
sum-to-one, although they are still nonnegative. Transformations of the data are required to improve the match of the model to reality.  If a true mapping from units of radiance to reflectance can be found, then that transformation is sufficient.  However, estimating that mapping can be difficult problem or impossible.  Other methods can be applied to to ensure that the  sum-to-one constraint is a better model, such as the following:

\begin{figure}
    \centering
  \includegraphics[width=8cm,angle=0]{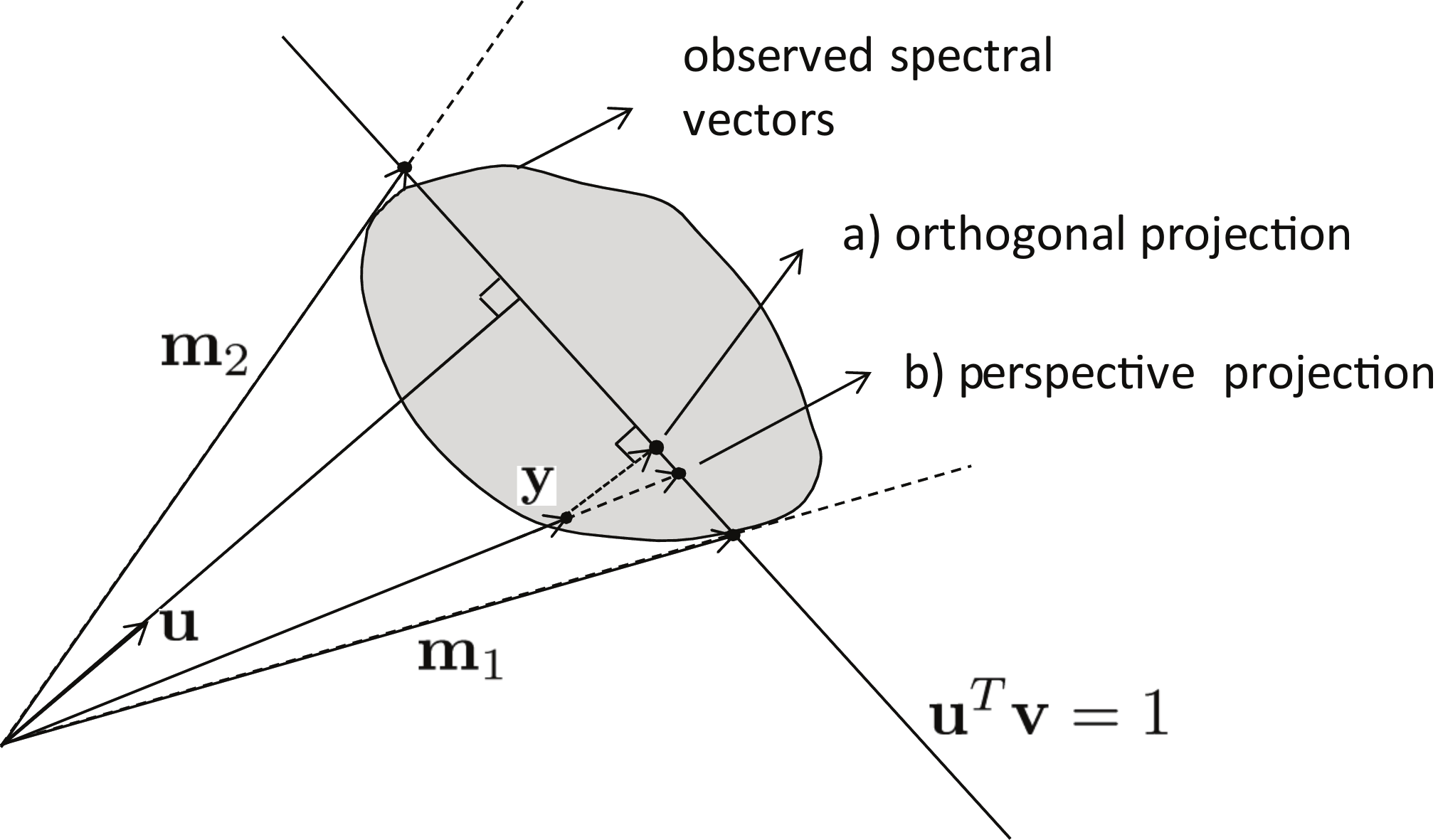}
  \caption{Projections of the observed data onto an hyperplane: a) Orthogonal projection on an  hyperplane
  (the projected vectors suffers a rotation); b) Perspective projection (the scaling
  $\bo{y}/(\bo{y}^T\bo{u})$ brings them to  the hyperplane defined by  $\bo{y}'^T\bo{u}=1)$.
   }
  \label{fig:perspective_proj}
\end{figure}

\begin{enumerate}
\item[a)]  \textbf{Orthogonal projection}: Use PCA to identify the affine set that
            best represent the observed data in the least squares sense and then compute the orthogonal
            projection of the observed vectors onto this set (see \cite{art:Tsung-Han_etal:TSP:09} for details).
            This projection is illustrated in  Fig. \ref{fig:perspective_proj}.

  \item[b)] \textbf{Perspective projection}: This is the so-called  dark point fixed transform (DPFT)
        proposed in \cite{bi:Craig_94}.  For a given observed vector $\bf y$,
         this projection, illustrated in Fig. \ref{fig:perspective_proj},
         amounts to  rescale  $\bf y$  according to  ${\bf y}/({\bf y}^T{\bf u})$,
         where $\bf u$ is chosen such that ${\bf y}^T{\bf u} > 0$ for every  $\bf y$ in the data set.  The   hyperplane containing the projected vectors is defined by  ${\bf v}^T{\bf u}=1$, for any ${\bf v}\in\mathbb{R}^p$.

\end{enumerate}

\begin{figure}
    \centering
  \includegraphics[width=8cm,angle=0]{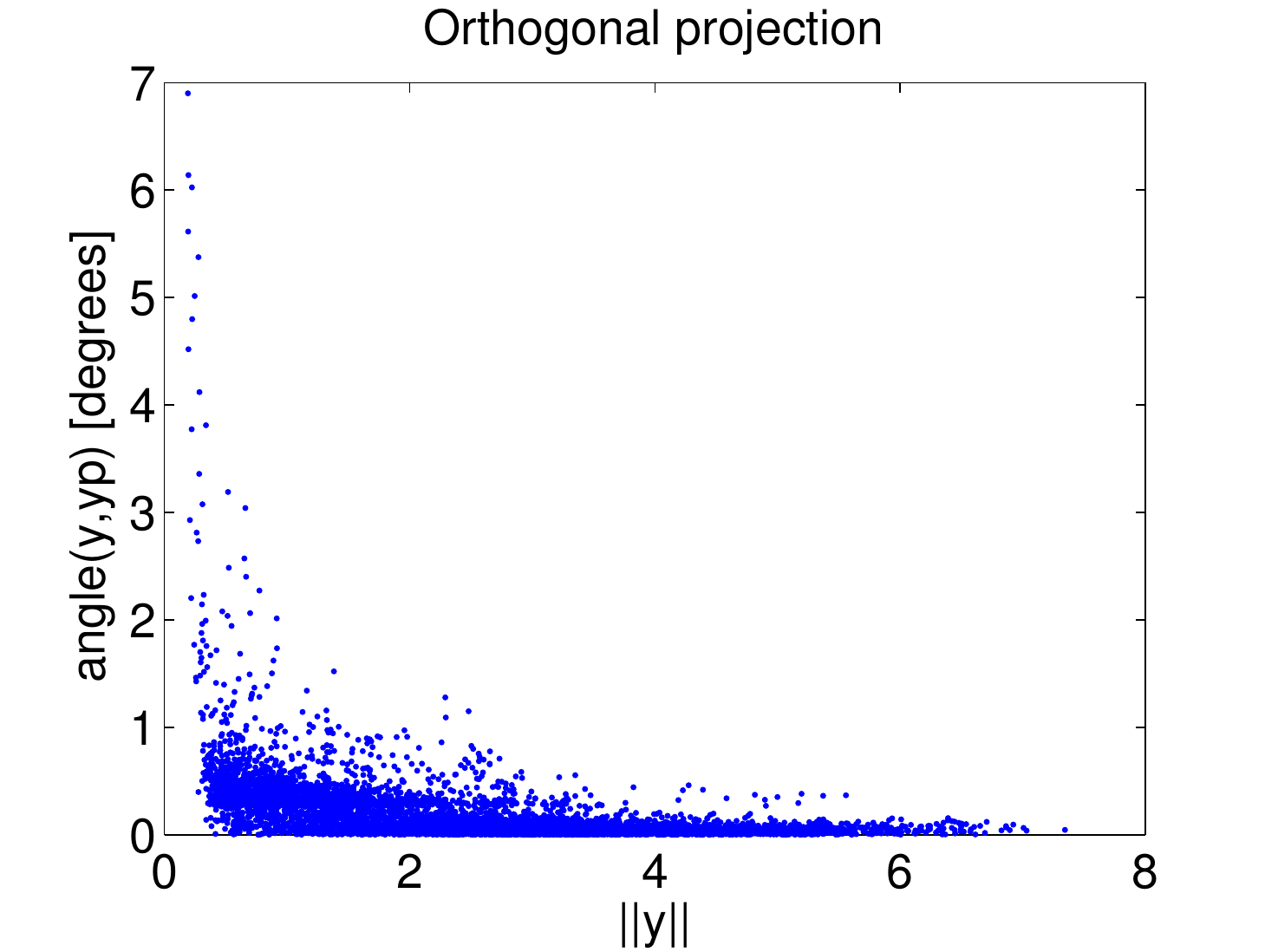}
  \includegraphics[width=8cm,angle=0]{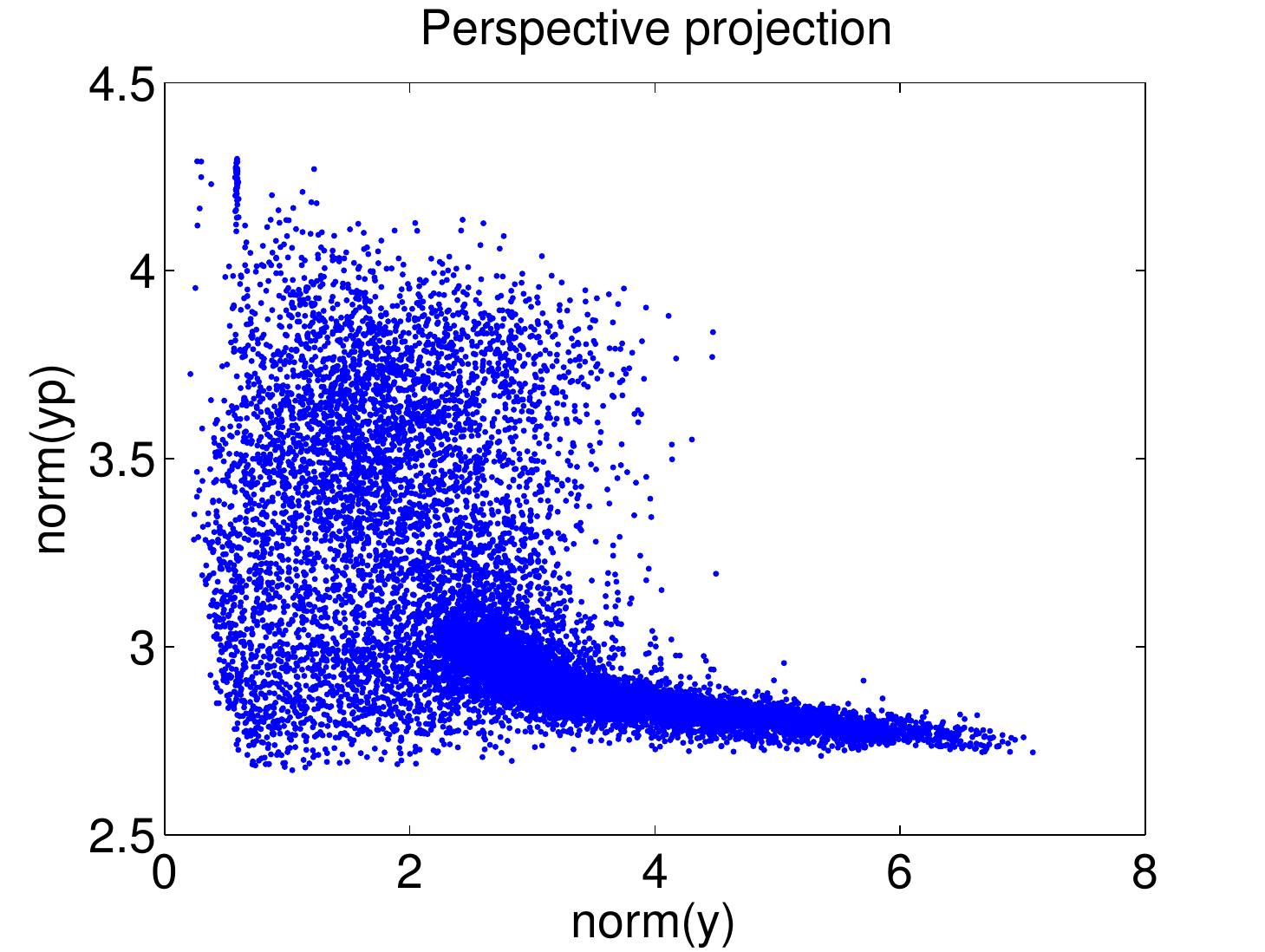}
  \caption{Left (orthogonal projection): angles between  projected and unprojected vectors.
           Right (perspective projection): scale factors  between  projected and unprojected vectors.}
  \label{fig:ortho_persp_effects}
\end{figure}

Notice that the orthogonal projection modifies the direction of the spectral vectors whereas the
perspective projection does not. On the other hand, the perspective projection introduces large
scale factors, which may become negative, for spectral vectors close to being orthogonal to $\bf u$.
Furthermore,  vectors $\bf u$  with different angles  produce non-parallel affine sets and thus
different fractional abundances, which implies that the choice of $\bf u$ is a critical issue for accurate estimation.

These effects are illustrated in Fig. \ref{fig:ortho_persp_effects} for the \textbf{Rterrain}
data set\footnote{http://www.agc.army.mil/hypercube}. This is a publicly available hyperspectral data cube distributed by the Army Geospatial Center, United States Army Corps of Engineers, and was collected by the hyperspectral image data collection experiment (HYDICE). Its dimensions are 307 pixels by 500 lines and 210 spectral bands. The figure on the left hand side plots the angles between the unprojected and
the orthogonally projected vectors, as a function of the  norm of the unprojected vectors.
The higher angles, of the order of $1-7^\circ$, occur for vectors of small norm, which usually correspond to shadowed areas.  The figure on the right hand side plots the norm of  the projected vectors as a function of the  norm of the unprojected vectors. The corresponding scale factors varies between, approximately, between 1/3 and 10.

A possible way of mitigating these projection errors is discarding the problematic
projections, which are vectors with angles between projected  and unprojected vectors larger than a given small threshold, in the case of the perspective projection, and vectors with   very  small or  negative scale factors  $\bo{y}^T\bo{u}$, in the case of the orthogonal projection.

%% file: geo_unmixing.tex


\section{Geometrical based approaches to linear spectral unmixing}

\label{sec:unmix}

The geometrical-based approaches are categorized into two main categories: Pure Pixel (PP) based and Minimum Volume (MV) based.  There are a few other approaches that will also be discussed.

\subsection{Geometrical based approaches: pure pixel based  algorithms}

\label{sec:pure_pixel_algs}

The pure pixel based algorithms still belong to the MV  class but assume  the presence in
the data of at  least one pure pixel per  endmember, meaning that there is at least one
spectral vector on each vertex of the data simplex.  This assumption, though  enabling
the design of very efficient algorithms from the computational point of view, is a strong
requisite that may not hold in many  datasets. In any case, these algorithms find the set
of most pure pixels in the data.  They have probably been the most often used in linear hyperspectral
unmixing applications, perhaps because of their light computational burden and clear conceptual meaning. Representative  algorithms of this class are the following:
\begin{itemize}

\item The  pixel purity index (PPI) algorithm \cite{bi:Boardman_93, bi:Boardman_95} uses MNF as
a preprocessing step to reduce dimensionality and to improve the SNR. PPI projects every
spectral vector onto \emph{skewers}, defined as a large set of  random vectors. The points corresponding
to extremes, for each skewer direction, are stored. A cumulative account records the
number of times each pixel (i.e., a given spectral vector) is found to be an extreme. The
pixels with the highest scores are the purest ones.

\item N-FINDR
\cite{bi:Winter_99} is based on the fact that in spectral dimensions, the volume defined
by a simplex formed by the purest pixels is  larger than any other volume defined by any
other combination of pixels. This algorithm finds the set of pixels defining the largest
volume by inflating a simplex inside the data.

\item The iterative error analysis (IEA)
algorithm \cite{conf:neville:99} implements a series of linear constrained unmixings,
each time choosing as endmembers  those pixels which minimize the remaining error in the
unmixed image.

\item The vertex component analysis (VCA) algorithm
\cite{art:nascimento_bioucas:VCA:05} iteratively projects data onto a direction
orthogonal to the subspace spanned by the endmembers already determined. The new
endmember signature corresponds to the  extreme of the projection. The algorithm iterates
until all endmembers are exhausted.

\item The simplex growing algorithm (SGA) \cite{SGAchang06} iteratively grows a simplex by
finding the vertices corresponding to the maximum volume.

\item The sequential maximum angle convex cone (SMACC) algorithm \cite{conf:Gruninger:SPIE:04}
is based on a convex cone for representing the spectral vectors. The algorithm  starts
with a single endmember and increases incrementally in dimension.  A new endmember is
identified based on the angle it makes with the existing cone. The data vector making the
maximum angle with the existing cone is chosen as the next endmember  to enlarge the
endmember set.  The algorithm terminates when all of the data vectors are within the
convex cone, to some tolerance.

\item The \emph{alternating volume maximization} (AVMAX) \cite{Chan:MAC:11},
inspired by N-FINDR,  maximizes, in a cyclic fashion, the volume of the simplex
defined by the endmembers with respect to only one endmember   at one time.
AVMAX is quite  similar to the SC-N-FINDR variation of N-FINDR introduced in
\cite{art:wuChuChang08}.

\item The \emph{successive volume maximization} (SVMAX) \cite{Chan:MAC:11} is similar to
VCA. The main difference concerns the way data is projected   onto  a direction orthogonal
the subspace spanned  by the endmembers already determined. VCA considers a random direction
in these subspace, whereas SVMAX considers the complete subspace.

\item The  \emph{collaborative convex framework} \cite{mollerEsser2010}  factorizes the  data matrix $\bf Y$
into a nonnegative mixing matrix $\bf M$ and a sparse and also nonnegative abundance  matrix  $\bf S$. The columns of the mixing matrix $\bf M$ are constrained to be columns of the data $\bf Y$.

\item \emph{Lattice Associative Memories} (LAM) \cite{RitterUrcid2009, RitterUrcid2011, Grana2011} model sets of spectra as elements of the lattice of partially ordered  real-valued vectors.  Lattice operations are used to nonlinearly construct LAMS.  Endmembers are found by constructing so-called min and max LAMs from spectral pixels. These LAMs contain maximum and minimum coordinates of spectral pixels (after appropriate additive scaling) and are candidate endmembers.  Endmembers are selected from the LAMS using the notions of affine independence and similarity measures such as spectral angle, correlation, mutual information, or Chebyschev distance.
\end{itemize}

Algorithms AVMAX and SVMAX  were derived in \cite{Chan:MAC:11} under  a continuous optimization framework inspired  by Winter's maximum volume  criterium \cite{bi:Winter_99}, which underlies N-FINDR. Following a rigorous  approach,
Chan {\em et al.}  not only derived AVMAX and SVMAX, but have also unveiled  a number of links  between
apparently disparate algorithms such as  N-FINDR and VCA.

\subsection{Geometrical based approaches: Minimum volume based algorithms}

\label{sec:geo_unmixing}

\begin{figure*}[h]
\centering
\includegraphics[width=6cm]{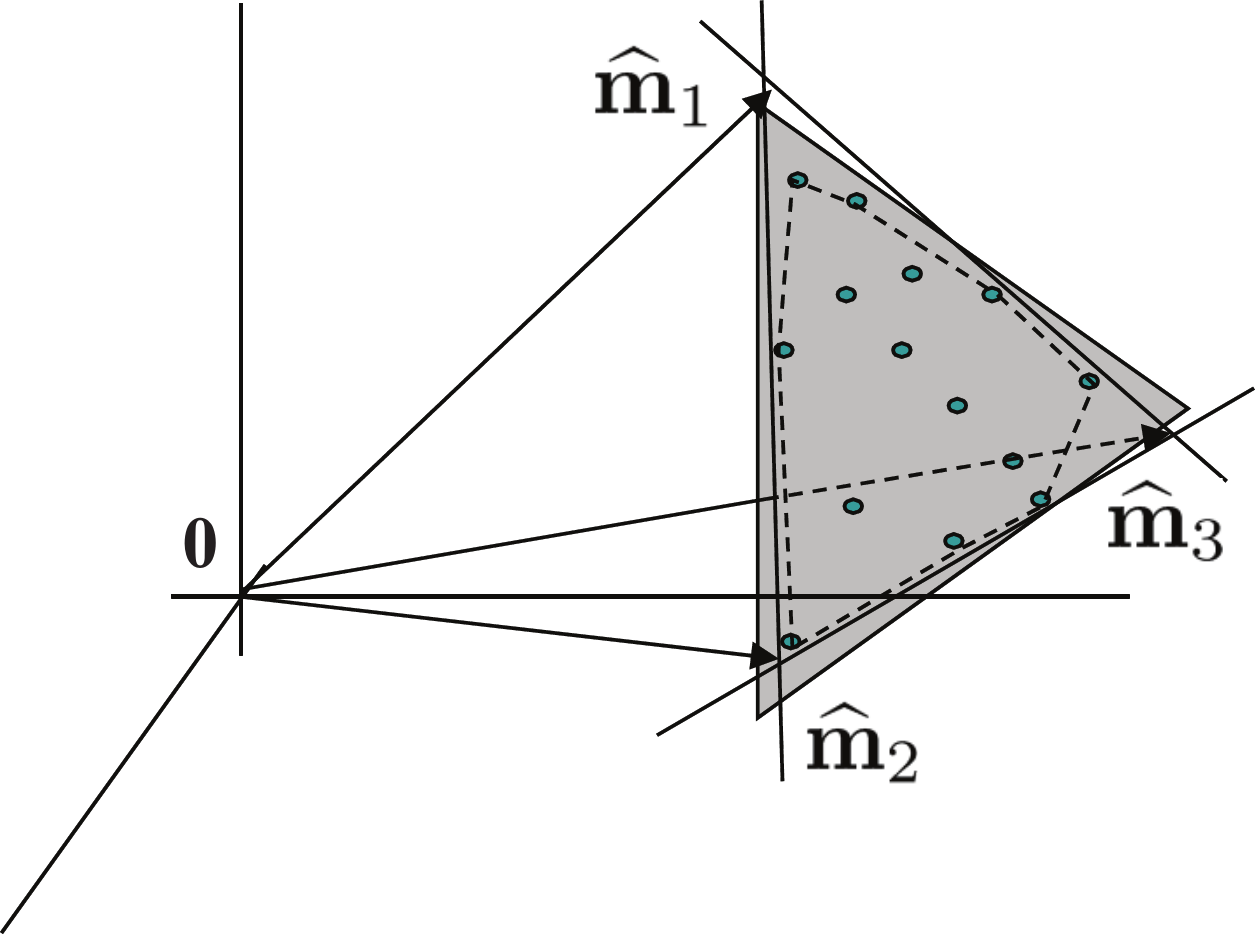}
\caption{Illustration of the concept of simplex of minimum volume containing the data.}
\label{fig:min_vol}
\end{figure*}

The MV approaches seek a mixing matrix $\bo{M}$ that minimizes the volume of the
simplex defined by its columns, referred to as $\mbox{conv}(\bo{M})$,  subject to the constraint that $\mbox{conv}(\bo{M})$ contains the observed spectral vectors.  The constraint can be soft or hard.  The pure pixel constraint is no longer enforced, resulting in a much harder nonconvex optimization problem. Fig. \ref{fig:min_vol}
further illustrates the concept of simplex of minimum size
containing the data.  The  estimated  mixing matrix $\widehat{\bf
M}\equiv[\widehat{\bo{m}}_1,\widehat{\bo{m}}_2 ,\widehat{\bo{m}}_3]$  differs slightly
from the true mixing matrix because there are not enough data points per facet
(necessarily $p-1$ per facet) to define the true simplex.

Let us assume that the data set has been projected onto the signal subspace ${\cal S}$,
of dimension $p$, and that the vectors $\bo{m}_i \in\mathbb{R}^p$, for $i=1,\dots,p$, are
affinely independent ({\em i.e.}, $\bo{m}_i-\bo{m}_1 $, for $i=2,\dots,p$ , are linearly
independent). The dimensionality of the simplex $\mbox{conv}(\bo{M})$ is therefore $p-1$ so the volume of  $\mbox{conv}(\bo{M})$ is zero in $\mathbb{R}^p$.   To obtain a nonzero volume, the
extended simplex  $\bo{M}_0\equiv \bo{[\bo{0},\bf{M}]}$, containing the origin, is
usually considered. We recall that  the  volume  of $\mbox{conv}(\bo{M}_0)$, the
convex hull of  $\bo{M}_0$, is given by
\begin{equation}
    V(\bo{M}_0) \equiv \frac{|\mbox{det}(\bo{M})|}{p!}.
    \label{eq:vol_M_Rp}
\end{equation}

An alternative to (\ref{eq:vol_M_Rp}) consists of shifting the data set to the origin
and working in the subspace of dimension $p-1$. In this case, the  volume of the simplex
is given by
\begin{eqnarray*}
    V(\bo{M}) & = & \frac{1}{(p-1)!}\left|\mbox{det}\left[
              \begin{array}{lll}
                1 & \cdots & 1 \\
                \bo{m}_1 & \cdots & \bo{m}_p
              \end{array}
              \right]\right|.
\end{eqnarray*}

Craig's work\cite{bi:Craig_94}, published in 1994, put forward the seminal concepts
regarding the algorithms {\color{black} of MV type}.  After  identifying the subspace and
applying  projective projection (DPFT), the algorithm iteratively changes one facet of the simplex at a time,  holding the others fixed, such that the volume
  $$
     V(\widehat{\bo{M}}_0) \equiv \frac{\mbox{abs}(|\widehat{\bo{M}}|)}{p!}
  $$
  is minimized and all spectral vectors belong to this simplex; {\em i.e.},
  $ \widehat{\bo{M}}^{-1}\bo{y}_i\succeq 0$ and ${\bf 1}_p^T\widehat{\bo{M}}^{-1}\bo{y}_i=1$ (respectively, ANC and ASC constraints\footnote{The notation
  ${\bf 1}_p$ stands for a column vector of ones with size $p$.})
  for $i=1,\dots,n$. In a more formal way:
  \begin{eqnarray*}
      {\bf for} \;t=1,\dots,&&\\
         \widehat{\bo{M}}^{t+1}_0 & = & \arg\min_{{\bo{M}}_0} V(\bo{M}_0) \\
         \mbox{s.t.:} &&  \mbox{facets} (\bo{M}_0) =  \mbox{facets}(\widehat{\bo{M}}^t_o), \; \; \mbox{except for facet}\;\;
                         i= (t\;\mbox{mod}\; p)\\
         \mbox{s.t.:} &&  \bo{M}^{-1}\bo{y}_i\succeq 0,\;\;{\bf 1}_p^T\widehat{\bo{M}}^{-1}\bo{y}_i=1,\;\; \mbox{for}\;\; i=1,\dots,n.
  \end{eqnarray*}

\begin{figure*}[h]
\centering
\includegraphics[width=5cm]{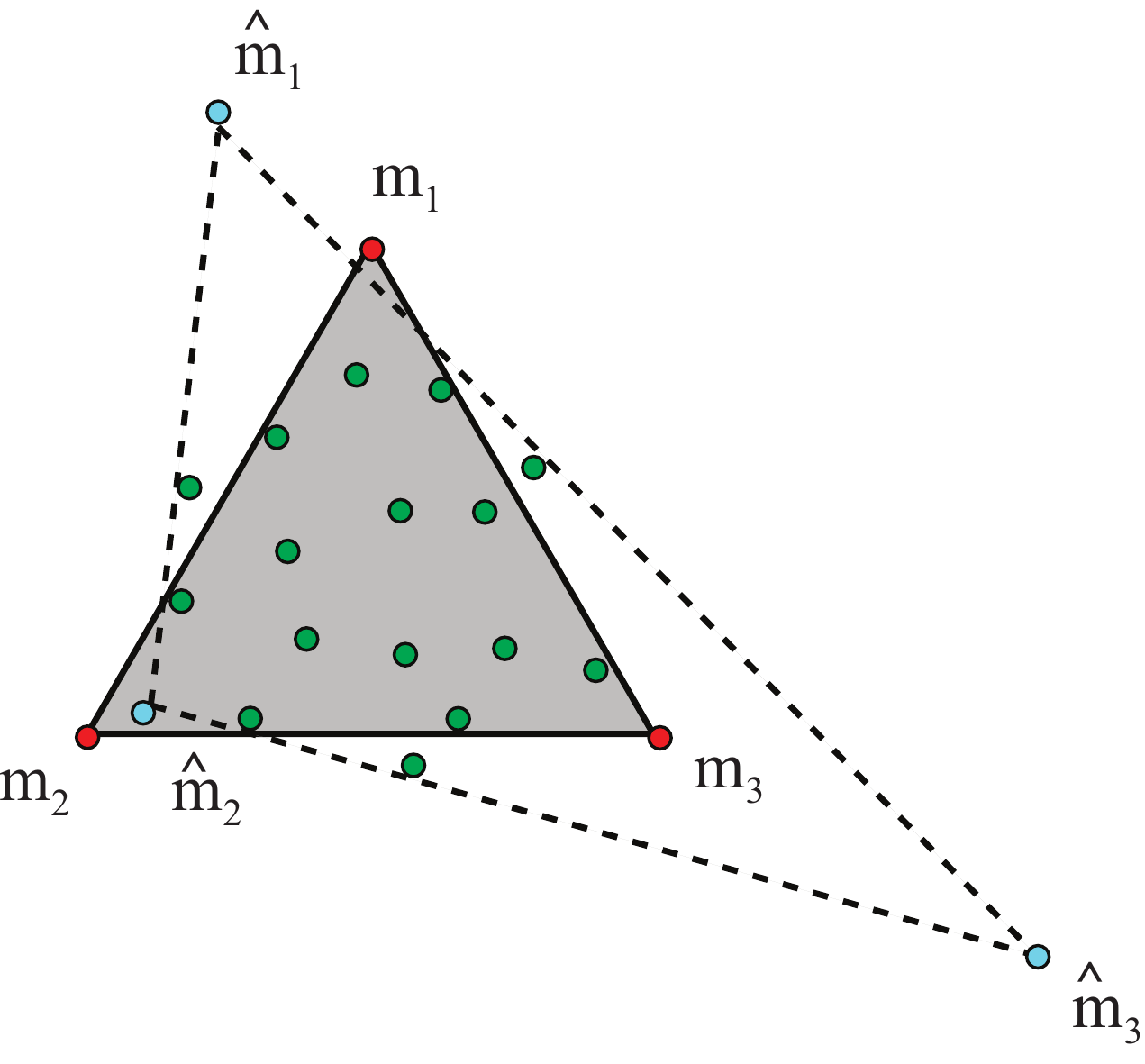}
\caption{Noisy data. The dashed simplex represents the simplex of minimum volume required to contain all the data; by
allowing violations to the positivity constraint, the MVSA and SISAL algorithms yield a
simplex very close to the true one.}
\label{fig:noisy_lmm}
\end{figure*}

The minimum volume simplex analysis (MVSA) \cite{li_bioucas08} and  the simplex identification via variable splitting and augmented Lagrangian (SISAL) \cite{conf:bioucasSISAL09} algorithms implement a robust
version of the MV concept. The robustness is introduced by allowing the positivity
constraint to be violated.  To grasp the relevance of this modification, noisy spectral vectors are depicted in Fig.
\ref{fig:noisy_lmm}. Due to the presence of noise, or
any other perturbation source, the spectral vectors may lie outside the true data simplex.
The application of a MV algorithm would lead to the dashed estimate, which is
far from the original.

In order to estimate endmembers more accurately, MVSA/SISAL allows
violations to the positivity constraint. Violations are penalized using the \emph{hinge} function ($\mbox{hinge}(x) = 0$ if $x\geq 0$ and $-x$ if $x<0$). MVSA/SISAL project the data onto a signal subspace.  Thus the representation of section \ref{sec:aff_proj} is used.  Consequently, the matrix $\bf M$ is square and theoretically invertible (ill-conditioning can make it difficult to compute the inverse numerically).  Furthermore,

\begin{equation}
\bf{M}^{-1}\bf{y}=\bf{M}^{-1}( \bf{M}\bm{\alpha}+\bf{w})=\bm{\alpha}+\bf{M}^{-1}\bf{w}.
\end{equation}

MVSA/SISAL aims at solving the following optimization
problem:
\begin{eqnarray}
         \widehat{\bo{Q}} & = & \arg\max_{{\bo{Q}}} \; \log(|\mbox{det}(\bo{Q})|) - \lambda \bo{1}^T_p\mbox{hinge}(\bo{QY})\bo{1}_n\\
         & = &\arg\min_{{\bo{M}}} \; \log(|\mbox{det}(\bo{M})|) + \lambda \bo{1}^T_p\mbox{hinge}(\bm{\alpha}+\bf{M}^{-1}\bf{w})\bo{1}_n
         \label{eq:mvsa_sisal}\\
         \ \ \mbox{s.t.:} &&  \bo{1}^T_p\bo{Q} = \bo{q}_m,
         \nonumber
  \end{eqnarray}
  where $\bo{Q}\equiv \bo{M}^{-1}$,
  $\bo{q}_m \equiv \bo{1}^T_p\bo{Y}_p^{-1}$  with $\bo{Y}_p$ being any set of of linearly independent spectral vectors taken from the data set $\bo{Y}\equiv[{\bf y}_1,\dots,{\bf y}_n]$,  $\lambda$ is a regularization  parameter,
  and $n$ stands for the number of spectral vectors.

 We make the following two remarks: a) maximizing $\log(|\mbox{det}(\bo{Q})|)$ is  equivalent to minimizing
 $V(\bo{M}_0)$; b)  the term $-\lambda \bo{1}^T_p\mbox{hinge}(\bo{QY})\bo{1}_n$ weights the ANC
 violations. As $\lambda$ approaches infinity, the soft constraint approaches the hard constraint.  MVSA/SISAL optimizes by solving a sequence of convex optimization problems using the method of augmented Lagrange multipliers, resulting in a  computationally efficient algorithm.
%
%

The minimum volume enclosing simplex (MVES) \cite{chan-convex:TIP:09} aims at solving the optimization problem  (\ref{eq:mvsa_sisal}) with $\lambda=\infty$, {\em i.e.}, for hard positivity constraints. MVES implements a cyclic minimization   using linear programs (LPs). Although the optimization problem (\ref{eq:mvsa_sisal}) is nonconvex, it is proved  in  (\ref{eq:mvsa_sisal}) that the existence of pure pixels is  a sufficient condition for MVES to identify the true endmembers.

A robust  version of MVES (RMVES) was recently introduced in \cite{ambikapathichanChan:TGRS:11}.
RMVES accounts for the noise effects in the observations by employing chance constraints, which act as
soft constraints  on the fractional abundances. The chance constraints  control the volume of the resulting simplex. Under the Gaussian noise assumption, RMVES infers the mixing matrix and the  fractional
abundances  via alternating optimization involving  quadratic programming solvers.

The minimum volume transform-nonnegative matrix factorization (MVC-NMF)  \cite{miao07}
solves the following optimization problem applied to the original data set, {\em i.e.},
without dimensionality reduction:
\begin{eqnarray}
  \label{eq:mvt_NMF}
  (\widehat{\bo{M}},\widehat{\bo{S}}) &=& \arg\min_{\bo{M}\in \mathbb{R}^{B\times p},\bo{S}\in \mathbb{R}^{p\times n}} \frac{1}{2}\|\bo{Y}-\bo{M}\bo{S}\|_F^2+\lambda V^2(\bo{M})\\
                  \mbox{s.t.:} &=& \bo{M}\succeq 0,\;\;  \bo{S}\succeq 0, \;\;\bo{1}^T\bo{S} = \bo{1}^T_n,
                  \nonumber
\end{eqnarray}
where $\bo{S}\equiv [\bm{\alpha}_1,\dots,\bm{\alpha}_n]\in\mathbb{R}^{p\times n}$ is a
matrix containing the fractional abundances $\|\bo{A}\|_F^2\equiv
\mbox{tr}(\bo{A}^T\bo{A})$ is the Frobenius norm of matrix $\bo{A}$ and $\lambda$ is a
regularization parameter. The optimization (\ref{eq:mvt_NMF}) minimizes a two term objective function, where the  term $(1/2)\|\bo{Y}-\bo{M}\bo{S}\|_F^2$ measures the  approximation error and the term
$V^2(\bo{M})$ measures the square of the volume of the simplex defined by the columns of $\bf M$. The regularization parameter $\lambda$ controls the tradeoff between  the reconstruction errors and  simplex volumes. MVC-NMF implements a sequence of alternate minimizations with respect to ${\bf S}$
(quadratic programming problem) and  with respect to ${\bf M}$ (nonconvex programming
problem).  The major difference  between  MVC-NMF and MVSA/SISAL/RMVES algorithms is that the
latter allows  violations of the ANC, thus bringing robustness to the SU inverse problem, whereas the former does not.

\begin{figure*}[h]
\centering
\includegraphics[width=8cm]{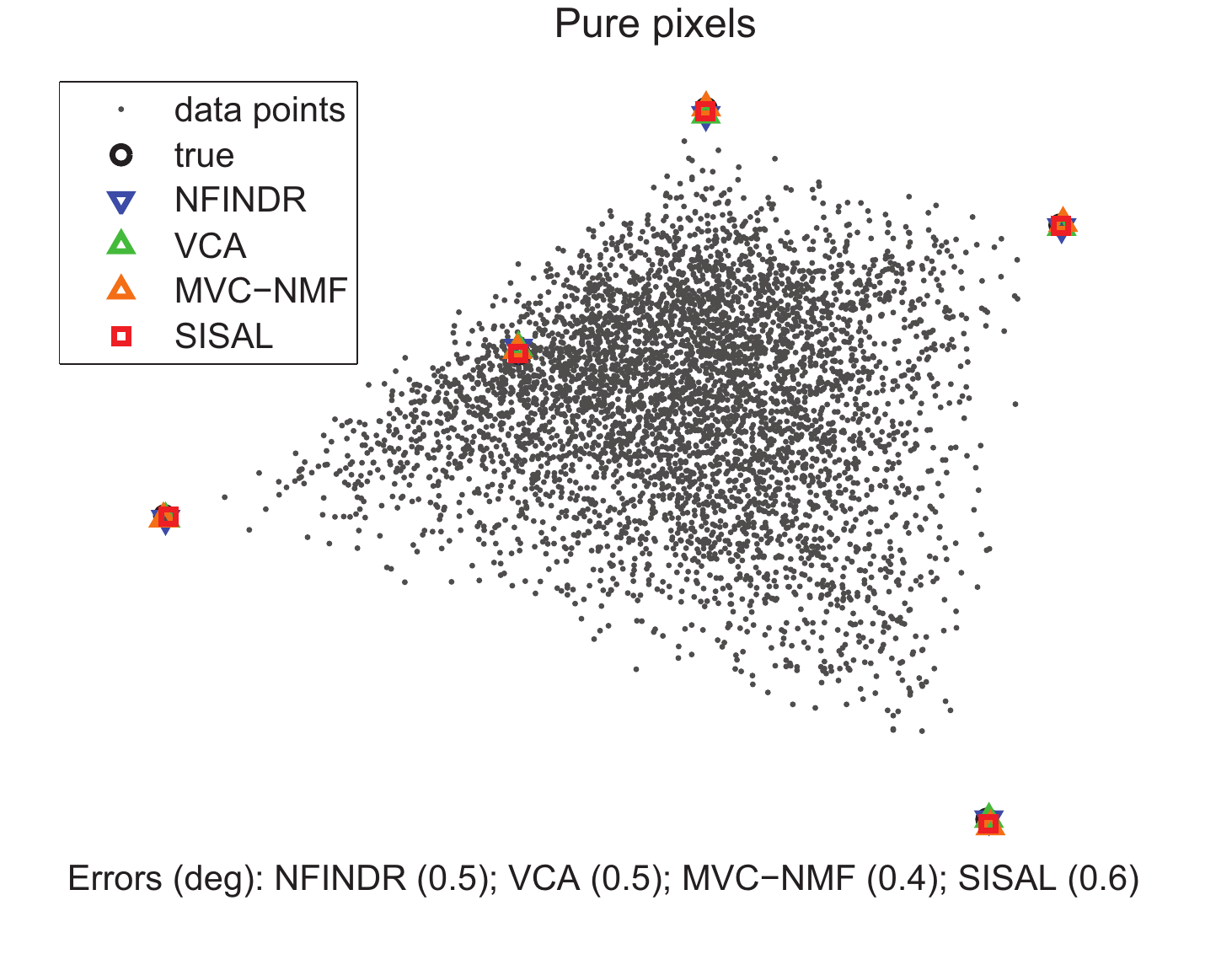}
\includegraphics[width=8cm]{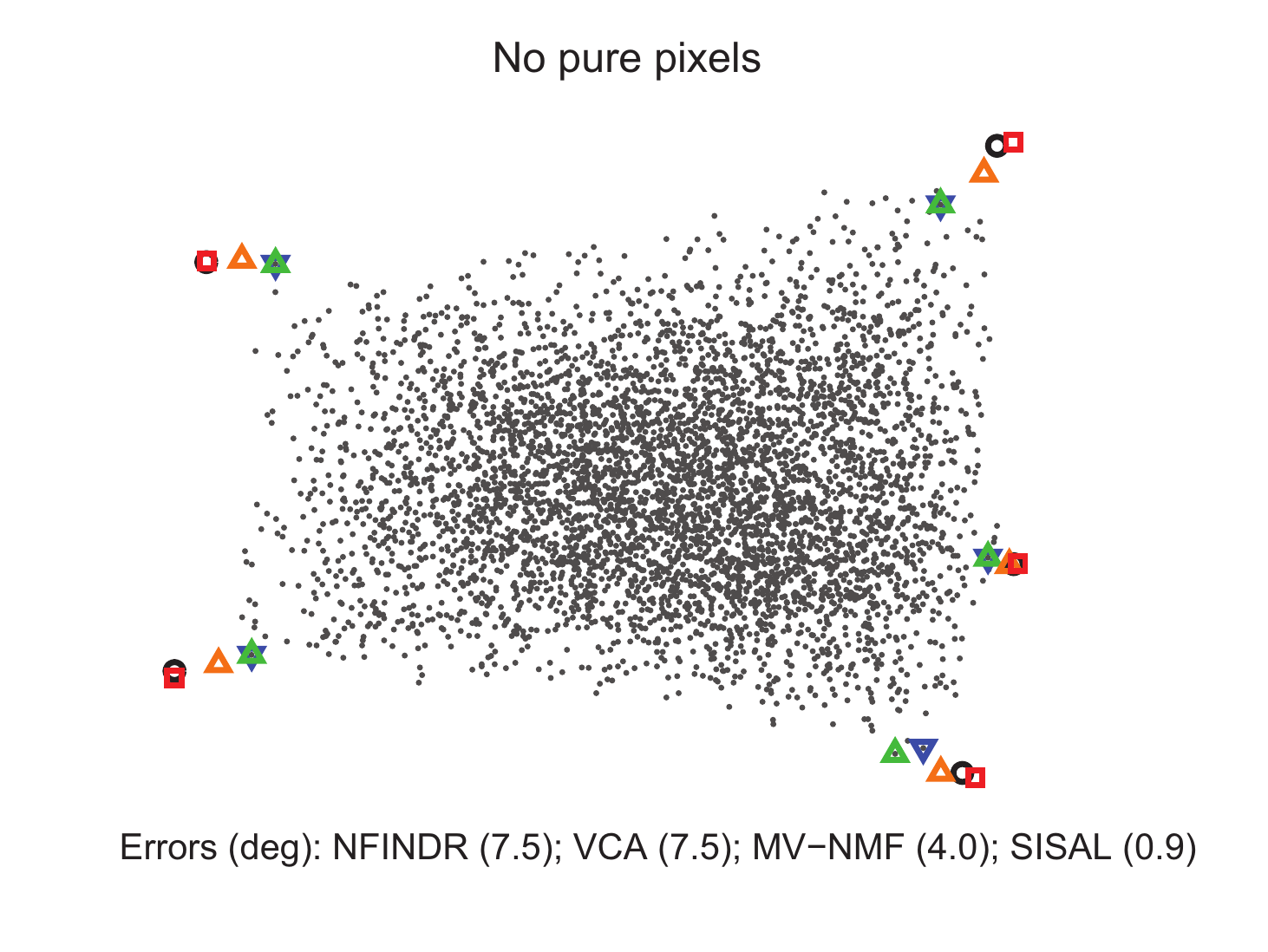}\\
\includegraphics[width=8cm]{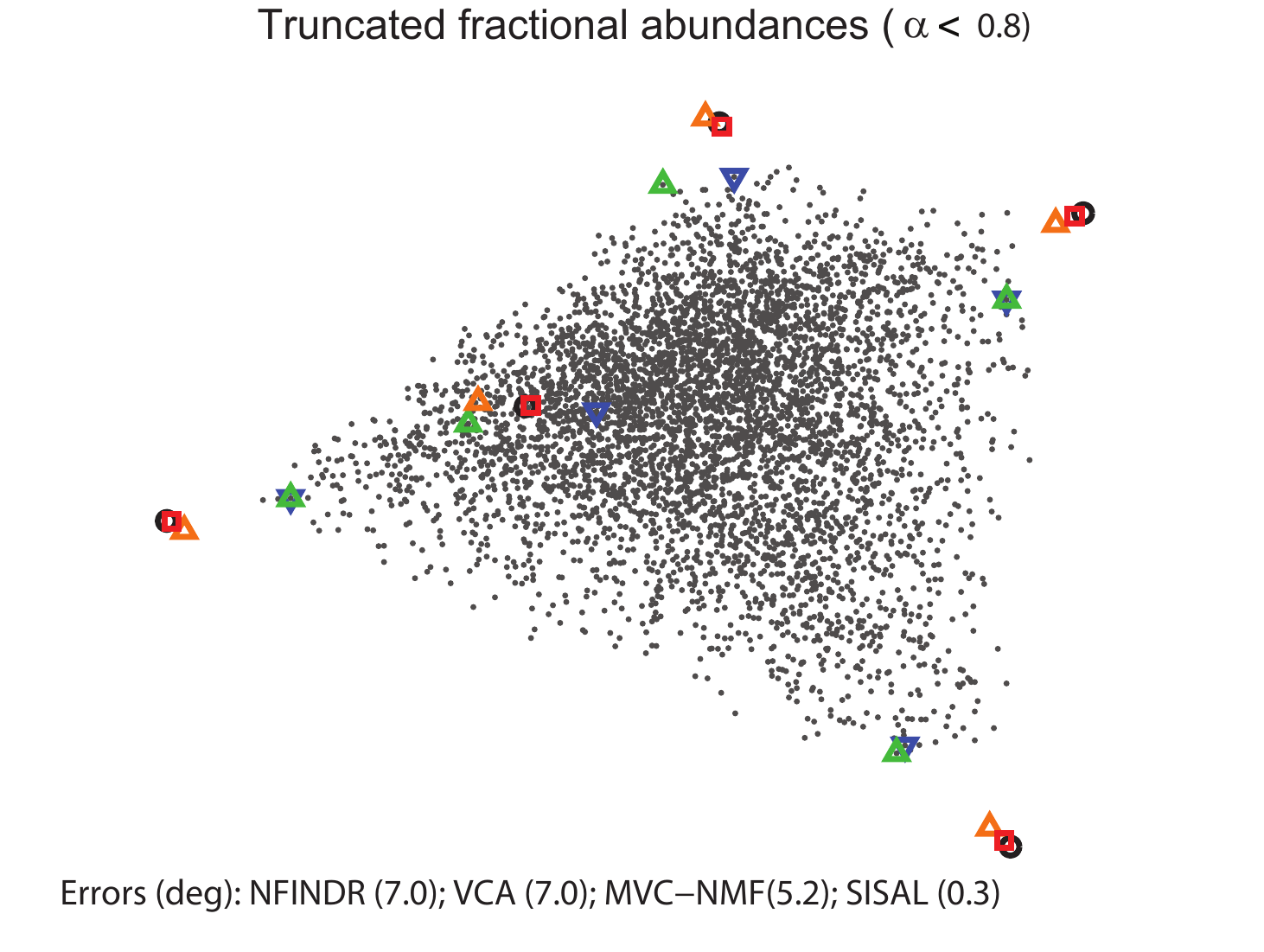}
\includegraphics[width=8cm]{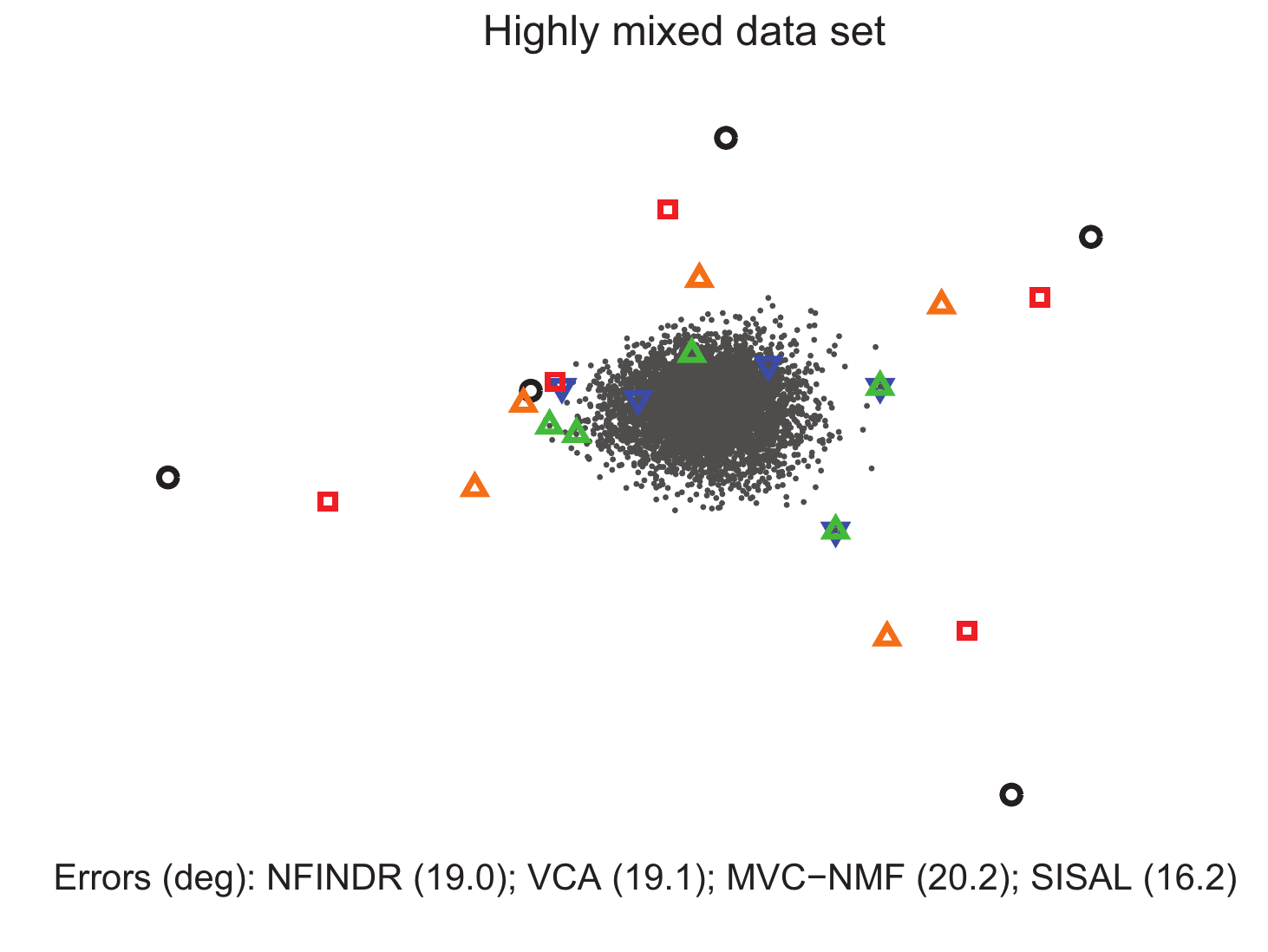}
\caption{Unmixing results of N-FINDR, VCA, MVC-NMF, and SISAL on different data sets: SusgsP5PPSNR30 - pure-pixel (top-left);  SusgsP5SNR30 - non pure pixel (top right); SusgsP5MP08SNR30 -  truncated fractional abundances (bottom left);  SusgsP5XS10SNR30 - and highly mixed (bottom tight).}
\label{fig:min_vol_results}
\end{figure*}

The iterative constrained endmembers (ICE) algorithm \cite{art:berman:TGRS:04} aims at
solving an optimization problem similar to that of MVC-NMF, where the volume of the
simplex is replaced by a much more manageable approximation: the sum of squared distances between all
the simplex vertices. This volume regularizer is quadratic and  well defined in any ambient
dimension and in degenerated simplexes.  These are relevant advantages over the  $|\text{det}({\bf M})|$
regularizer, which is non-convex  and prone to complications when the HU problem is badly conditioned or
if the number of endmembers is not exactly known. Variations of these ideas have recently been proposed in \cite{conf:Arngren:MLSP:09}, \cite{art:arngrenn}, \cite{art:arngrenJSPSS:09}, \cite{Col_NMF_li_igarss12}.
ICE implements a sequence of alternate minimizations with
respect to ${\bf S}$ and with respect to $\bf M$. An advantage of ICE  over MVC-NMF,
resulting from the use of a quadratic volume regularizer,
is that in the former one minimization is a quadratic programming problem while the other is a least squares problem that can be solved analytically, whereas in the MVC-NMF the
optimization with respect to $\bo{M}$ is a nonconvex problem. The sparsity-promoting ICE
(SPICE) \cite{art:zareGader:TGRS:07} is an extension of the ICE algorithm that
incorporates sparsity-promoting priors aiming at finding  the correct number of
endmembers.  Linear terms are added to the quadratic objective function, one for all the proportions associated with one endmember.  The linear term corresponds to an exponential prior.  A large number of endmembers are used in the initialization.  The prior tends to push all the proportions associated with particular endmembers to zero.  If all the proportions corresponding to an endmember go to zero, then that endmember can be discarded.  The addition of the sparsity promoting prior does not incur additional complexity to the model as the minimization still involves a quadratic program.

The quadratic volume regularizer used in the ICE and SPICE algorithms also provides robustness in the sense of allowing data points to be outside of the simplex $\mbox{conv}(\bo{M})$.  It has been shown that the ICE objective function can be written in the following way:

\begin{eqnarray}
\mathcal{I}\left(\bo{M}, \bo{S} \right) &=& \frac{1-\mu}{N}\|{\bf Y}-\bo{M}\bf{S}\|^2_F+{\mu}\, \text{trace}\left( \Sigma_{\bo{M}}\right)\\
\nonumber
&=&\frac{1-\mu}{N}\|\bf{Y}-\bo{M}\bf{S}\|^2_F+\mu\sum_{b=1}^B\sigma_b^2\\
\nonumber
\end{eqnarray}
where $\Sigma_{\bo{M}}$ is the sample covariance matrix of the endmembers and $\mu \in \left[0,1 \right]$ is a regularization parameter that controls the tradeoff between error and smaller simplexes.  If $\mu = 1$, then the best solution is to shrink all the endmembers to a single point, so all the data will be outside of the simplex.  If $\mu = 0$, then the best solution is one that yields no error, regardless of the size of the simplex.  The solution can be sensitive to the choice of $\mu$.  The SPICE algorithm has the same properties.  $L_1$ versions also exist \cite{ZareGaderIGARSS2010}.

It is worth noting the both Heylen et al. \cite{HeylenFCLS} and Silv‡n-C‡rdenas \cite{FuzzyAnalytic} have reported geometric-based methods that can either search for or analytically solve for the fully constrained least squares solution.

The $L_{1/2}$-NMF method introduced in \cite{special-issue-qian} formulates a nonnegative matrix factorization
problem similar to  (\ref{eq:mvt_NMF}),  where the volume regularizer is replaced with the
sparsity-enforcing regularizer $\|{\bf S}\|_{1/2}\equiv\sum_{i=1}^p\sum_{j=1}^n |\alpha_{ij}|^{1/2}$.
By promoting  zero or small abundance  fractions, this regularizer pulls endmember facets towards
the data cloud having an effect similar to  the volume regularizer.  The estimates of the endmembers
and of the fractional abundances are obtained by a modification of the multiplicative update rules
introduced in \cite{lee2001algorithms}.


Convex cone analysis (CCA) \cite{art:Ifarraguerri:TGRS:99}, finds  the boundary points of the data convex cone (it does not apply affine projection), what is very close to MV concept. CCA starts by selecting the   eigenvectors corresponding to the largest eigenvalues. These eigenvectors are then used as a basis to form linear combinations that have only nonnegative elements, thus belonging to a convex cone. The vertices of the convex cone
correspond to spectral vectors contains as many zero elements as the number of
eigenvectors minus one.

Geometric methods can be extended to piecewise linear mixing models.  Imagine the following scenario:  An airborne hyperspectral imaging sensor acquires data over an area.  Part of the area consists of farmland containing alternating rows of two types of crops (crop A and crop B) separated by soil whereas the other part consists of a village with paved roads, buildings (all with the same types of roofs), and non-deciduous trees.  Spectra measured from farmland are almost all linear mixtures of endmember spectra associated with crop A, crop B, and soil.   Spectra over the village are almost all linear mixtures of endmember spectra associated with pavement, roofs, and non-deciduous trees.  Some pixels from the boundary of the village and farmland may be mixtures of all six endmember spectra.  The set of all pixels from the image will then consist of two simplexes. Linear unmixing may find some, perhaps all, of the endmembers.   However, the model does not accurately represent the true state of nature.  There are two convex regions and the vertices (endmembers) from one of the convex regions may be in the interior of the convex hull of the set of all pixels.  In that case, an algorithm designed to find extremal points on or outside the convex hull of the data will not find those endmembers (unless it fails to do what it was designed to do, which can happen).  Relying on an algorithm failing to do what it is designed to do is not a desirable strategy.  Thus, there is a need to devise methods for identifying multiple simplexes in hyperspectral data.  One can refer to this class of algorithms as piecewise convex or piecewise linear unmixing.

One approach to designing such algorithms is to represent the convex regions as clusters.  This approach has been taken in \cite{ZareGaderFuzzPop,BchirWhisp2010,ZareWhips2010DetProb,PCE,ZareWhisp2010}.  The latter methods are Bayesian and will therefore be discussed in the next section.  The first two rely on algorithms derived from fuzzy and possibilistic clustering. Crisp clustering algorithms (such as k-means) assign every data point to one and only one cluster.  Fuzzy clustering algorithms allow every data point to be assigned to every cluster to some degree.  Fuzzy clusters are defined by these assignments, referred to as membership functions.  In the example above, there should be two clusters.  Most points should be assigned to one of the two clusters with high degree.  Points on the boundary, however, should be assigned to both clusters.

Assuming that there are $C$ simplexes in the data, then the following objective function can be used to attempt to find endmember spectra and abundances for each simplex:

\begin{equation}
J  =  \sum_{i=1}^{C}\left( \sum_{n=1}^{N}u_{in}^{2}\Vert\mathbf{y}_{n}-\mathbf{M}_{i}\bm{\alpha}_{in}\Vert_2^2 + \lambda\sum_{k=1}^{p-1}\sum_{j=k+1}^{p}\Vert\mathbf{e}_{ik}-\mathbf{e}_{ij}\Vert_2^2 \right)
\label{eqn:obj_function}
\end{equation}
such that
\begin{eqnarray}
& \alpha_{ikn}\geq 0 \quad \forall k = 1, \ldots, p; \quad \sum_{k=1}^p \alpha_{ikn} = 1\nonumber\\
& u_{ik}\geq 0 \quad \forall k = 1, \ldots, p; \quad \sum_{k=1}^p u_{ik} = 1.\nonumber
\label{eqn:PropMemConstraints}
\end{eqnarray}

Here, $u_{in}$ represents the membership of the $n^{th}$ data point in the $i^{th}$ simplex.  The other terms are very similar to those used in the ICE/SPICE algorithms except that there are $C$ endmember matrices and $NC$ abundance vectors.  Analytic update formulas can be derived for the memberships,  the endmember updates, and the Lagrange multipliers.  An update formula can be used to update the fractional abundances but they are sometimes negative and are then clipped at the boundary of the feasible region.  One can still use quadratic programming to solve for them.  As is the case for almost all clustering algorithms, there are local minima. However, the algorithm using all update formulas is computationally efficient.  A robust version also exists that uses a combination of fuzzy and possibilistic clustering \cite{ZareWhips2010DetProb}.

Fig. \ref{fig:min_vol_results}   shows  results of pure pixel based algorithms  (N-FINDR and VCA) and
MV based algorithms (MVC-NMF and SISAL) in simulated data sets representative of the classes of problems illustrated in
Fig. \ref{fig:simplices}.  These data sets have  $n = 5000$ pixels and SNR = 30$\,$dB and the following characteristics:  \textbf{SusgsP5PPSNR30} - pure pixels and abundances uniformly   distributed over the simplex (top left);
\textbf{SusgsP5SNR30} non pure pixels and abundances uniformly   distributed over the simplex (top right);
\textbf{SusgsP5MP08SNR30} abundances uniformly   distributed over the simplex  but truncated to 0.8 (bottom left);
\textbf{SusgsP5XS10SNR30} abundances with Dirichlet distributed with concentration parameter set to 10, thus yielding a highly mixed data set.

In the top left data set all algorithm produced very good results because pure pixels are present.
In the top right SISAL and MVC-NMF  produce good results but
VCA and N-FINDR shows a degradation in performance because there are no pure pixels.
In the bottom left  SISAL and MVC-NMF still produce good results but
VCA and N-FINDR show a significant degradation in performance because  the pixels close to the vertices were removed. Finally, in the bottom right all algorithm  produce unacceptable results because there are no pixels
in the vertex of the simplex neither  on its facets.   These  data sets are beyond the reach of geometrical based
algorithms.

%% file: stat_unmixing_ND.tex


\section{Statistical methods}

\label{sec:statistical_methods}

When the spectral mixtures are highly mixed, the geometrical based
methods yields poor results because there are not enough spectral
vectors in the  simplex facets. In these cases, the statistical
methods are a powerful alternative, which, usually, comes with
price:  higher computational complexity, when compared with the
geometrical based approaches.  Statistical methods also provide a natural framework for representing variability in endmembers. Under the statistical framework, spectral unmixing is formulated as
a statistical inference problem.

Since, in most cases, the number of substances and their reflectances are not known,
hyperspectral unmixing falls into the class of blind source separation problems
\cite{bi:Common_94}. Independent Component Analysis (ICA), a well-known tool in blind source separation, has been proposed as a
tool to blindly unmix hyperspectral data
\cite{bi:Bayliss_97,bi:Chen_99,bi:Tu_00}. Unfortunately, ICA is based on the assumption of
mutually independent sources (abundance fractions), which is not the case of
hyperspectral data, since the sum of abundance fractions is constant, implying
statistical dependence among them. This dependence compromises ICA applicability to
hyperspectral data as shown in \cite{art:nascimeto_bioucas_ICA_05,bi:Keshava_00}. In
fact, ICA finds the endmember signatures by multiplying the spectral vectors with an
unmixing matrix which minimizes the mutual information among channels. If sources are
independent, ICA provides the correct unmixing, since the minimum of the mutual
information corresponds to and only to independent sources. This is no longer true for
dependent fractional abundances. Nevertheless, some endmembers may be approximately
unmixed. These aspects are addressed in \cite{art:nascimeto_bioucas_ICA_05}.

Bayesian approaches have the ability to model statistical variability and to impose priors that can constrain solutions to physically meaningful ranges and regularize solutions.  The latter property is generally considered to be a requirement for solving ill-posed problems.  Adopting a Bayesian framework, the
inference engine is the posterior density of the random quantities
to be estimated. When the unknown mixing matrix $\bf M$ and the
abundance fraction matrix ${\bf S}$ are assumed to be a priori
independent, the Bayes paradigm allows the joint posterior of $\bf
M$ and ${\bf S}$ to be computed as
\begin{equation}
\label{eq:posterior_distribution}
     p_{M,S|Y}({\bf M}, {\bf S}|{\bf Y}) =  p_{Y|M,S}({\bf Y}|{\bf M}, {\bf S})p_M({\bf M})p_S({\bf S})/p_Y({\bf Y}),
\end{equation}
where the notation $p_A$ and $p_{A|B}$ stands for the probability
density function (pdf) of $\bf A$  and of $\bf A$ given $\bf B$,
respectively. In \eqref{eq:posterior_distribution}, $p_{Y|M,S}({\bf
Y}|{\bf M}, {\bf S})$ is the likelihood function depending on the
observation model and the prior distribution $p_M({\bf M})$ and
$p_S({\bf S})$ summarize the prior knowledge regarding these unknown
parameters.

A popular Bayesian estimator is \cite{book:bernardo:94} the joint
{\em maximum a posteriori} (MAP) estimator given by
\begin{eqnarray}
    \label{eq:map_estimate}
    (\widehat{\bf M}, \widehat{\bf S})_{\mathrm{MAP}}& \equiv &\arg\max_{{\bf M}, {\bf S}} p_{M,S|Y}({\bf M}, {\bf S}|{\bf
    Y})\\
                  &=& \arg\min - \log p_{Y|M,S}({\bf Y}|{\bf M}, {\bf S}) -\log p_M({\bf M}) -\log p_S({\bf S}).
\end{eqnarray}
Under the linear mixing model and assuming the
noise random vector $\bf w$ is Gaussian with  covariance matrix
$\sigma^2\bf I$, then, we have $ -\log p_{Y|M,S}({\bf Y}|{\bf M, S})
= (1/(2\sigma^2))\|{\bf Y}-{\bf MS}\|^2_F+\text{const}$. It is then
clear that ICE/SPICE \cite{art:zareGader:TGRS:07} and MVC-NMF
\cite{miao07} algorithms, which have been classified as geometrical,
can also be classified as statistical, yielding joint MAP estimates
in \eqref{eq:map_estimate}. In all these algorithms, the estimates
are obtained by minimizing a two-term objective function: $-\log
p_{Y|M,S}({\bf Y}|{\bf M}, {\bf S})$ plays the role of a data
fitting criterion and $-\log p_M({\bf M})-\log p_S({\bf S})$
consists of a penalization. Conversely, from a Bayesian perspective,
assigning prior distributions $p_M({\bf M})$ and $p_S({\bf S})$ to
the endmember and abundance matrices ${\bf M}$ and ${\bf A}$,
respectively, is a convenient way to ensure physical constraints
inherent to the observation model.

The work \cite{art:parra:00} introduces a Bayesian approach where
the linear mixing model with
zero-mean white Gaussian noise of covariance $\sigma^2\bf I$ is
assumed, the  fractional abundances are uniformly distributed on the
simplex, and the prior on $\bf M$ is an autoregressive model.
Maximization of the negative log-posterior distribution is then
conducted in an iterative scheme. Maximization with respect to the
abundance coefficients is formulated as $n$ weighted least square
problems with linear constraints that are solved separately.
Optimization with respect to $\bf M$ is conducted using a
gradient-based descent.

The Bayesian approaches introduced in
\cite{art:Moussaoui:06,Dobigeon2009sp,art:Dobigeon:TGRS:09,Arngren2011jsps}
have all the same flavor. The posterior distribution of the
parameters of interest is computed from the linear mixing model
 within a hierarchical Bayesian model, where conjugate
prior distributions are chosen for some unknown parameters to
account for physical constraints. The hyperparameters involved in
the definition of the parameter priors are then assigned
non-informative priors and are jointly estimated from the full
posterior of the parameters and hyperparameters. Due to the
complexity of the resulting joint posterior, deriving closed-form
expressions of the MAP estimates or designing an optimization scheme
to approximate them remain impossible. As an alternative, Markov
chain Monte Carlo algorithms are proposed to generate samples that
are asymptotically distributed according to the target posterior
distribution. These samples are then used to approximate the {\em
minimum mean square error} (MMSE) (or posterior mean) estimators of
the unknown parameters
\begin{eqnarray}
    \label{eq:mmmse_estimate}
    \widehat{\bf M}_{\mathrm{MMSE}}& \equiv \mathbb{E}[{\bf M}|{\bf Y}] & =\int {\bf M} p_{{ M}|{ Y}}({\bf M}|{\bf Y}) d{\bf M}\\
    \widehat{\bf S}_{\mathrm{MMSE}}& \equiv \mathbb{E}[{\bf S}|{\bf Y}] & =\int {\bf S} p_{{ S}|{ Y}}({\bf S}|{\bf Y}) d{\bf S}.
\end{eqnarray}
These algorithms mainly differ by the choice of the priors assigned
to the unknown parameters. More precisely, in
\cite{art:Moussaoui:06,art:Moussaoui:06b}, spectral unmixing is
conducted for spectrochemical analysis. Because of the sparse nature
of the chemical spectral components, independent Gamma distributions
are elected as priors for the spectra. The mixing coefficients are
assumed to be non-negative without any sum-to-one constraint.
Interest of including this additivity constraint for this specific
application is investigated in \cite{Dobigeon2009sp} where uniform
distributions over the admissible simplex are assigned as priors for
the abundance vectors. Note that efficient implementations of both
algorithms for operational applications are presented in
\cite{Moussaoui2008neurocomp} and \cite{Schmidt2010tgrs},
respectively.

\begin{figure}[h]
    \centering
  \includegraphics[width=8cm,angle=0]{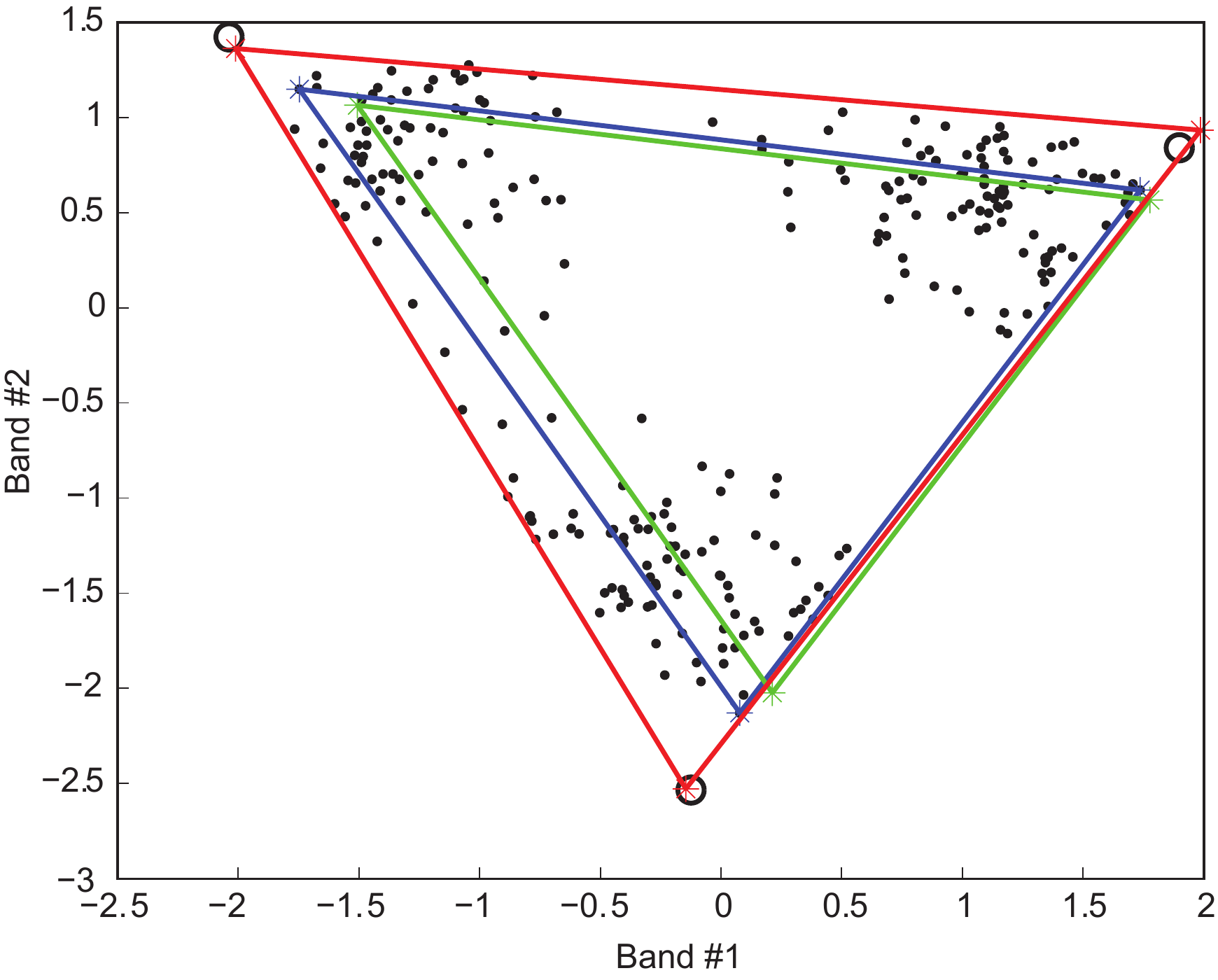}
  \caption{Projected pixels (black points), actual endmembers (black circles), endmembers estimated by
N-FINDR (blue stars), endmembers estimated by VCA (green stars) and
endmembers estimated by the algorithm in \cite{art:Dobigeon:TGRS:09}
(red stars}.
  \label{fig:stat_BLU_vs_NFINDR}
\end{figure}

In \cite{art:Dobigeon:TGRS:09}, instead of estimating the endmember
spectra in the full hyperspectral space, Dobigeon \emph{et al.}
propose to estimate their projections onto an appropriate lower
dimensional subspace that has been previously identified by one of
the dimension reduction technique described in paragraph
\ref{sect:background}. The main advantage of this approach is to
reduce the number of degrees of freedom of the model parameters
relative to other approaches, e.g.,
\cite{art:Moussaoui:06,art:Moussaoui:06b,Dobigeon2009sp}. Accuracy
and performance of this Bayesian unmixing algorithm when compared to
standard geometrical based approaches is depicted in
Fig.~\ref{fig:stat_BLU_vs_NFINDR} where a synthetic toy example has
been considered. This example is particularly illustrative since it
is composed of a small dataset where the pure pixel assumption is
not fulfilled. Consequently, the geometrical based approaches that
attempt to maximize the simplex volume (e.g., VCA and N-FINDR) fail
to recover the endmembers correctly, contrary to the statistical
algorithm that does not require such hypothesis.

Note that in \cite{Dobigeon2009sp}, \cite{art:Dobigeon:TGRS:09} and
\cite{Arngren2011jsps} independent uniform distributions over the
admissible simplex are chosen as prior distributions for the
abundance vectors. This assumption, which is equivalent of choosing
Dirichlet distributions with all hyperparameters equal to $1$, could
seem to be very weak. However, as demonstrated in
\cite{art:Dobigeon:TGRS:09}, this choice favors estimated endmembers
that span a simplex of minimum volume, which is precisely the
founding characteristics of some geometrical based unmixing
approaches detailed in paragraph \ref{sec:geo_unmixing}.

Explicitly constraining the volume of the simplex formed by the
estimated endmembers has also been considered in
\cite{Arngren2011jsps}. According to the optimization perspective
suggested above, penalizing the volume of the recovered simplex can
be conducted by choosing an appropriate negative log-prior $- \log
p_M(\bf M)$. Arngren \emph{et al.} have investigated three measures
of this volume: exact simplex volume, distance between vertices,
volume of a corresponding parallelepiped. The resulting techniques
can thus be considered as stochastic implementations of the MVC-NMF
algorithm \cite{miao07}.

All the Bayesian unmixing algorithms introduced above rely on the
assumption of an independent and identically Gaussian distributed
noise, leading to a covariance matrix $\sigma^2\bf I$ of the noise
vector $\bf w$. Note that the case of a colored Gaussian noise with
unknown covariance matrix has been handled in
\cite{Dobigeon2008icassp}. However, in many applications, the
additive noise term may neglected because the noise power is very
small. When that is not the case but the signal subspace has much
lower dimension than the number of bands, then, as seen in Section
\ref{sect:background}, the projection onto the signal subspace
largely reduces the noise power. Under this circumstance, and
assuming that ${\bf M}\in\mathbb{R}^{p\times p}$ is invertible and
the observed spectral vectors are independent, then we can write
\[
    p_{Y|M}({\bf Y}|{\bo M}) =  \left(\prod_{i=1}^n  p_\alpha ({\bf M}^{-1}{\bf y}_i)\right)|\text{det}({\bf M}^{-1})|^n,
\]
where $p_\alpha$ is the fractional abundance pdf,  and compute the
{em maximum likelihood} (ML) estimate of ${\bf W}\equiv {\bf
M}^{-1}$. This is precisely the ICA line of attack,  under the
assumption that the fractional abundances are independent, {\em
i.e.}, $p_\alpha=\prod_{k=1}^{p}p_{\alpha_k}$.  The fact that  this
assumption is not valid in hyperspectral applications
\cite{art:nascimeto_bioucas_ICA_05} has promoted research on
suitable statistical models for hyperspectral fractional abundances
and  in effective algorithms to infer the mixing matrices. This is
the case with DECA  \cite{deca_igars07, art:deca:TGRS:10}; the
abundance fractions are modeled as mixtures of Dirichlet densities,
thus, automatically enforcing the constraints on abundance fractions
imposed by the acquisition process, namely nonnegativity and
constant sum. A cyclic minimization algorithm is developed where: 1)
the number of Dirichlet modes is inferred based on the minimum
description length (MDL) principle; 2) a generalized expectation
maximization (GEM) algorithm is derived to infer the model
parameters; 3) a sequence of augmented Lagrangian based
optimizations are used to compute the signatures of the endmembers.

Piecewise convex unmixing, mentioned in the geometrical approaches section, has also been investigated using a Bayesian approach\footnote{It is an interesting to remark that by taking the negative of the logarithm of a fuzzy clustering objective function, such as in Eq. \ref{eqn:obj_function}, one can represent a fuzzy clustering objective as a Bayesian MAP objective.  One interesting difference is that the precisions on the likelihood functions are the memberships and are data point dependent.}  In \cite{ZareGaderMaxEnt2010} the normal compositional model is used to represent each convex set as a set of samples from a collection of random variables.  The endmembers are represented as Gaussians.  Abundance multinomials are represented by Dirichlet distributions.  To form a Bayesian model, priors are used for the parameters of the distributions.  Thus, the data generation model consists of two stages.  In the first stage, endmembers are sampled from their respective Gaussians.  In the second stage, for each pixel, an abundance multinomial is sampled from a Dirichlet distribution.   Since the number of convex sets is unknown, the Dirichlet process mixture model is used to identify the number of clusters while simultaneously learning the parameters of the endmember and abundance distributions.  This model is very general and can represent very complex data sets.  The Dirichlet process uses a Metropolis-within-Gibbs method to estimate the parameters, which can be quite time consuming.   The advantage is that the sampler will converge to the joint distribution of the parameters, which means that one can select the maximum a-posterior estimates from the estimated joint distributions.  Although Gibbs samplers seem inherently sequential, some surprising new theoretical results by \cite{VikPhD} show that theoretically correct sampling samplers can be implemented in parallel, which offers the promise of dramatic speed-ups of algorithms such as this and other probabilistic algorithms mentioned here that rely on sampling.

\begin{figure}[!t]
  \centering
     \includegraphics[width=80mm]{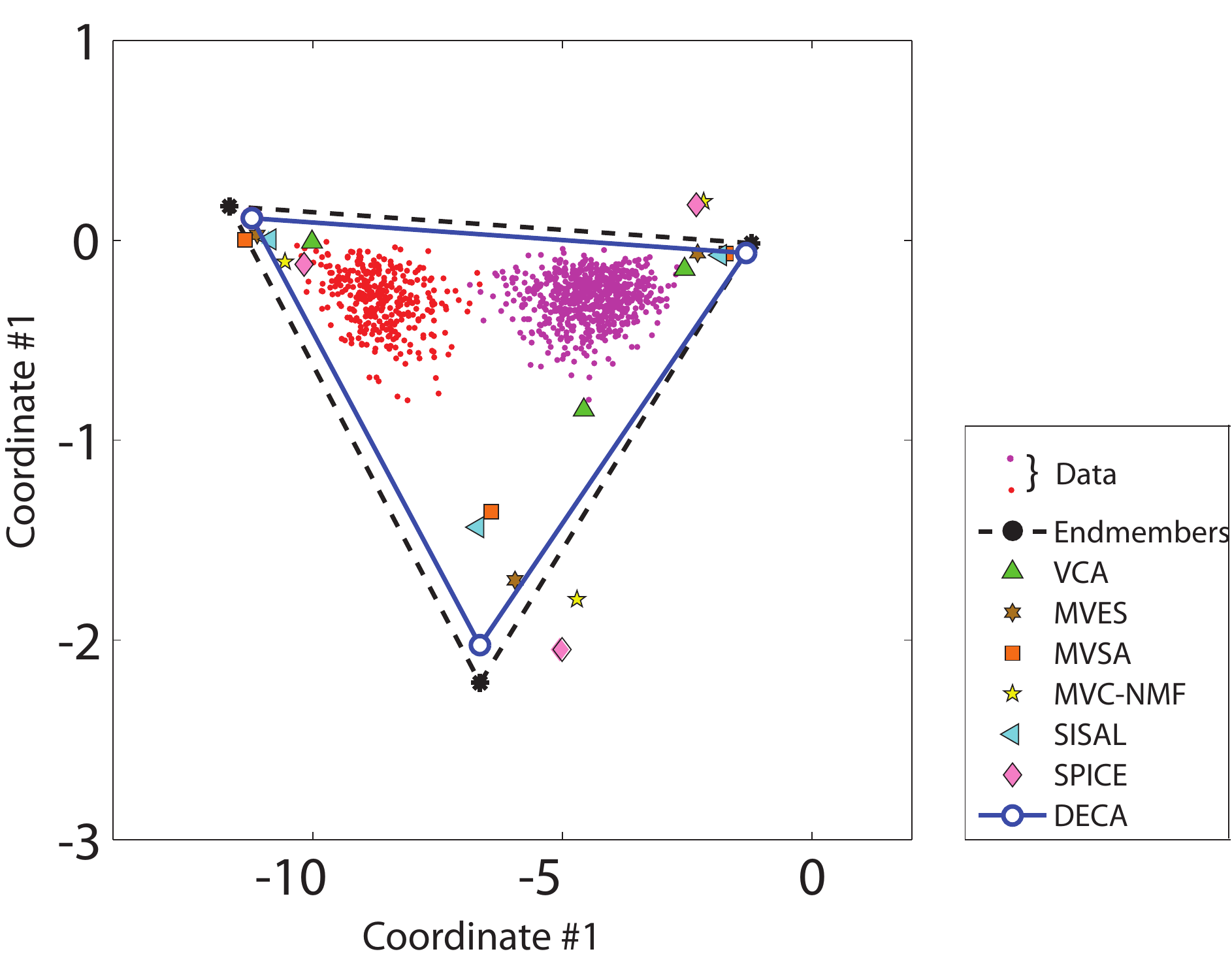}
     \includegraphics[width=87mm]{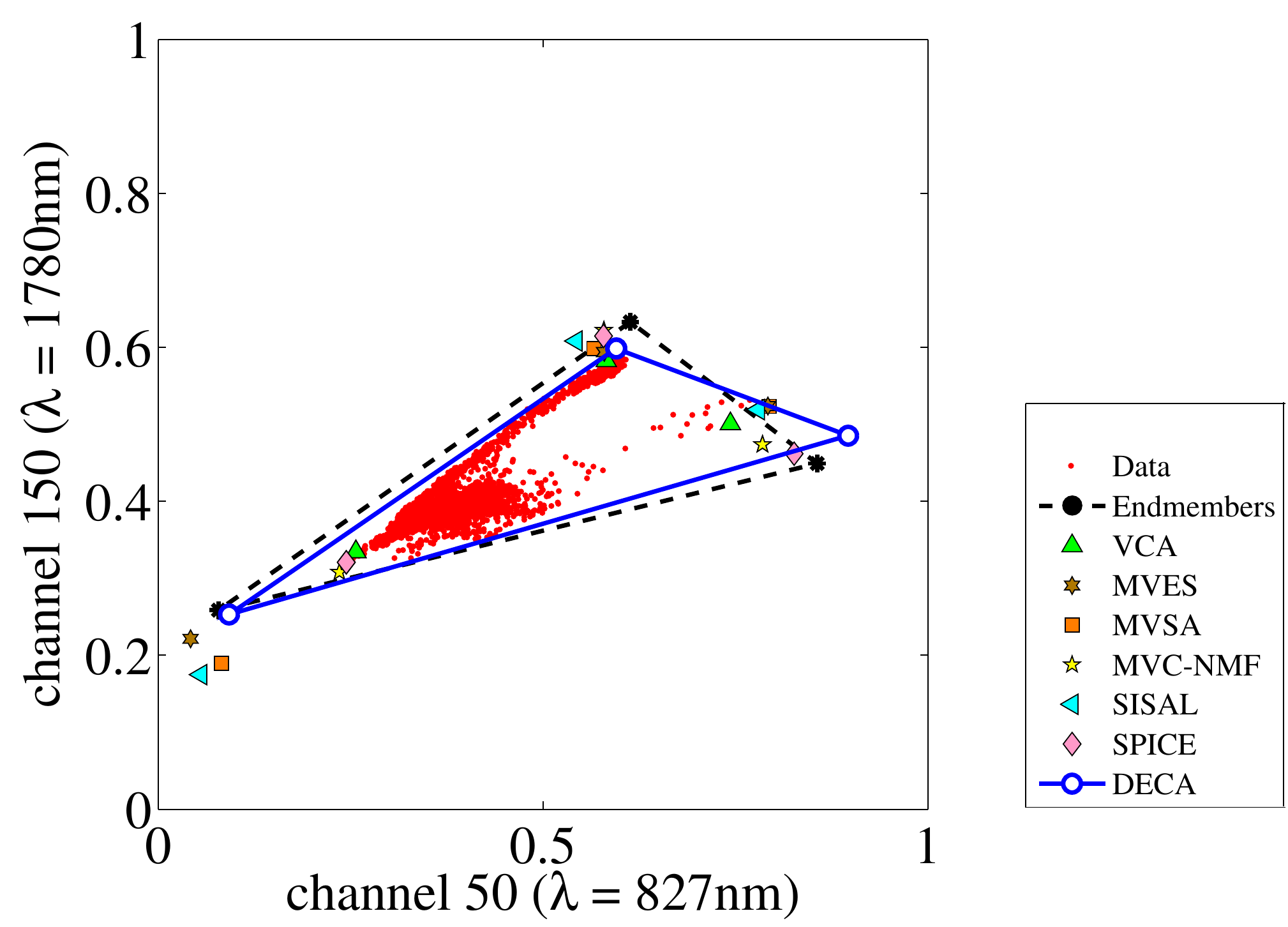}
     \caption{ Left: Scatterplot  of the \textbf{SusgsP3SNRinfXSmix}   dataset jointly    with the
                     true   and estimated endmembers. Right: Scatterplot of a Cuprite
                     data subset jointly with the  projections of Montmorillonite, Desert Varnish, and Alunite,
                     witch are known to dominate this subset,
                     and estimated endmembers.
           }
  \label{fig:deca_scatter_mix}
\end{figure}

Fig.~\ref{fig:deca_scatter_mix}, left,  presents a scatterplot  of the simulated data set {\bf SusgsP3SNRinfXSmix}
and the endmember estimates produced by VCA, MVES, MVSA, MVC-NMF, SISAL, and DECA algorithms. This data set
is generated with a mixing matrix $\bf M$ sampled from the USGS library and with   $p=3$ endmembers,
$n=10000$  spectral vectors, and fractional abundances given by mixtures of  two Dirichlet
modes with parameters $[6,\,25,\,9]$ and $[7,\,8,\,23]$ and  mode weights of 0.67 and 0.33, respectively.
DECA Dirichlet parameters were  randomly initialized and the mixing probabilities
uniformly.  This setting reflects a situation in which no knowledge
of the size and the number of regions in the scene exists.
The parameters of the remaining methods were hand tuned
for optimal performance\footnote{MVC-MNF regularization parameter: $\tau = 1^{-4}$;
MVES tolerance: $10^{-6}$; SISAL regularization parameter $\lambda = 10$;
SPICE regularization parameter $\mu = 10^{-3}$;
SPICE sparsity parameter $\Gamma = 0.5$; SPICE stopping parameter $10^{-6}$.}.
See [170] for more details.
 The considered data set corresponds to a highly mixed scenario, where  the geometrical based algorithms performs poorly, as explained in Section \ref{sec:unmix}. On the contrary,  DECA  yields useful estimates.

\begin{figure}[!t]
      \centering
        \includegraphics[width=62mm]{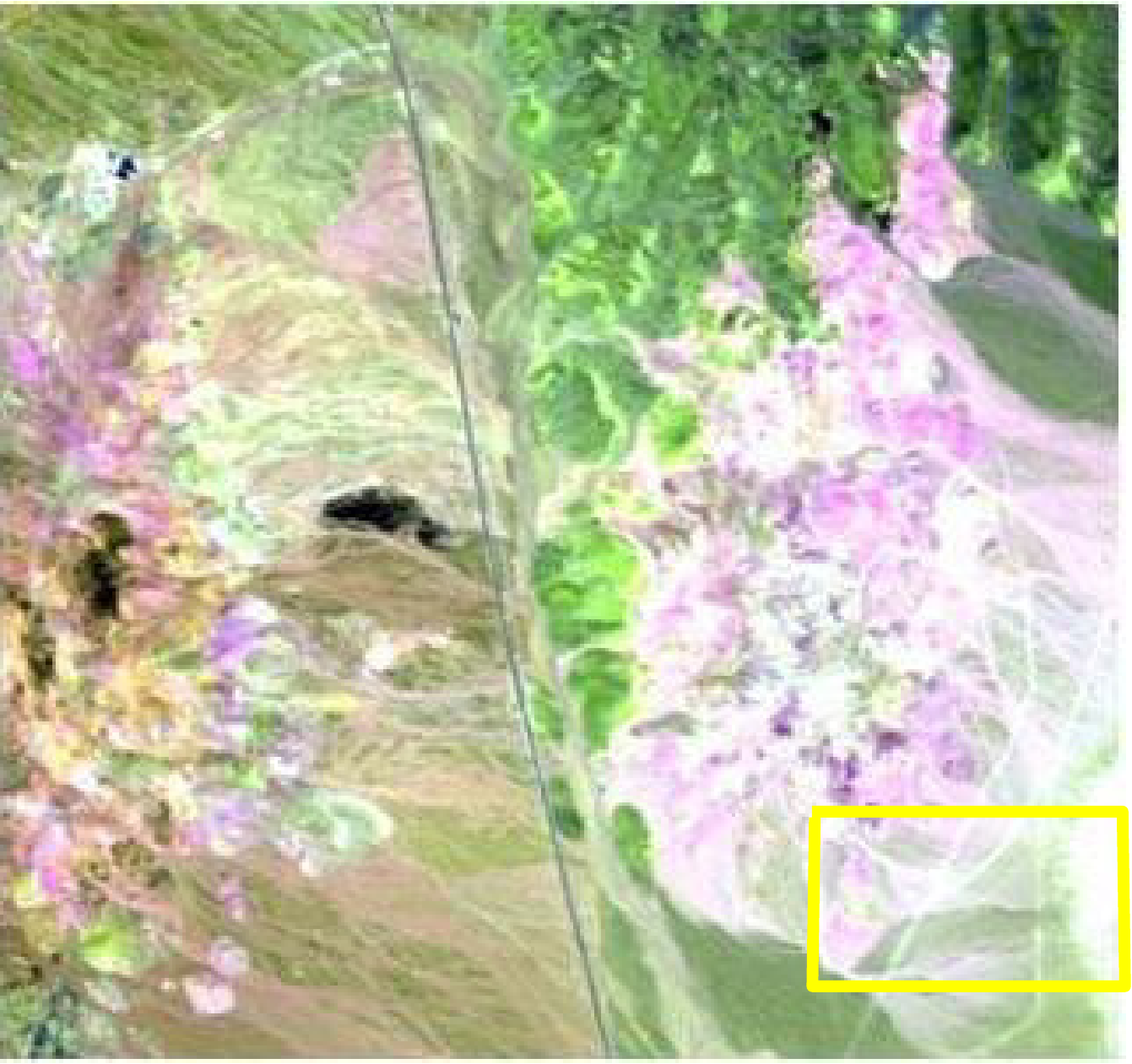}
        \caption{AVIRIS subset and of  $30$ (wavelength $\lambda=667.3nm$)
        used to compute  the results plotted in Fig. \ref{fig:deca_scatter_mix}, right.
      }
      \label{fig:cuprite_image_subset}
\end{figure}

Fig.~\ref{fig:deca_scatter_mix}, right, is similar the one in the left hand side for a  Cuprite
data subset of size $50 \times 90$ pixels shown in  Fig. \ref{fig:cuprite_image_subset}. This subset is
dominated by Montmorillonite, Desert Varnish, and Alunite, which are known to
dominate the considered subset image \cite{art:clark:93}. The projections of this endmembers are  represented
by black circles. DECA  identified $k=5$ modes, with parameters $\bm{\theta}_1 =[1.5,\,
4.1,\, 2.9]$, $\bm{\theta}_2 =[23.4,\, 51.3,\, 15.5]$,
$\bm{\theta}_3 =[27.2,\, 26.6,\, 4.3]$, $\bm{\theta}_4 =[17.5,\,3.6,\, 2.5]$, and $\bm{\theta}_5 =[10.3,\, 8.0,\, 7.3]$, and mode weights $\epsilon_1 = 0.04$, $\epsilon_2 = 0.69$, $\epsilon_3
= 0.07$, $\epsilon_4 = 0.10$, and $\epsilon_5 = 0.10$. These parameters correspond to a highly non-uniform distribution over the simplex as could be inferred from the scatterplot.  Although the estimation results are  more difficult to judge in the case of real data than in the case on simulated data, as we not really sure about the true endmembers,
it is reasonable to conclude that the statistical approach is producing  similar to or better estimates  than  the
geometrical based algorithms.

The examples shown Fig.~\ref{fig:deca_scatter_mix} illustrates the  potential and flexibility of the Bayesian methodology. As already referred to above, these advantages come at  a price:  computational complexity linked to the posterior computation and to the inference of the estimates.

%% file: sparse_unmixing.tex


\section{Sparse regression based unmixing}
\label{sec:sparse_regress}

The spectral unmixing problem has recently been  approached in a semi-supervised fashion,
by assuming that the observed image signatures can be expressed in the form of linear
combinations of a number of pure spectral signatures known in advance\cite{art:sparseUnmix:TGARS:10, art:isma:06, conf:sparseUnmix:whispers:10} ({\em e.g.},  spectra
collected on the ground by a field spectro-radiometer). Unmixing then amounts to finding
the optimal subset of signatures in a (potentially very large) spectral library that can
best model each mixed pixel in the scene. In practice, this is a combinatorial problem
which calls for efficient linear sparse regression techniques based on sparsity-inducing
regularizers, since the number of endmembers participating in a mixed pixel is usually
very small compared with the (ever-growing) dimensionality – and availability – of
spectral libraries \cite{bi:Keshava_02}. Linear sparse regression is an area of very active research with strong links to compressed sensing \cite{art:candes:CPAM:06,art:donoho06,baraniuk2007compressive},
least angle regression \cite{efron2004least}, basis pursuit, basis pursuit denoising \cite{ChenDonohooSounders01}, and matching pursuit \cite{mallat1993matching}, \cite{pati03}.

Let us assume then that the spectral endmembers used to solve the mixture problem are no
longer extracted nor generated using the original hyperspectral data as input, but
instead selected from a library ${\bf A}\in\mathbb{R}^{B\times m}$ containing a large number of  spectral samples, say $m$,  available a priori. In this case, unmixing amounts to finding the optimal subset of samples in the library that can best model each mixed pixel in the scene.  Usually, we have $m > B$ and therefore the linear problem at hands is  underdetermined. Let ${\bf x}\in \mathbb{R}^m$ denote the
fractional abundance vector with regards to the library $\bf A$.  With these
definitions in place, we can now write our sparse regression problem as
\begin{align}
   &\min_{\bf x}\|{\bf x}\|_0\;\;\mbox{subject to} \;\; \|{\bf Ax}-{\bf y}\|_2\leq \delta,\;\; {\bf x}\succeq {\bf 0},
   \label{eq:LSR_}
\end{align}
where $\|{\bf x}\|_0$  denotes the number of non-zero components of $\bf x$ and
$\delta\geq 0$ is  the error tolerance due to noise and modeling errors. Assume for a while that
$\delta = 0$.  If the system of linear equations ${\bf Ax = y}$ has a solution satisfying $2\,\|{\bf x}\|_0< \text{spark}({\bf A})$, where $\textrm{spark}({\bf A})\leq \text{rank}{\bf A} +1$  is the smallest number of linearly dependent columns of $\bf A$, it is necessarily the unique solution of $(\ref{eq:LSR_})$ \cite{donoho2003optimally}. For $\delta > 0$, the concept of uniqueness of the sparsest solution is  replaced with the concept of stability
\cite{art:candes:CPAM:06}.

In most HU applications, we do have $2\,\|{\bf x}\|_0 \ll  \text{spark}({\bf A})$ and therefore,
at least in noiseless scenarios, the solutions of  (\ref{eq:LSR_}) are unique.  However, problem (\ref{eq:LSR_}) is NP-hard \cite{Natarajan95} and therefore there is no hope in solving it in a straightforward way.  Greedy algorithms such as the orthogonal matching  pursuit (OMP) \cite{pati03} and convex approximations replacing the
the $\ell_0$ norm  with the $\ell_1$ norm, termed  basis pursuit (BP), if $\delta = 0$, and  basis pursuit denoising (BPDN )\cite{ChenDonohooSounders01}, if $\delta > 0$, are alternative approaches to compute the sparsest solution.  If we add the ANC to BP and BPDN problems, we have the constrained basis pursuit (CBP) and the  constrained basis pursuit denoising (CBPDN)  problems \cite{conf:sunsal:10}, respectively.  The  CBPDN optimization problem is thus
\begin{align}
   &\min_{\bf x}\|{\bf x}\|_1\;\;\mbox{subject to} \;\; \|{\bf Ax}-{\bf y}\|_2\leq \delta,\;\; {\bf x}\succeq {\bf 0}.
   \label{eq:P1}
\end{align}
An equivalent formulation to (\ref{eq:P1}), termed constrained sparse regression (CSR) (see \cite{conf:sunsal:10}), is
\begin{align}
   &\min_{\bf x} \, (1/2)\|{\bf Ax}-{\bf y}\|_2+\lambda\|{\bf x}\|_1\;\;\mbox{subject to} \;\;  {\bf x}\succeq {\bf 0},
   \label{eq:UP1}
\end{align}
where $\lambda > 0$ is related with the Lagrange multiplier of the inequality
$\|{\bf Ax}-{\bf y}\|_2\leq \delta$, also interpretable as a regularization parameter.

Contrary to the  problem (\ref{eq:LSR_}), problems (\ref{eq:P1}) and (\ref{eq:UP1}) are convex and  can be solved efficiently \cite{conf:sunsal:10, conf:guospie:09}.   What is, perhaps,  totally
unexpected is that sparse vector of fractional abundances can be reconstructed by
solving (\ref{eq:P1}) or (\ref{eq:UP1}) provided that the columns of  matrix $\bf A$
are incoherent in a given sense \cite{candes2007sparsity}. The applicability of  sparse regression to HU was  studied in detail in  \cite{art:sparseUnmix:TGARS:10}.
Two main conclusions were drawn:
\begin{description}
  \item[a)] hyperspectral signatures tend to be highly correlated what imposes limits to the
            quality of the results provided by  solving CBPDN or CSR optimization problems.

  \item[b)] The  limitation imposed  by the  highly correlation  of the  spectral signatures is
            mitigated by the high level of sparsity most often observed in the hyperspectral
            mixtures.
\end{description}

At this point, we make a brief comment about the role of ASC in the
context of CBPDN and of CSR problems. Notice that  if  $\bf x$ belongs to the unit simplex
({\em i.e.},  $x_i\geq 0$ for $i=1,\dots,m$,  and $\sum_{i=1}^m x_i =1 $), we have $\| {\bf x}\|_1 = 1$.  Therefore,  if we add the sum-to-one constraint to   (\ref{eq:P1}) and (\ref{eq:UP1}),
the  corresponding optimization problems do not  depend on the $\ell_1$ norm  $\| {\bf x}\|_1$.
In this case, the  optimization  (\ref{eq:UP1}) is converted into  the well known fully constrained  least squares (FCLS) problem and (\ref{eq:P1}) into a feasibility problem, which for
$\delta = 0$ is
\begin{align}
  \label{eq:feas_under_nonneg}
  \text{solution} \, \{\bf Ax = y\}\;\; \text{subject to} \;\;{\bf x \succeq 0}.
\end{align}
The uniqueness of  sparse solutions to (\ref{eq:feas_under_nonneg}) when the system is  underdetermined  is addressed in \cite{bruckstein2008uniqueness}. The main finding is that
for matrices $\bf A$ with a row-span intersecting the positive orthant (this is the case of hyperspectral libraries), if this problem admits a sufficiently sparse solution, it is necessarily unique. It is remarkable how   the  ANC    alone  acts as
a sparsity-inducing regularizer.

In practice, and for the reasons pointed Section \ref{sec:aff_proj}, the ASC is rarely satisfied.  For this reason, and also  due to the presence of noise and model mismatches, we have  observed that
the CBPDN and CSR often yields  better  unmixing results than  CLS and FCLS.

\begin{figure}[!t]
      \centering
        \includegraphics[width=80mm]{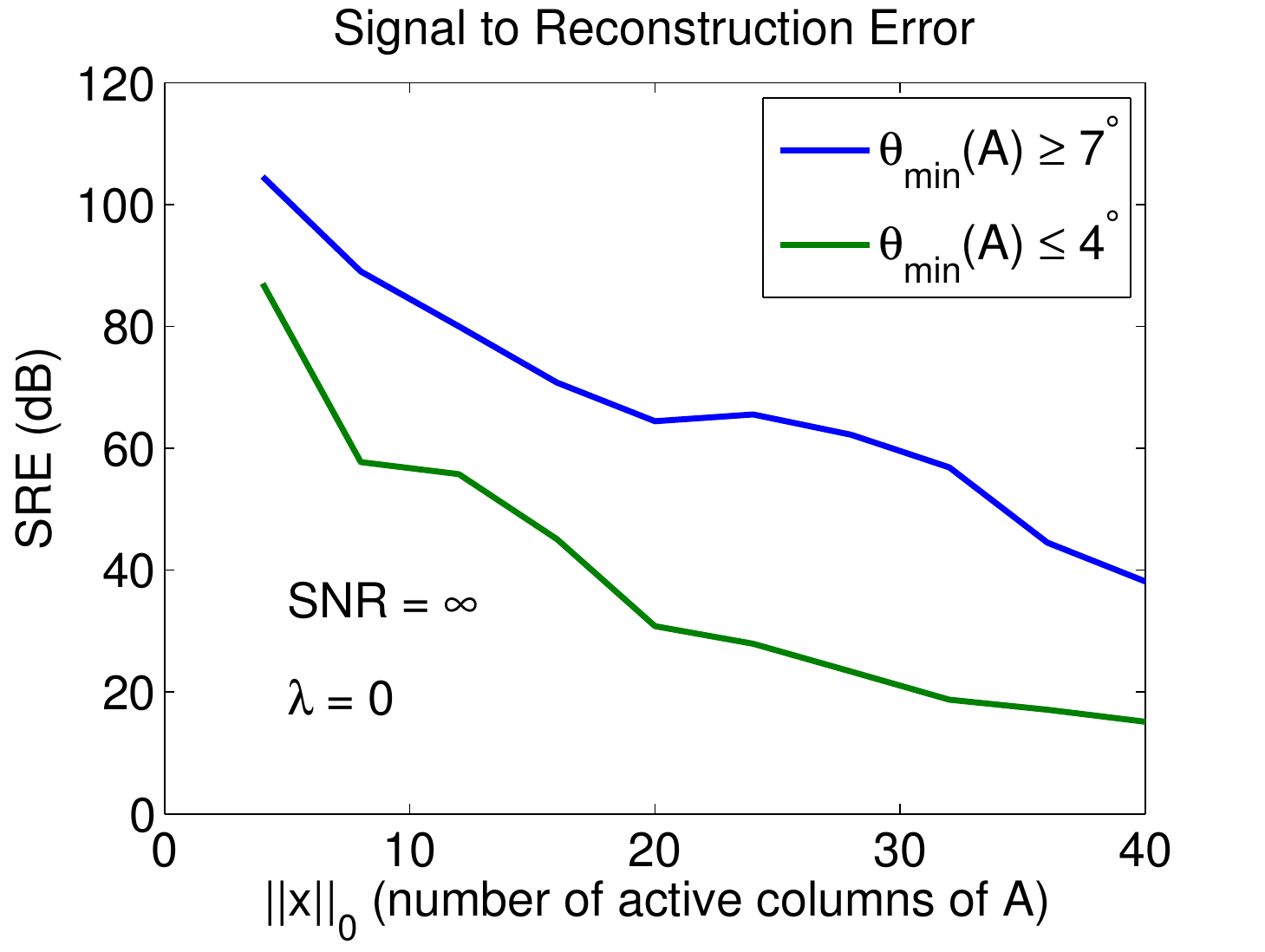}
        \includegraphics[width=80mm]{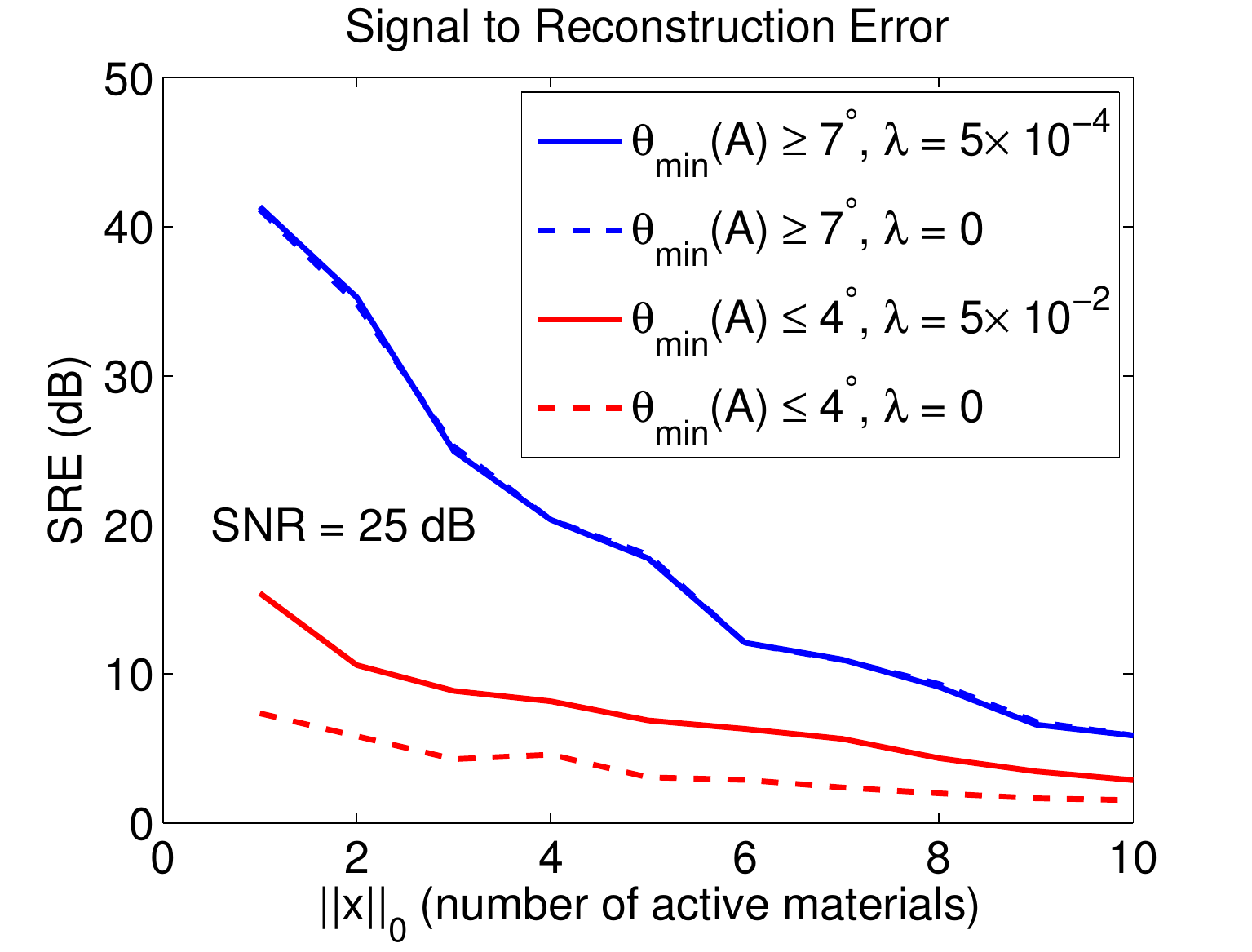}
        \includegraphics[width=80mm]{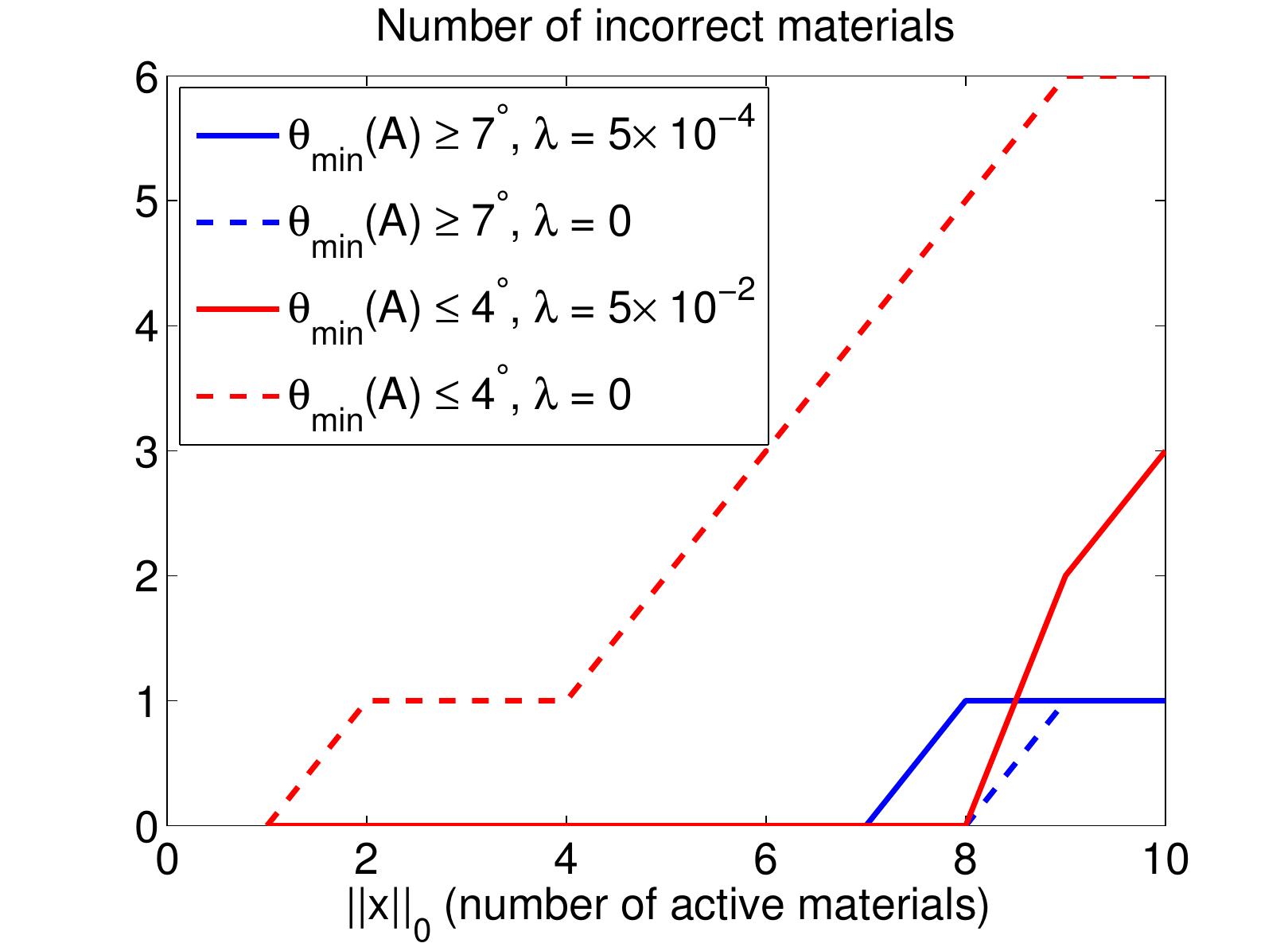}\\
        \caption{Sparse reconstruction results for a simulated data set generated from the USGS library.
                 Top: Signal to reconstruction error (SRE) as a function of the number of active materials.
                 Bottom: Number of  incorrect selected material as a  function of the number of active materials.}
      \label{fig:SR}
\end{figure}

In order to illustrate the potential of the sparse regression methods, we run an experiment with
simulated data. The hyperspectral  library  $\bf A$, of size $L=224$ and $m=342$, is a pruned  version of the USGS library in which the angle between any two spectral signatures is no less than $0.05\,$rad (approximately 3 degrees).
The abundance fractions  follows a Dirichlet  distribution with constant parameter of value 2,  yielding a  mixed data set beyond the reach of geometrical algorithms. In order to put in evidence the impact of the angles between the library vectors, and therefore the mutual coherence of the library \cite{bruckstein2008uniqueness},  in the unmixing results, we organize the library into two subsets; the  minimum angle between any two spectral signatures is higher the $7^\circ$ degrees in the first set and  lower than  $4^\circ$  in the second set.

Fig. \ref{fig:SR} top, plots  unmixing results obtained by solving the CSR problem (\ref{eq:UP1}) with the SUNSAL algorithm introduced in \cite{conf:sunsal:10}. The regularization parameter $\lambda$ was hand tuned for optimal
performance. For each value of the abscissa $\|{\bf x}\|_0$, representing the number of active columns of $\bf A$, we select $\|{\bf x}\|_0$  elements of one of the subsets above referred to and generate $n=1000$   Dirichlet distributed mixtures. From the sparse regression results, we  estimate the {\em signal-to-reconstruction error} (SRE) as
\[
   \text{SRE (dB)} \equiv 10\,\text{log}_{10}\left(\frac{\langle \|{\bf x} \|^2 \rangle}
                          {{\langle \|\widehat{\bf x}-{\bf x} \|^2 \rangle}}\right),
\]
where $\widehat{\bf x}$ and $\langle\cdot\rangle$  stand for estimated abundance fraction vector and  sample average,
respectively.

The curves on the top left hand side were obtained with the noise set to zero.  As expected there is a degradation
of performance as $\|{\bf x}\|_0$ increases and $\theta_m$ decreases.  Anyway, the obtained  values of
SRE  correspond to an almost perfect reconstruction  for $\theta_m({\bf A})\geq 7^\circ$.
For $\theta_m({\bf A})\leq 3^\circ$    the reconstruction is almost perfect
for $\|{\bf x}\|_0\leq 20$, as well, and of good quality for most unmixing purposes for  for $\|{\bf x}\|_0 > 20$.

The curves on the top right  hand side were obtained with SNR$\,=25\,$dB.  This  scenario is  much
more challenging than the previous one. Anyway, even for $\theta_{min} \leq 4^\circ$, we get SRE$\,\gtrsim 10\,$ dB
for, $\|{\bf x}\|_0\leq 5$, corresponding to a useful performance in HU applications. Notice that best values of SRE
for $\theta_{min} \leq 4^\circ$ are obtained  with $\lambda = 5\times 10^{-2}$, putting in evidence the regularization
effect of the $\ell_1$ norm in the CSR problem (\ref{eq:UP1}), namely when the spectral are strongly coherent.

The curves on the bottom  plot the number of incorrect selected materials for SNR$\,=25\,$dB. This  number is  zero for SNR$\,=\infty$. For each value of $\|{\bf x}\|_0$, we compare the $\|{\bf x}\|_0$ larger elements of $\widehat{\bf x}$ with the true  ones and count the number of mismatches.  We conclude that a suitable  setting of the regularization
parameter yields a  correct selection of the materials for $\|{\bf x}\|_0\lesssim 8$.\\

The success of  hyperspectral sparse regression relies crucially  on  the availability  of  suitable
hyperspectral  libraries. The acquisition of these libraries is often a time consuming
and expensive procedure. Furthermore, because the libraries are hardly  acquired under the same conditions
of the data sets under consideration, a delicate calibration procedure  have to be carried
out to  adapt either the library to the data set or vice versa \cite{art:sparseUnmix:TGARS:10}. A way to sidestep these difficulties is the  learning of the libraries  directly from the  dataset with no other a priori information involved. For the application of these ideas, frequetly termed {\em dictionary learning}, in signal and image processing see, {\em e.g.,}, \cite{elad2006image,aharon2006rm} and references therein).  Charles {\em et al.} have recently applied  this line of attack to   sparse  HU in \cite{charlessparse}.  They have modified an existing unsupervised
learning algorithm  to learn an optimal  library under the sparse representation moldel. Using this learned library they have shown that the sparse representation model learns spectral signatures of materials in the scene and locally approximates nonlinear manifolds for individual materials.

%% file: spatial_ND_v2.tex


\section{Spatial-spectral contextual information}


Most of the unmixing strategies presented in the previous paragraphs
are based on a objective criterion generally defined in the
hyperspectral space. When formulated as an optimization problem
(e.g., implemented by the geometrical-based algorithms detailed in
Section \ref{sec:unmix}), spectral unmixing usually relies on
algebraic constraints that are inherent to the observation space
$\mathbb{R}^B$: positivity, additivity and minimum volume.
Similarly, the statistical- and sparsity-based algorithms of
Sections \ref{sec:statistical_methods} and \ref{sec:sparse_regress}
exploit similar geometric constraints to penalize a standard
data-fitting term (expressed as a likelihood function or quadratic
error term). As a direct consequence, all these algorithms ignore
any additional contextual information that could improve the
unmixing process. However, such valuable information can be of great
benefit for analyzing hyperspectral data. Indeed, as a prototypal
task, thematic classification of hyperspectral images has recently
motivated the development of a new class of algorithms that exploit
both the spatial and spectral features contained in image. Pixels
are no longer processed individually but the intrinsic $3$D nature
of the hyperspectral data cube is capitalized by taking advantage of
the correlations between spatial and spectral neighbors (see, e.g.
\cite{Fauvel2008,Tarabalka2009tgrs,Tarabalka2010,Tarabalka2010b, liSpectralSpatial2012,
li2011active,li2010semisupervised, borges2011bayesian}).\\

Following this idea, some unmixing methods have targeted the
integration of contextual information to guide the endmember
extraction and/or the abundance estimation steps. In particular, the
Bayesian estimation setting introduced in Section
\ref{sec:statistical_methods} provides a relevant framework for
exploiting spatial information. Anecdotally, one of the earliest
work dealing with linear unmixing of multi-band images (casted as a
soft classification problem) explicitly attempts to highlight
spatial correlations between neighboring pixels. In \cite{Kent1988},
abundance dependencies are modeled using Gaussian Markov random
fields, which makes this approach particularly well adapted to unmix
images with smooth abundance transition throughout the observed
scene.

In a similar fashion, Eches \emph{et al.} have proposed to exploit
the pixel correlations by using an underlying membership process.
The image is partitioned into regions where the statistical
properties of the abundance coefficients are homogeneous
\cite{Eches2011tgrs}. A Potts-Markov random field has been assigned
to hidden labeling variables to model spatial dependencies between
pixels within any region. It is worthy to note that, conditionally
upon a given class, unmixing is performed on each pixel individually
and thus generalizes the Bayesian algorithms of \cite{Dobigeon2008}.
In \cite{Eches2011tgrs}, the number of homogeneous regions that
compose the image must be chosen and fixed \emph{a priori}. An
extension to a fully unsupervised method, based on nonparametric
hidden Markov models, have been suggested by Mittelman \emph{et al.}
in \cite{Mittelman2012}.\\

Several attempts to exploit spatial information have been also made
when designing appropriate criteria to be optimized. In addition to
the classical positivity, full additivity and minimum volume
constraints, other penalizing terms can be included in the objective
function to take advantage of the spatial structures in the image.
In \cite{Jia2007tgrs}, the spatial autocorrelation of each abundance
is described by a measure of spatial complexity, promoting these
fractions to vary smoothly from one pixel to its neighbors (as in
\cite{Kent1988}). Similarly, in \cite{Zare2011igarss}, spatial
information has been incorporated within the criterion by including
a regularization term that takes into account a weighted combination
of the abundances related to the neighboring pixels. Other
optimization algorithms operate following the same strategy (see for
examples \cite{Zare2010,Zare2011fuzzy,Huck2010whispers}).\\

Extended morphological operations have been also used as a
baseline to develop an automatic morphological endmember extraction
(AMEE) algorithm \cite{plaza:amee:02} for spatial-spectral endmember
extraction. Spatial averaging of spectrally similar endmember
candidates found via singular value decomposition (SVD) was used in
the spatial spectral endmember extraction (SSEE) algorithm
\cite{rogge2007iss}. Recently, a spatial preprocessing (SPP)
algorithm \cite{zorteaplaza09} has been proposed which estimates,
for each pixel vector in the scene, a spatially-derived factor that
is used to weight the importance of the spectral information
associated to each pixel in terms of its spatial context. The SPP is
intended as a preprocessing module that can be used in combination
with an existing spectral-based endmember extraction algorithm.

Finally, we mention very recent research directions  aiming at exploiting
contextual information under the sparse  regression framework.
Work  \cite{conf:guospie:09} assumes that the endmembers are known  and
formulates a deconvolution problem,  where a Total Variation  regularizer \cite{rudin1992nonlinear}
is applied to the spatial bands to enhance their resolution. Work \cite{zymnis2007hyperspectral}
formulates the HU problem as   nonconvex optimization problem
similar to the  nonnegative matrix factorization (\ref{eq:mvt_NMF}), where the
volume regularization term is replaced with an $\ell_1$ regularizer applied to
differences between  neighboring  vectors of abundance fractions.
The limitation imposed  to the  sparse regression methods by the usual high correlation
of the hyperspectral signatures is  mitigated in  \cite{iordache2011total, art:spaseTV:TGARS:12} by adding the Total Variation \cite{rudin1992nonlinear} regularization term, applied to the individual bands,
to CSR problem (\ref{eq:UP1}).  A related approach is followed in
\cite{iordache:collaborative:igarss:12}; here  a  collaborative  regularization term
\cite{sprechmann2011c} is added to CSR problem (\ref{eq:UP1}) to enforce the same set of active materials in all
pixels of the  data set.